\documentclass[twocolumn,superscriptaddress,longbibliography,aps,pra,preprintnumbers]{revtex4-1}

\usepackage{graphicx}
\usepackage{bm}
\usepackage{color}
\usepackage{epstopdf}
\usepackage{amsmath}
\usepackage{amssymb}
\usepackage{epstopdf}

\usepackage[urlcolor=blue,colorlinks=true,citecolor=blue,linkcolor=blue,pdfstartview={FitH},bookmarks=false]{hyperref}

\graphicspath{{fig/}{./fig/}{.}}

\begin{document}

\title{On controlling the bound states in quantum-dot hybrid-nanowire}

\author{Andrzej Ptok}
\email[e-mail: ]{aptok@mmj.pl}
\affiliation{Institute of Nuclear Physics, Polish Academy of Sciences, \\ 
ul. E. Radzikowskiego 152, PL-31342 Krak\'{o}w, Poland}
\affiliation{Institute of Physics, M. Curie-Sk\l{}odowska University, \\ 
pl. M. Sk\l{}odowskiej-Curie 1, PL-20031 Lublin, Poland}

\author{Aksel Kobia\l{}ka}
\email[e-mail: ]{akob@kft.umcs.lublin.pl}
\affiliation{Institute of Physics, M. Curie-Sk\l{}odowska University, \\ 
pl. M. Sk\l{}odowskiej-Curie 1, PL-20031 Lublin, Poland}

\author{Tadeusz Doma\'{n}ski}
\email[e-mail: ]{doman@kft.umcs.lublin.pl}
\affiliation{Institute of Physics, M. Curie-Sk\l{}odowska University, \\ 
pl. M. Sk\l{}odowskiej-Curie 1, PL-20031 Lublin, Poland}

\date{\today} 

\begin{abstract}
Recent experiments using the quantum dot coupled to the topological superconducting nanowire [M.T. Deng {\it et al.}, 
\href{http://doi.org/10.1126/science.aaf3961}{Science {\bf 354}, 1557 (2016)}]
revealed that zero-energy bound state coalesces from the Andreev bound states.
Such quasiparticle states, present in the quantum dot, can be controlled  by the magnetic and electrostatic means.
We use microscopic model of the quantum-dot--nanowire structure to reproduce the experimental results, applying the Bogoliubov--de~Gennes technique. 
This is done by studying the gate voltage dependence of the various types of bound states and mutual influence between them.
We show that the zero energy bound states can emerge from the Andreev bound states in topologically trivial phase and can be controlled using various means.
In non-trivial topological phase we show the possible resonance between this zero energy levels with Majorana bound states.
We discuss and explain this phenomena as a result of dominant spin character of discussed bound states.
Presented results can be applied in experimental studies by using the proposed nanodevice.
\end{abstract}


\maketitle

\section{Experimental introduction}
\label{sec.intro}

Boundaries of the low dimensional topological superconductors can host the zero-energy Majorana 
bound states (MBS)~\cite{read.green.00,kitaev.01,fu.kane.08}. 
Topological protection and non-Abelian 
statistics obeyed by such exotic guasiparticles make them appealing candidates for realization 
of stable qubits which could be useful for quantum computing~\cite{nayak.simon.08,sau.tewari.10,
alicea.oreg.11,rainis.loss.12,vanheck.akhmerov.12,sticlet.bena.12,
aasen.hell.16}. Intensive 
studies of the topological superconductors provided evidence for the MBS in various nano-devices~\cite{mourik.zuo.12,
das.ronen.12,deng.yu.12,rokhinson.liu.12,churchill.fatemi.13,
finck.vanharlingen.13,feldman.randeria.17,niechele.drachmann.17,
nadjperge.drozdov.14,krogstrup.ziino.15,
chang.albrecht.15,albrecht.higginbotham.16,gul.zhang.17,deng.vaitiekenas.16} which are tunable by the gate 
potentials and magnetic field, as have been demonstrated by 
M.T. Deng {\it et al.} in Ref.~\cite{deng.vaitiekenas.16}.

In practice the topologically non-trivial phase can be induced in nanoscopic systems 
via the superconducting proximity effect in cooperation with some additional effects, 
e.g.the spin orbit coupling (SOC) and Zeeman splitting for semiconducting 
nanowires~\cite{krogstrup.ziino.15,chang.albrecht.15}.
Such phenomena have been indeed reported for InAs-Al semiconductor--superconductor 
nanostructures~\cite{chang.albrecht.15} or at interface between the semiconducting 
InSb nanowire and NbTiN superconductor~\cite{gul.zhang.17}. 
Another possible setup for this phenomena is a nanowire with proximity induced superconducting gap, due to the deposited on the surface of superconductor~\cite{yazdani.jones.97}. 
This has been reported i.e. in the case of Fe~\cite{nadjperge.drozdov.14,pawlak.kisiel.16} or Co~\cite{ruby.heinrich.17} atoms on the Pb surface.

The Andreev bound states (ABS) induced in the nanowire spectrum can be varied by the external magnetic field~\cite{chevallier.sticlet.12,chevallier.simon.13}.
In some range of parameters~\cite{sato.fujimoto.09,sato.takahashi.10,chen.yu.16}, above critical magnetic field transition from trivial to non-trivial topological phase occurs.
One pair of such ABS merge at the zero-energy, giving rise to the (double degenerate)  MBS,
which is localized 
near the nanowire ends.

\begin{figure}[!b]
\centering
\includegraphics[width=0.6\linewidth]{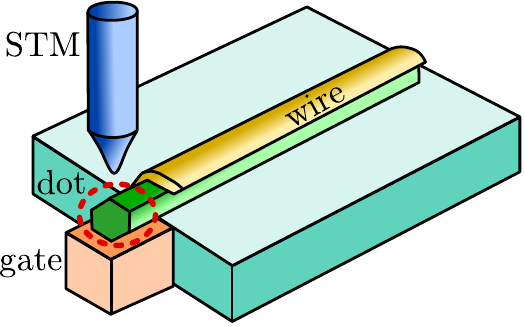}
\caption{
Schematic representation of the experimental system discussed in Ref.~\cite{deng.vaitiekenas.16}. 
InAs wire (green) is epitaxially covered by the superconducting Al (yellow). Quantum dot (InAs) 
is formed between the normal contact (dark orange) and the epitaxial Al shell (inside dashed circle). 
Magnetic field applied parallel to the wire axis can control the bound states. Measurements of the
differential conductance has been done using STM tip (blue), whereas the quantum dot energy levels 
have been tuned by the gate potential.
\label{fig.schem}
}
\end{figure}

Recent experimental results of the Copenhagen group~\cite{deng.vaitiekenas.16}, 
showed that the ABS/MBS can be induced in a controllable way in the quantum dot
region side-coupled to the semiconductor-superconductor hybrid-nanowire. 
Schematic of this structure is displayed in Fig.~\ref{fig.schem}. The semiconducting InAs wire was epitaxially covered by the conventional Al superconductor~\cite{gazibegovic.car.17}, except 
of a small piece of wire which was interpreted as the quantum dot (QD).
The thickness of the superconducting shell should be comparable to its coherence length, as some non-trivial finite-size effects can occur if this condition is not met~\cite{reeg.loss.2017}.
Upon varying the magnetic field and the gate potential there have been induced
the bound states of either the Andreev (Shiba) or the exotic Majorana type, 
as shown by peaks in the differential conductance of the tunneling 
current~\cite{prada.sanjose.12,chevallier.klinovaja.16,stenger.stanescu.17}. 
In particular, the QD energy levels can be varied by the gate voltage 
eventually leading to emergence of the zero-energy Majorana mode.

The main purpose of this paper is to explore the Andreev and Majorana bound states of the single and multiple quantum dots coupled to the hybrid-nanowire. 
We study their evolution with respect to the electrostatic (gate) potential, magnetic field and the chemical potential.
This paper is organized as follows. In Sec.~\ref{sec.model} we introduce the model and present some computational details concerning the Bogoliubov--de~Gennes technique. 
Next, in Sec.~\ref{sec.wires_noQD} we describe basic properties and main terminology in relation to studied problem.
Thorough discussion of the quasiparticle spectrum of the single QD is presented in Sec.~\ref{sec.num.1dot}.
Revision of more general system with higher number of sites in QD, is performed in Sec.~\ref{sec.num.2dots} and Sec.~\ref{sec.num.multi-dot}, which are devoted to the double and multi-site QDs cases, respectively. 
In Sec.~\ref{sec.num.device} we propose a feasible quantum device, which could enable an experimental realization of various tunable  bound states.
Finally, in Sec.~\ref{sec.sum} we summarize the results.

\section{Model and methods}
\label{sec.model}

\begin{figure}[!b]
\centering
\includegraphics[width=0.7\linewidth]{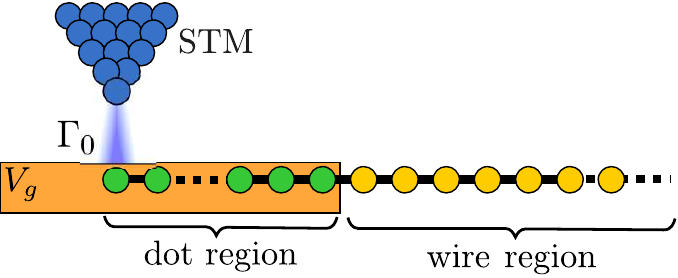}
\caption{
Schematic idea of the described system. The sites of the quantum dot (green) are side-attached 
to the superconducting nanowire (yellow) with the proximity-induced electron pairing. Using the STM tip 
(blue) we can measure the LDOS at each site of the system. Parameter $\Gamma_{0}$ denotes the coupling 
strength between the STM tip and the probed atom.
}
\label{fig.schem2}
\end{figure}

For description of the nanostructure shown in Fig.~\ref{fig.schem}, we will use microscopic model in real space with Hamiltonian  $\mathcal{H} = \mathcal{H}_{w} + \mathcal{H}_{prox} + \mathcal{H}_{soc} + \mathcal{H}_{dot}$. The first term describes mobile electrons in the wire
\begin{eqnarray}
\mathcal{H}_{w} = \sum_{ij\sigma} \left\lbrace - t \delta_{\langle i,j \rangle} - \left( \mu + \sigma h \right) \delta_{ij} \right\rbrace c_{i\sigma}^{\dagger} c_{j\sigma} ,
\end{eqnarray}
where $t$ denotes a hopping integral between the nearest-neighbor sites, $\mu$ is a chemical potential, and $h$ denotes  a magnetic field parallel to the whole wire.
Here $c_{i\sigma}^{\dagger}$ ($c_{i\sigma}$) describes creation (annihilation) operator in site {\it i}-th with spin $\sigma$.
The second term accounts for the proximity effect
\begin{eqnarray}
\mathcal{H}_{prox} = \sum_{i} \Delta \left( c_{i\downarrow} c_{i\uparrow} +   c_{i\uparrow}^{\dagger} c_{i\downarrow}^{\dagger}  \right) 
\end{eqnarray}
and we assume the uniform energy gap $\Delta$ induced by the epitaxially covered classical superconductor.
The spin-orbit coupling (SOC) term is given by
\begin{eqnarray}
\mathcal{H}_{soc} &=& - i \lambda \sum_{i \sigma\sigma'} c_{i\sigma}^{\dagger} ( \sigma_{y} )_{\sigma\sigma'} c_{i+1,\sigma'} ,
\end{eqnarray}
where $\sigma_{y}$ stands for {\it y}-component of the Pauli matrix and $\lambda$ is the SOC coupling along the chain. 
The we treat the QD as part of a nanowire not covered by superconductor.
The last part
\begin{eqnarray}
\mathcal{H}_{dot} &=& \sum_{i \in \text{dot},\sigma} V_{g} c_{i\sigma}^{\dagger} c_{i\sigma} 
\end{eqnarray}
describes the electrostatic energy contributed by the gate potential $V_{g}$ (see Fig.~\ref{fig.schem2}).
In what follows we shall consider the quantum dot region comprising one, two and multiple sites coupled 
to the superconducting nanowire.

Hamiltonian $\mathcal{H}$ of the entire chain can be diagonalized by the Bogoliubov--Valatin transformation~\cite{degennes.89}
\begin{eqnarray}
c_{i\sigma} &=& \sum_{n} \left( u_{in\sigma} \gamma_{n} - \sigma v_{in\sigma}^{\ast} \gamma_{n}^{\dagger} \right) ,
\end{eqnarray}
where $\gamma_{n}$, $\gamma_{n}^{\dagger}$ are the  quasiparticle fermionic operators and $u_{in\sigma}$ and $v_{in\sigma}$ are the Bogoliubov--de~Gennes (BdG) eigenvectors, respectively. Such unitary transformation implies  
\begin{eqnarray}
\label{eq.bdg}  && \mathcal{E}_{n}
\left(
\begin{array}{c}
u_{in\uparrow} \\ 
v_{in\downarrow} \\ 
u_{in\downarrow} \\ 
v_{in\uparrow}
\end{array} 
\right) =\\
\nonumber &=& \sum_{j} \left(
\begin{array}{cccc}
H_{ij\uparrow} & D_{ij} & S_{ij}^{\uparrow\downarrow} & 0 \\ 
D_{ij}^{\ast} & -H_{ij\downarrow}^{\ast} & 0 & S_{ij}^{\downarrow\uparrow} \\ 
S_{ij}^{\downarrow\uparrow} & 0 & H_{ij\downarrow} & D_{ij} \\ 
0 & S_{ij}^{\uparrow\downarrow} & D_{ij}^{\ast} & -H_{ij\uparrow}^{\ast}
\end{array} 
\right)
\left(
\begin{array}{c}
u_{jn\uparrow} \\ 
v_{jn\downarrow} \\ 
u_{jn\downarrow} \\ 
v_{jn\uparrow}
\end{array} 
\right) ,
\end{eqnarray}
where $H_{ij\sigma} = - t \delta_{\langle i,j \rangle} - ( \mu + \sigma h - V_{G} \delta_{i\in\text{dot}} ) \delta_{ij}$ is the single-particle term, 
$D_{ij} = \Delta \delta_{ij}$ refers to the induced on-site pairing, 
and the SOC term (mixing the particles with different spins) is given by 
$S_{ij}^{\sigma\sigma'} = - i \lambda ( \sigma_{y} )_{\sigma\sigma'} \delta_{\langle i,j \rangle}$, 
where $S_{ij}^{\downarrow\uparrow} = ( S_{ji}^{\uparrow\downarrow} )^{\ast}$.

To study our system, we will use the local density 
of states (LDOS) defined as $\rho_{i} ( \omega ) = - \frac{1}{\pi} \sum_{\sigma} \mbox{Im} \langle 
\langle c_{i\sigma} | c_{i\sigma}^{\dagger} \rangle \rangle$.
From numerical solution of the BdG equations~(\ref{eq.bdg}) we obtain the Green's function
$\langle \langle c_{i\sigma} | c_{i\sigma}^{\dagger} \rangle \rangle$,
which formally gives
\begin{eqnarray}
\rho_{i} ( \omega ) = \sum_{n\sigma} \left[ | u_{in\sigma} |^{2} \delta \left( \omega - \mathcal{E}_{n} \right) + | v_{in\sigma} |^{2} \delta \left( \omega + \mathcal{E}_{n} \right) \right] .
\label{eq.ldos}
\end{eqnarray}
This physical quantities can be measured experimentally in relatively simply way~\cite{matsui.sato.03,figgins.morr.10}.
In practice this spatially and energy dependent spectrum can be also probed by a differential conductance 
$G_{i}(V) = dI_{i}(V)/dV$ of the tunneling current $I_{i}(V)$, 
which depends on the coupling  between $i$-th atom of the wire and the STM tip~\cite{tersoff.hamann.85} (indicated by $\Gamma_{0}$ in Fig.~\ref{fig.schem2}).

We have solved the BdG equations~(\ref{eq.bdg}) for a chain with $N=200$ sites, choosing $\Delta/t = 0.2$, $\lambda/t=0.15$, $\mu/t=-2$. 
For numerical purposes we have also replaced the Dirac delta functions appearing in Eq.~(\ref{eq.ldos}) by a Lorentzian 
$\delta ( \omega ) = \zeta / [ \pi ( \omega^{2} + \zeta^{2} ) ]$
with a small broadening $\zeta = 0.0025t$.

\section{Basic properties} 
\label{sec.wires_noQD}

\begin{figure}[!b]
\centering
\includegraphics[width=\linewidth]{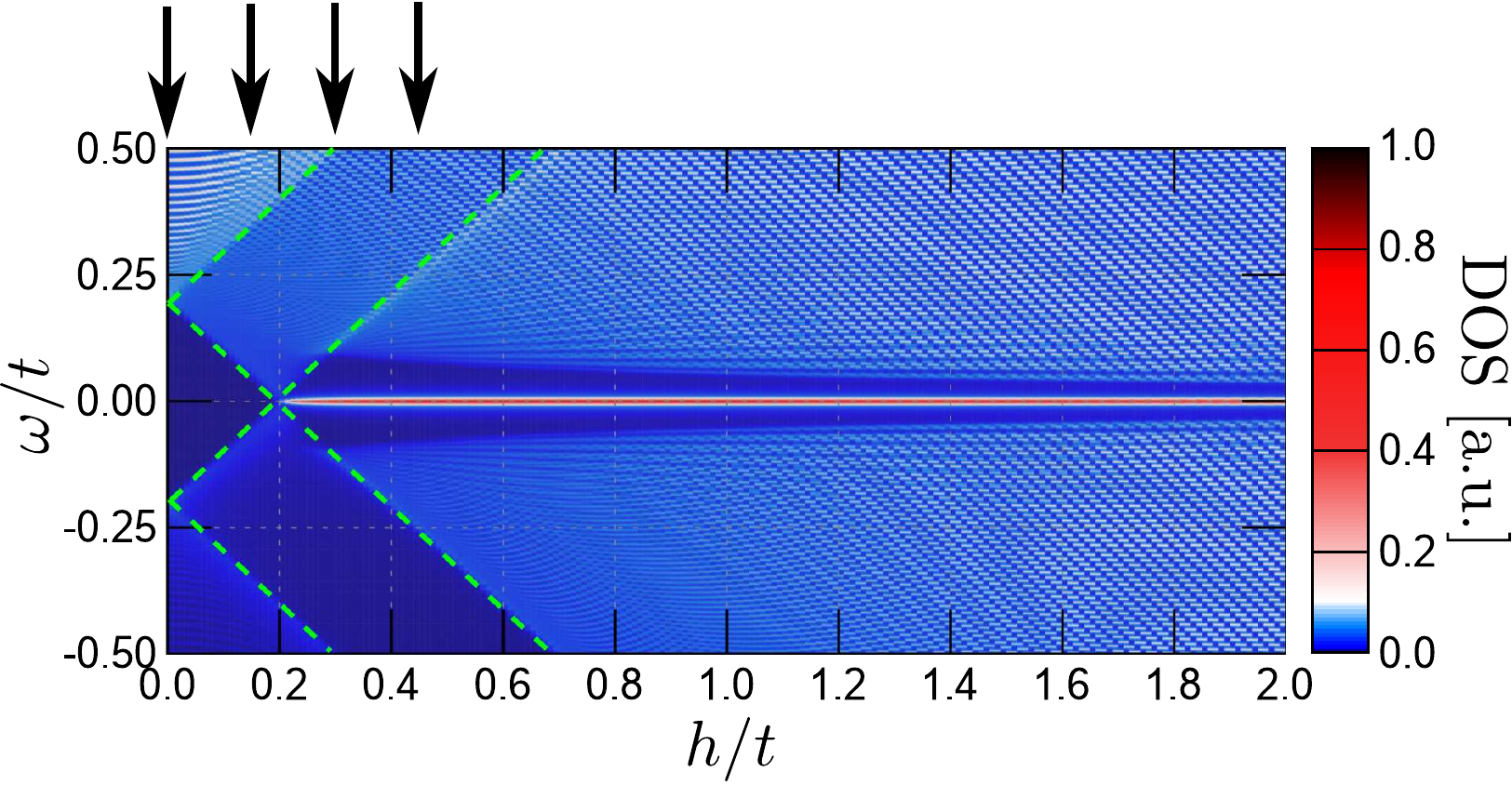}
\caption{
Total DOS for chain in magnetic field. Result for $k_{B}T=0t$, $\mu = -2t$, 
$\lambda = 0.15t$ and $\Delta = 0.2t$ with $V_{g}=0t$.
Black arrows represent specific values of $h$, indicated for further analysis.
Green lines specify the regions of Zeeman shifted induced superconducting gap by proximity effect.
\label{fig.dosnodot}
}
\end{figure}

In this section, we will briefly describe basic physical properties of the nanowire without coupled QD. 
We will also define terminology which will be used in the further sections of manuscript.

As we mentioned in Sec.~\ref{sec.intro}, in a case of wires with SOC and superconductivity induced by the proximity effect, for some magnetic field $h_{c}$ phase transition from trivial to non-trivial topological phase occurs.
In a case of one dimensional chain described by the Hamiltonian $\mathcal{H}$ defined in Sec.~\ref{sec.model} we have $h_{c} = \sqrt{ \Delta^{2} + ( 2 t \pm \mu )^{2} }$~\cite{sato.fujimoto.09,sato.takahashi.10}.
For chosen parameters we have $h_{c} / t = 0.2$.

Change in magnetic field $h$ leads to the typical evolution of the total density of states (DOS) for this case.
Numerical calculation for chosen parameters is shown in Fig.~\ref{fig.dosnodot}.
In consequence, due to the finite size effect, we can observe a separate line in the DOS. This line correspond to singular state of wire~\cite{chevallier.simon.13}.
Characteristic structure of the DOS restricted by the asymptotic line (shown by dashed green line) will be explained below.
As we can see, when magnetic field crosses the critical value $h_{c}$, previously closed superconducting gap is reopened partly as a {\it topological gap}~\cite{klinovaja.loss.12}.

\begin{figure}[!b]
\centering
\includegraphics[width=\linewidth]{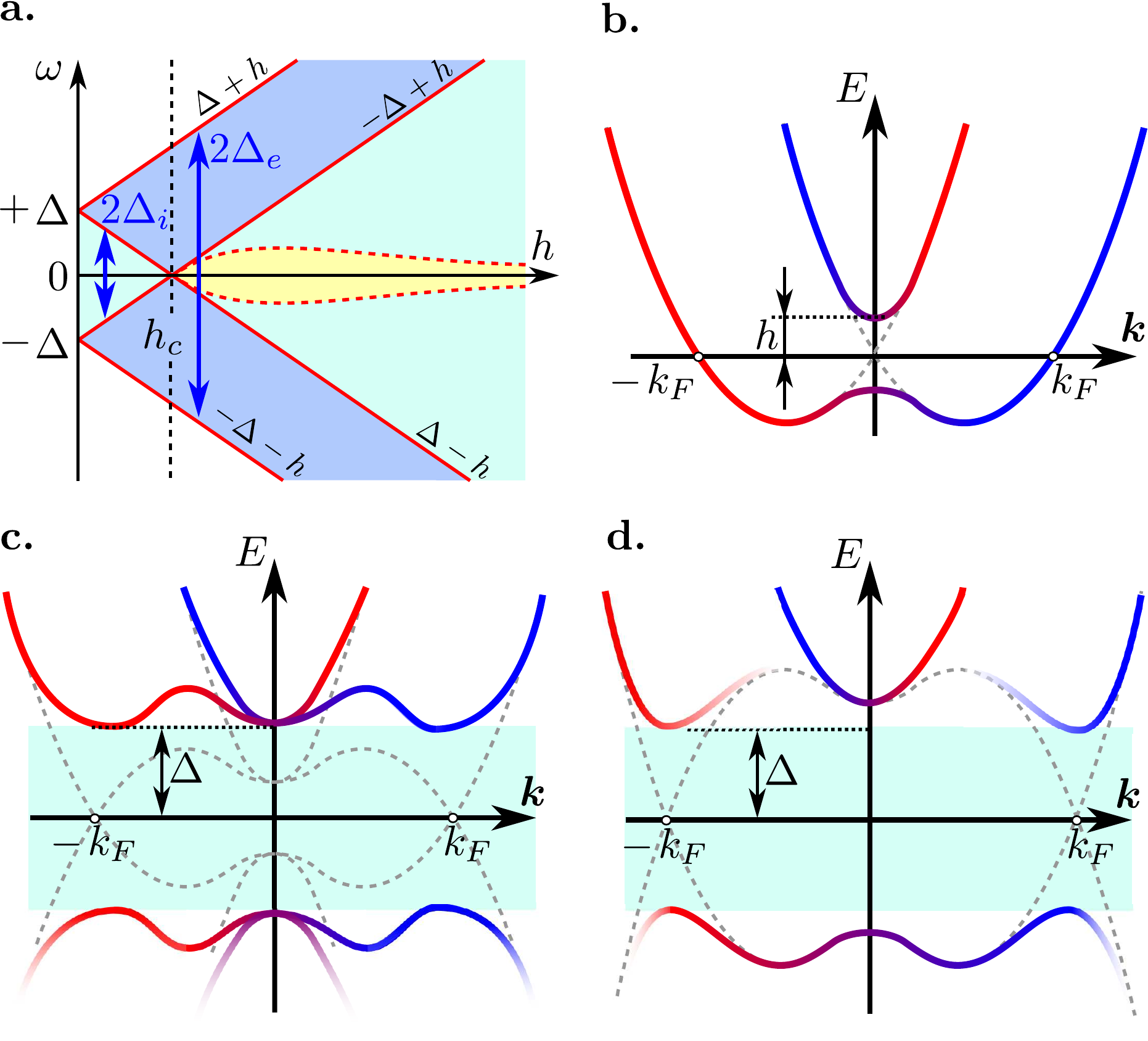}
\caption{
On the panel a we show schematic representation of shifted superconducting gaps by magnetic field $h$. 
$2\Delta_{e}$ and $2\Delta_{i}$  denote external and internal gaps respectively.
Region for magnetic fields smaller (bigger) than $h_{c}$ describe trivial (non-trivial) topological phase.
Panels b-d show band structure in the presence of magnetic field without (b) and with (c,d) superconductivity:
in case of trivial (c) and non-trivial (d) topological phases, where green region $2 \Delta$ represents the superconducting gap.
Grey dashed line (b-d) represents band structure in the absence of superconductivity.
\label{fig.schem6}
}
\end{figure}

Now we will introduce previously mentioned terminology, by referring to Fig.~\ref{fig.schem6}.a which schematically shows a change of the DOS by magnetic field $h$.
In described system, superconducting gap $\Delta$ in wire experimentally corresponds to {\it hard gap} induced by proximity effects~\cite{chang.albrecht.15,
albrecht.higginbotham.16,gul.zhang.17}, 
which value depends on the coupling between semiconductor wire with superconducting shell or base~\cite{cole.dassarma.15}.
In consequence of this, for $h = 0$ we observe a $2\Delta$ gap in the DOS.
Increasing $h$ leads to energy levels $\omega = \pm \Delta$ shift (red lines).
In this situation, similar like in Ref.~\cite{klinovaja.loss.12} we can define {\it exterior gap} $2 \Delta_{e} = 2 ( \Delta + h )$ and {\it interior gap} $2 \Delta_{i} = 2 ( \Delta - h )$ as a energy spacing between external and internal asymptotic line respectively (blue double-arrows).
For $0 < h < h_{c}$ the internal gap decreases, creating {\it soft gap} with value smaller than {\it hard gap}.
Finally, this superconducting gap is closed in $h_{c}$, while for $h > h_{c}$ {\it topological gap} reopens (yellow region between dashed red lines).
Note, that increasing the SOC leads to increased {\it topological gap}~\cite{sau.tewari.10}.

Experimentally observed {\it hard gap} depends on magnetic field~\cite{das.ronen.12}, which is approximately described by the BCS-like relation $\Delta ( h ) \simeq \Delta \sqrt{ 1 - ( h / h_{c2} )^{2} }$~\cite{liu.sau.17b}, where $h_{c2}$ denotes upper critical magnetic field of superconductor (magnetic field in which {\it hard gap} will be closed).
This dependence effectively leads to experimentally observed suppression of in-gap bound states.
However, we assume constant value of $\Delta$, which does not change interpretation of presented results.

Phase transition from trivial to non-trivial phase, characterized by $\mathbb{Z}_{2}$ topological invariant~\cite{kane.mele.05,qi.zhang.11,zhang.nori.16}, 
can be described in relation to band structure of infinite wire with periodic boundary conditions (Fig.~\ref{fig.schem6}.b-d)
~\cite{das.ronen.12,klinovaja.loss.12,oreg.refael.10,reeg.maslov.17}. 
In absence of the superconducting gap, the external magnetic filed $h$ leads to the gap opening and lifts spin degeneracy at momentum $k = 0$ (b).
Induction of the superconductivity in the wire, opens additional gap around the Fermi level $E = 0$ (horizontal axis).
Relation between $\Delta$ and $h$, corresponding to gap opening due to superconductivity and magnetic field respectively, defines the topologically trivial (c) and non-trivial (d) regimes.
In the trivial topological phase $h < h_{c}$ (c), new gap at the Fermi momentum $\pm k_{F}$ emerges and also increases gap at the $k = 0$ because $\Delta > h$ (in accord with ''positive'' value of the {\it interior gap} $2 \Delta_{i}$).
Situation looks differently in a non-trivial topological phase regime $h > h_{c}$ (d), when $\Delta < h$ (what corresponds to ''negative'' value of $2 \Delta_{i}$).
In this situation, opening of the superconducting gap at $\pm k_{F}$ does not change the character of gap at $k = 0$.
Moreover, from formal point of view, in our system a non-trivial {\it p-wave} pairing between quasiparticles from this same band is induced.
This possibility has been described before~\cite{sato.fujimoto.09,sato.takahashi.10,
gorkov.rashba.01,zhang.tewari.08,alicea.10,seo.han.12,yu.wu.16}.

However, in the absence of the boundary conditions (finite wire), discussion of the band structure is unreasonable because momentum is not a good quantum number.
Moreover, energy of {\it bound states} occurring at the {\it boundaries} of the wire, has symmetrical shape with respect to Fermi energy $\omega = 0$.
Non-zero magnetic field applied in the system leads to emergence of ABS in-gap states (with energies $\Delta > | \omega | > \Delta - h$) and ABS with lowest energy defines the boundary of the {\it interior gap}.
Increasing $h$ to value above $h_{c}$ allows the MBS to form from two lowest energy ABSs.
Simultaneously, when the lowest energy ABSs merged into MBS, the {\it topological gap} is created between new lowest energy ABSs.

In non-trivial topological phase ($h > h_{c}$) the zero energy MBS, can be experimentally observed i.e. in a form zero-bias peaks in the tunnelling conductance measurement~\cite{figgins.morr.10,gibertini.taddei.12,liu.potter.12,chevallier.klinovaja.16}.
In this type of experiments, the MBS is observed in the form of the zero-bias conductance peak $G_{0} = 2e^{2}/h$ at zero temperature.
However, in finite temperature regime conductance is significantly reduced, which has been observed experimentally~\cite{niechele.drachmann.17} and discussed theoretically~\cite{liu.potter.12,vanheck.lutchyn.16,liu.sau.17,danon.hansen.17,liu.sau.17b}.
Therefore, local density of states presented here is a good indicator for the differential conductance~\cite{stenger.stanescu.17}, however it strongly depends on temperature and coupling between tip and nanowire~\cite{liu.sau.17,niechele.drachmann.17,danon.hansen.17,devillard.chevallier.17,liu.setiawan.17,reeg.maslov.17,setiawan.liu.17}.

Moreover, the MBS are physically localized at the end of the wire.
Length of the wire plays important role in realization MBS wavefunction oscillation in space, which is connected to the MBS non-locality~\cite{prada.aguado.17}.
When considering a sufficiently short wire, overlapping of the two Majorana wavefunctions is too extensive and the ''true'' zero-energy MBS cannot be realised, as the MBS annihilate~\cite{potter.lee.10,dassarma.sau.12,liu.15}. 
This system requires a meticulously made nanowire~\cite{zhang.gul.17}, because any disorder has destructive role on the topological phase~\cite{more.stanescu.16,cole.sau.16,
hegde.vishveshwara.16,zhang.nori.16,awoga.bjornson.17}.
However, local impurity can lead to MBS separation into the pair of new MBS at the {\it newly} created boundaries of the homogeneous system in topological states~\cite{liu.drummond.12,xu.mao.14,liu.15,maska.gorczycagoraj.17,ptok.cichy.17}.

As we mentioned in Sec.~\ref{sec.intro}, the ABSs can be experimentally controlled. 
Moreover, for some experimental parameter the ABS can coalesce~\cite{deng.vaitiekenas.16} into {\it zero energy bound state} (ZEBS).
This feature is realised only in non-trivial topological phase ($h < h_{c}$).
Because the ZEBS and MBS are zero-energy states, we must mention the differences between those two similar kind of bound states.
Firstly, magnetic field in which ZEBS ($h < h_{c}$) coalesce is smaller than the one required for MBS to emerge ($h > h_{c}$).
Secondly, what is more important from practical point of view, ZEBS do not obey the non-Abelian statistics which is a consequence of different parity with respect to MBS~\cite{sticlet.bena.12,hegde.vishveshwara.16}.

\section{Single quantum dot}
\label{sec.num.1dot}

\begin{figure}[!b]
\centering
\includegraphics[width=\linewidth]{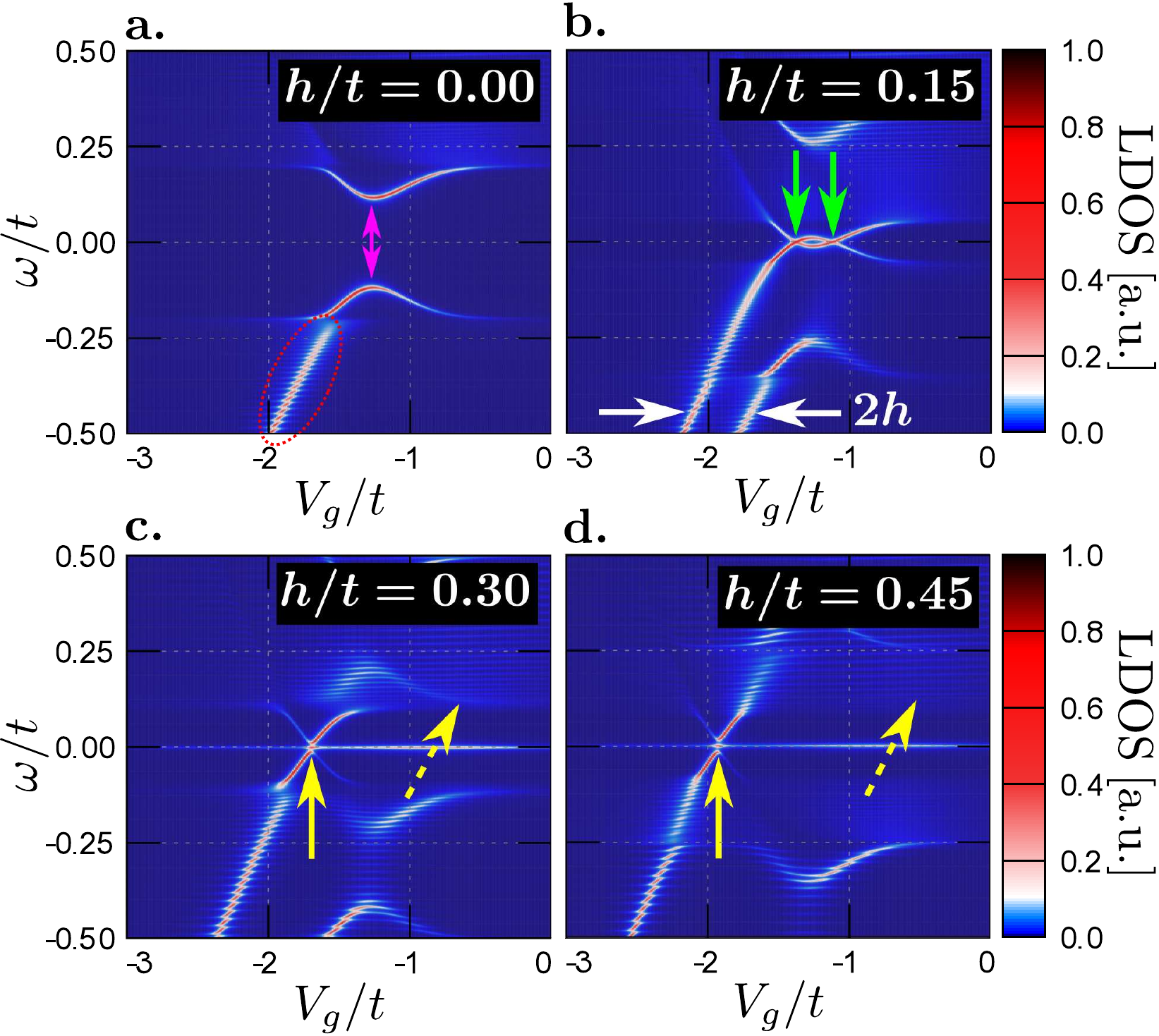}
\caption{
Evolution of the quantum dot spectral function with respect to $V_{g}$ for several magnetic fields, 
indicated by the black arrows in Fig.~\ref{fig.dosnodot}. Results are obtained for $k_{B}T=0t$, $\mu = -2t$, $\lambda = 0.15t$, $\Delta = 0.2t$. Red ellipse on the panel a indicates the {\it devil's staircase} structure.
\label{fig.dosdotvg}
}
\end{figure}

Let us now inspect the superconducting wire comprising $N = 200$ sites with one additional site, representing the normal QD.
Evolution of this QD spectrum with respect to the gate voltage $V_{g}$ is illustrated in Fig.~\ref{fig.dosdotvg} for several magnetic fields $h$.
In absence of the magnetic field (panel a) and for $V_{g}/t\leq -1.8$ the QD quasiparticles show up in LDOS as characteristic {\it devil's staircase} (red ellipse on panel a).
This avoided crossing structure occurs as a consequence of hybridization of the QD energy level with finite number of the nanowire energy levels.
In the regime $V_{g}/t \in \left( -1.8, 0.8\right)$ there appear two ABSs inside the {\it hard gap}, which never cross each other (as is indicated by the pink double-arrow).

For the $h < h_{c}$ in the trivial topological phase (panel b), we observe the Zeeman splitting of the initially single spin-degenerate QD levels (white arrows on panel b).
In consequence, majority spin character for both levels have been disjointed (character of ''left'' and ''right'' levels corresponds to majority spin $\downarrow$ and $\uparrow$ quasiparticles respectively).
Moreover, when magnetic filed is strong enough, the ABS can cross each other creating ZEBS at two different values 
of $V_{g}$, depending on $h$ (indicated by the green arrows).
Characteristic spin-split 
structure have been also observed~\cite{pillet.quay.10,dirks.hughes.11,pillet.joyez.13,
lee.jiang.14,chang.albrecht.15}.

For strong magnetic field $h > h_{c}$, at the non-trivial topological phase (panels c and d), the MBS emerge in the nanowire.
Let us remark, that such Majorana quasiparticles, for some range of parameters, coexist with the conventional ABS inside {\it topological gap}, whose spectral weights depend on $h$ and $V_{g}$.
Modification of the QD energy level with dominant $\sigma$-spin character by $V_{g}$ leads to two different kinds of \textit{resonance} with MBS.
In a case of the $\uparrow$-like state (in the region indicated by yellow dashed arrow) {\it leak} of the MBS into the QD has been observed, whereas for $\downarrow$-like state there is only a relatively weak resonance (yellow arrows).

\begin{figure}[!b]
\centering
\includegraphics[width=0.95\linewidth]{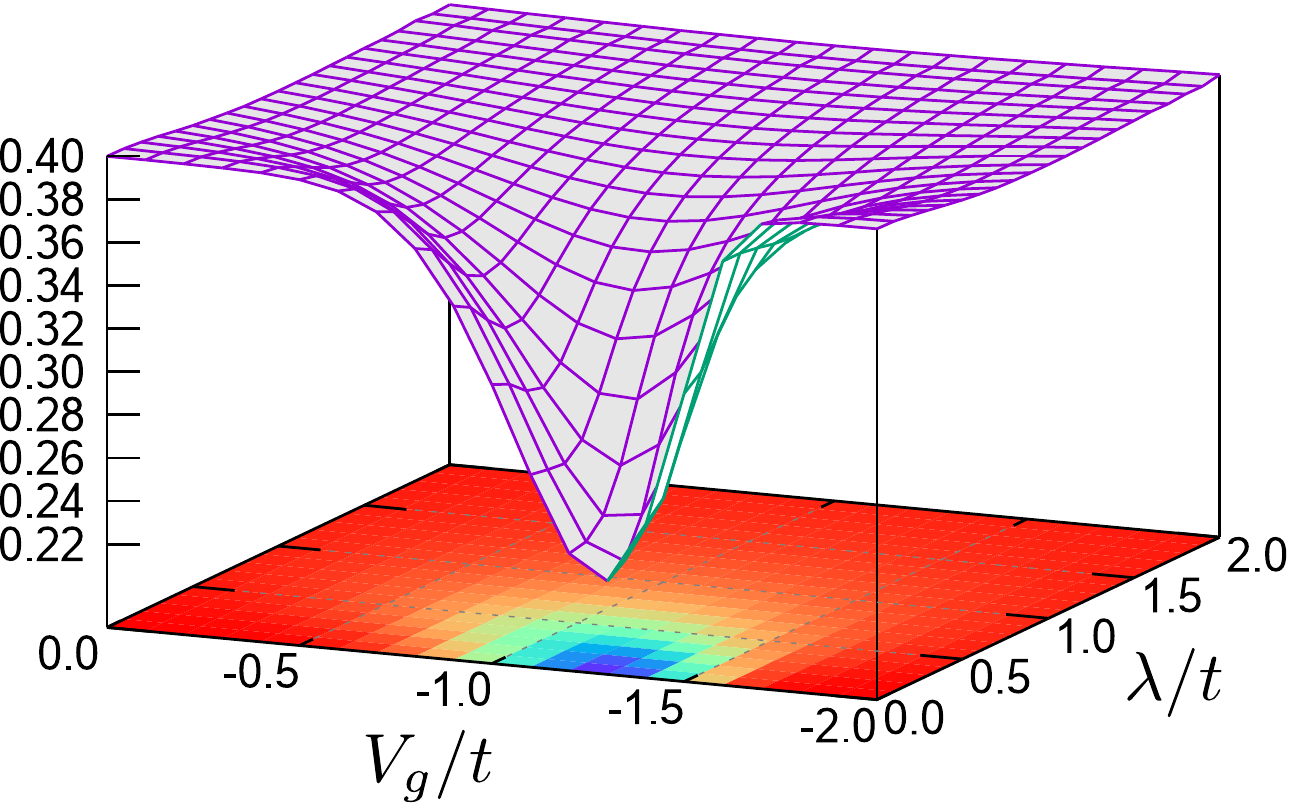}
\caption{
Effective gap between the ABS (inside the {\it hard gap}) versus the spin orbit coupling $\lambda$ and the gate potential $V_{g}$. 
Results are obtained for the single quantum dot at zero temperature for $\mu = - 2 t$, $\Delta = 0.2 t$ and $h = 0$.
\label{fig.absgap}
}
\end{figure}

It should be mentioned that, the possible crossing of the ABS in the absence of the magnetic field is possible when ratio coupling between the QD and nanowire would induce a {\it hard gap} (in our case $t / \Delta$) that is smaller than one~\cite{pillet.joyez.13}.
This scenario can be also realized at the quantum phase transition in the correlated quantum dot~\cite{pillet.joyez.13,baranski.domanski.13,lee.jiang.14,prada.aguado.17} but such issue is beyond a scope of the present study.
For parameters chosen in our system we have $ t / \Delta \gg 1$ and the gap between two ABS inside {\it hard gap} could not observed (Fig.~\ref{fig.absgap}).
In the case studied here, the minimum of gap mentioned above, occurs at $V_{g} \sim -1.3 t$, whereas its extreme value $2 \Delta$ is reached either away when gate potential is insignificant or for the strong SOC $\lambda$.
As we can see in Fig.~\ref{fig.dosdotvg}, the ABS crossing can be achieved
for some fixed gate potential $V_{g}$, in presence of the magnetic field $h$ which is equal to a half of the gap between the ABS when magnetic field is absent.


\subsection{Resonance of the quantum dot levels with Majorana bound states}
\label{sec.res.qd_mbs}

\begin{figure*}[!hp]
\centering
\includegraphics[width=0.8\linewidth]{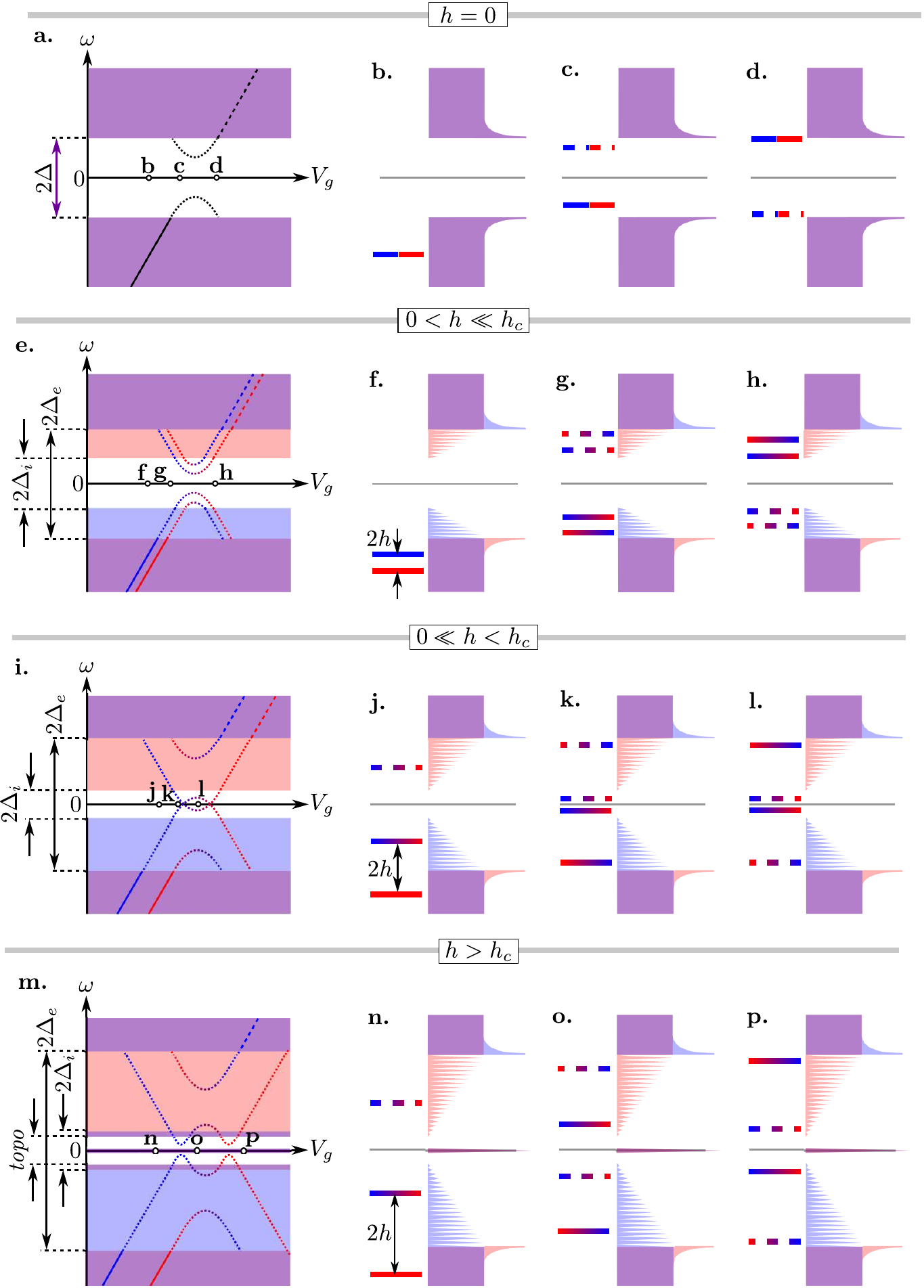}
\caption{
Schematic representation of the {\it resonance} of the dot energy levels and the nanowire with increasing magnetic field (from top to bottom). 
For every case in the most-left column, a solid, dashed and dotted lines represent occupied, unoccupied and Andreev bound states respectively. 
Moreover, gray dashed axis line shows the Fermi level and letters on those axes in most-left column denote the specific gate potential $V_{g}$ on quantum dot.
In the rest of the columns a solid (dashed) line indicate quasiparticles with dominant particle (hole) character.
Colors (red/blue/violet) illustrate the dominant spin ($\uparrow$/$\downarrow$/degenerate case) of 
energetic levels.
ABSs (f-h,j-l,n-p) inside the external gap become a mixture of spins, due to the spin orbit coupling, hence the transition in colors representing energy levels. Labels $2\Delta$, $2\Delta_{i}$ $2\Delta_{e}$ and {\it topo} represent the {\it hard gap}, {\it interior gap}, {\it exterior gap} and {\it topological gap} respectively, which have been introduced in Sec.~\ref{sec.wires_noQD}.
}
\label{fig.schem3c}
\end{figure*}

Let us now explain in detail the asymmetry in resonance of the QD energy levels with nonowire energy levels, presented previously in Fig.~\ref{fig.dosdotvg}.
We will do this using the schematic representation of the QD energy levels and nanowire total DOS shown in Fig.~\ref{fig.schem3c} and terminology introduced in Sec.~\ref{sec.wires_noQD}.
In absence of magnetic field (a-d), manipulation of gate voltage $V_{g}$ changes the spin degenerate dot energetic levels with respect to the Fermi level ($\omega = 0$).  
When states are localized below the superconducting {\it hard} gap (b) we can observe the {\it devil's staircase} structure.
This structure is formed as a consequence of coupling between the QD and nanowire energetic levels -- spin conserved ($t$) and spin-flip ($\lambda$) hoppings. 
ABS emerge, when the QD energy levels are near or inside the {\it hard gap} (c,d).
Spectral weight of ABS is leaking from the occupied to non occupied  (c $\rightarrow$ d) levels, converting its character (i.e. see also Fig.~\ref{fig.dosdotvg}.a). 
Initially, negative energy ABS is particle dominated, whereas the positive energy one is hole dominated (c). 
As the change in $V_{g}$ progresses, occupation of states is inverted (d).
For any non-zero magnetic field $h$ (e-p) spin degeneracy is lifted by Zeeman shift. 
When $0 < h < h_{c}$, sharp structures of the ABS are observed in the {\it hard gap}, creating {\it soft gap} which is equal to {\it interior gap} ($2\Delta_{i}$) for this value of magnetic field.
If the $h$ is sufficiently small (e), ABS does not cross Fermi level.
When both QD energy levels are localized below the {\it exterior gap} (f), then we observe two separate levels with different spin majority character (see Fig.~\ref{fig.dosdotvg}.b).
As we increase $V_{g}$, observed mirrored ABS resonances invert its dominant character from particle to hole (g $\rightarrow$ h). 
For high enough magnetic field (but still smaller than $h_{c}$) (i-l) the ABSs start to cross at zero-energy level. 
In consequence ABS coalesce into ZEBS at Fermi level and the {\it interior gap} narrows (i).
For some value of $V_{g}$ (j) only one pair of ABS exist inside {\it exterior gap} while the $\uparrow$ dominant spin level of the QD resides below this gap.
Before first coalescing of ABSs (k) we observe situation similar to (g). 
However, for $V_{g}$ between points of ABS coalescence (l) energy levels invert.
In a case of the $h > h_{c}$ (m-p) the {\it topological gap} opens and the MBS emerge at $\omega = 0$. 
In this non-trivial topological phase (with $h > \Delta$) the dot-energy level is shifted enough to treat it independently.
For $V_{g}$ at the point (n) the QD energy levels with $\uparrow$($\downarrow$) dominant spin character are located deep below (near) {\it topological gap}.
In consequence we observe ''in-topological-gap'' ABS detached from $\downarrow$-spin QD energy level, which suits minority spin in the whole system.
Increase of $V_{g}$ (o $\rightarrow$ p) leads to position of the QD energy level with majority $\uparrow$($\downarrow$)-spin character near (far above) the {\it topological gap} respectively.
Additionally, dominant spin component reverses during the topological phase transition~\cite{szumniak.chevallier.17}.

We must have in mind that quasiparticles with the $\uparrow$($\downarrow$)-spin character has dominant (inferior) role in the whole system due to the Zeeman splitting.
In this sense MBS at zero-energy level have $\uparrow$ spin polarization~\cite{sticlet.bena.12}.
Stemming from this, only remaining $\uparrow$ dominant spin character energy levels can resonate with MBS. 
Following this condition and keeping in mind that the SOC is sufficiently strong in the system, a characteristic structure of avoided crossing occurs, halting the ABS emerged from inferior $\downarrow$-spin QD energy level to cross zero energy level.
Simultaneously, $\uparrow$-spin dominant QD energy level can resonate with MBS, which can be clearly seen as an increasing of spectral weight of the MBS along dashed arrows in Fig.~\ref{fig.dosdotvg}.c and d.


\begin{figure}[!t]
\centering
\includegraphics[width=\linewidth]{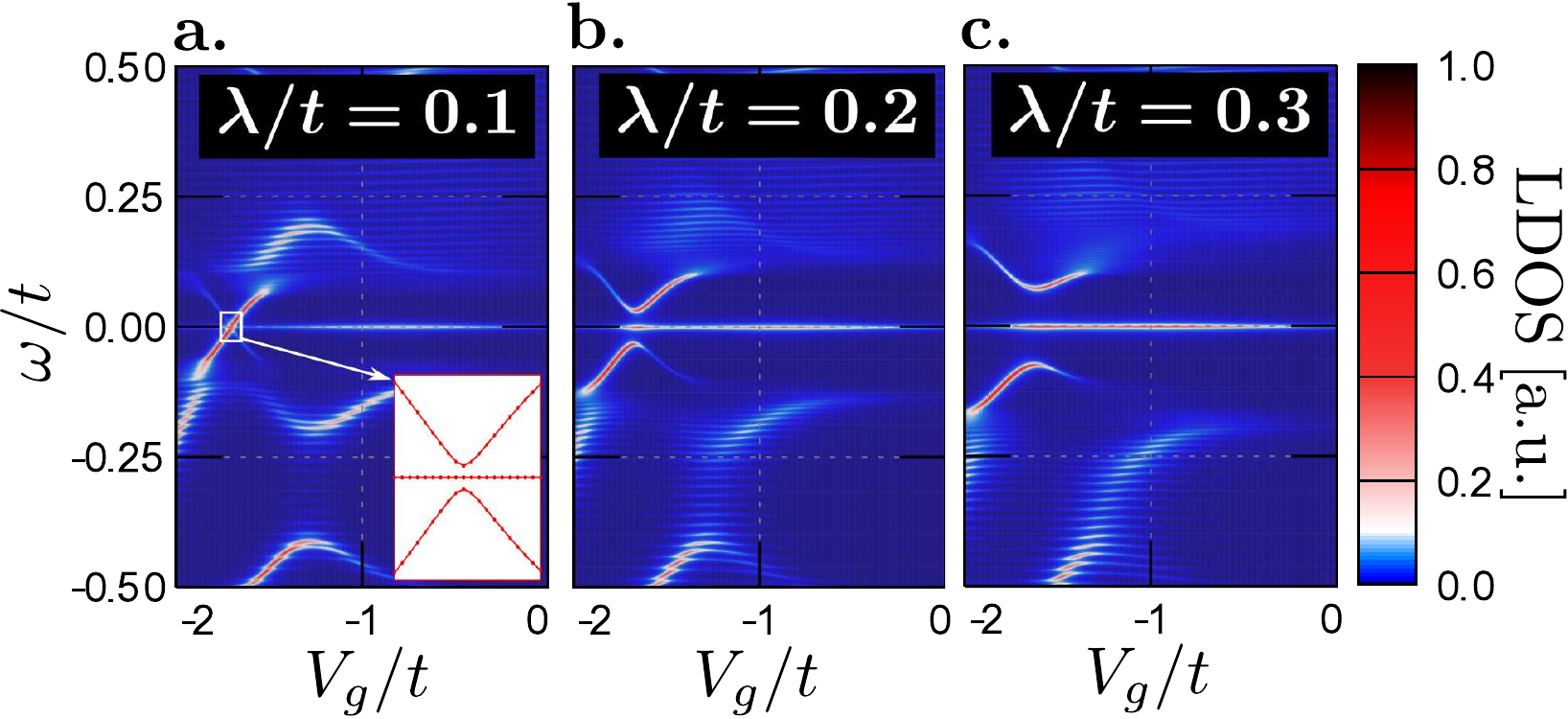}
\caption{
Effect of the spin orbit coupling $\lambda$ on the induced Majorana and Andreev 
bound states. Inset in panel a shows the {\it quasi}particle energies for the zoomed region.
Results are obtained for $k_{B}T=0t$, $\mu = -2 t$, $h = 0.3t$ and $\Delta = 0.2t$.
\label{fig.dotsoc}
}
\end{figure}

As a result of the QD coupling to the wire by spin-conserved $t$ and spin-flip $\lambda$ hopping, the resonance of the QD energy levels with minority $\downarrow$-spin character and MBS with $\uparrow$ polarization,  depends strongly on the spin orbit coupling.
Role of the spin-orbit influence on this behavior is shown Fig.~\ref{fig.dotsoc}, where we compare the resonance of the QD energy levels with the zero-energy MBS for several values of the SOC $\lambda$.
For any non-zero value of $\lambda$, system supports both the MBS and ABS, coexisting inside the {\it topological} gap.
It can be noticed, that the ABS become gapped (see the inset in panel a) and their avoided crossing behavior becomes significant with an increase of SOC strength $\lambda$. 
At the same time, the MBS gain more and more spectral weight.
Furthermore, we also observe constructive influence of the SOC $\lambda$ on the {\it devil's staircase} structure, existing outside the {\it topological} gap.
In relation to previous paragraph, this is a  consequence of spin-flip hybridization between QD and wire, supporting the resonance of the ABS and opposite spin character MBS.


\subsection{Different types of zero energy bound states}

\begin{figure}[!t]
\centering
\includegraphics[width=\linewidth]{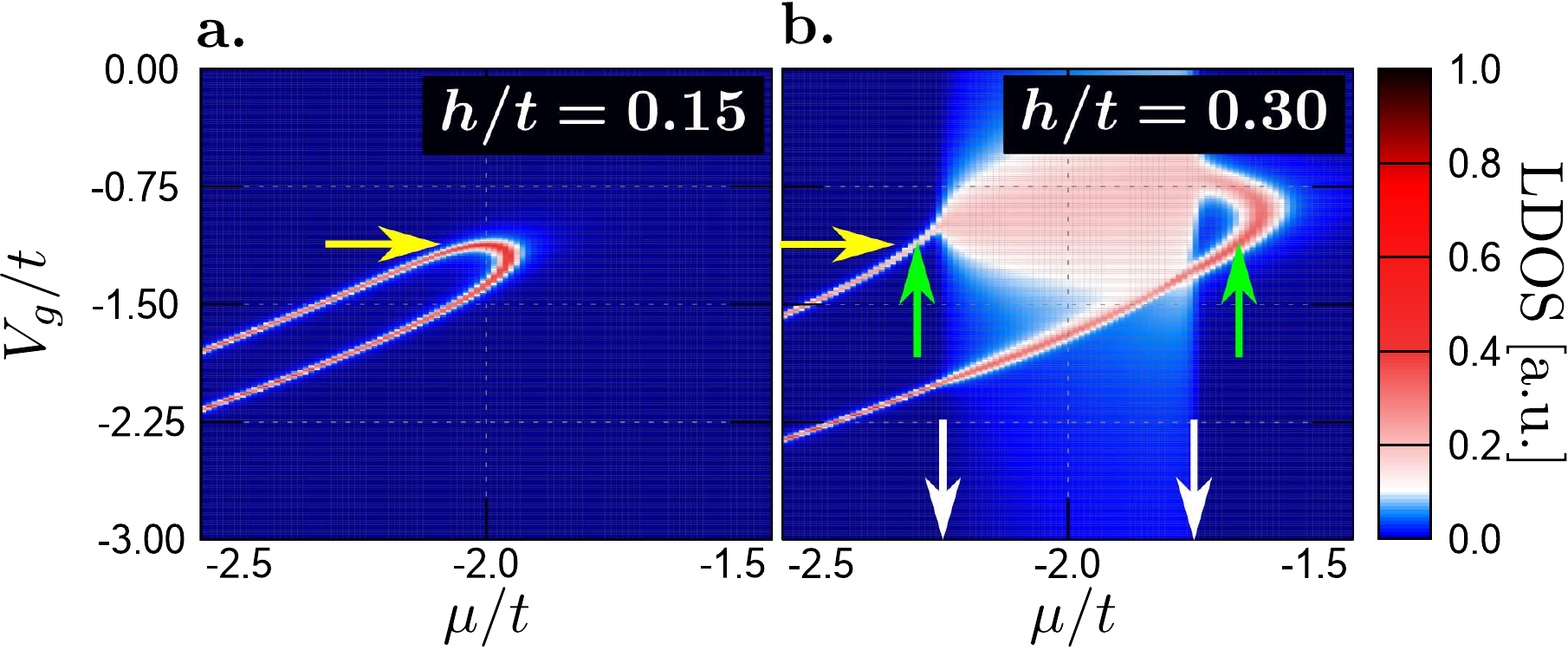}
\caption{
Modification of the zero-energy LDOS on the quantum dot site by change $\mu$ and $V_{g}$ in a cases phase not supported (a) and supported (b) realization of the MBS.
Result for $k_{B}T=0t$, $\lambda = 0.15t$ and $\Delta = 0.2t$
\label{fig.dos1dotmuzero}
}
\end{figure}

We have shown, that the  ABS  can coexist with MBS and sometimes their energies are 
identical (resonant). Such resonance depends on the quantum dot energy level,  which 
can by modified by the global Fermi level (i.e. the chemical potential $\mu$), the 
gate voltage $V_{g}$ and the magnetic field $h$. 
These quantities affect the ABS and for trivial topological phase ($h < h_{c}$) lead to  emergence of the ZEBS.
Here we should remind that the ZEBS and the MBS are zero-energy states, but emerge in different topological phase (trivial and non-trivial respectively).
It is illustrated in Fig.~\ref{fig.dos1dotmuzero}, 
where we plot the LDOS of the dot region for $\omega = 0$ versus $\mu$ and $V_{g}$.

\begin{figure}[!b]
\centering
\includegraphics[width=\linewidth]{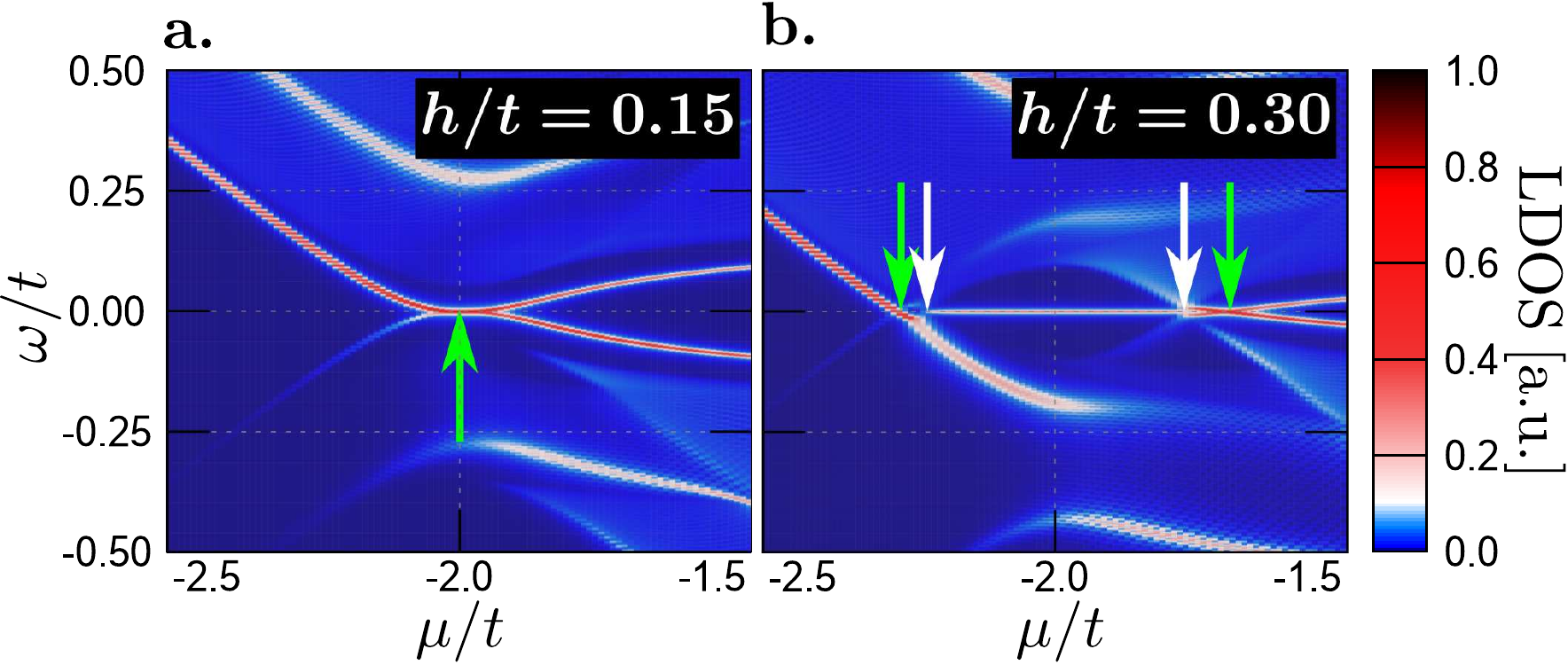}
\caption{
LDOS on the quantum dot versus the chemical potential $\mu$ for the cases not supporting 
(a) and supporting (b) realization of the MBS. The gate potential is $V_{g} = -1.125t$ 
as indicated by the right green arrow in Fig.~\ref{fig.dosdotvg}.b. 
Results are obtained for $k_{B}T=0t$, $\lambda = 0.15t$ and $\Delta = 0.2t$.
\label{fig.dos1dotmu}
}
\end{figure}

These results refer to the following cases: 
{\it (i)} $h < h_{c}$, when MBS are not realized for any parameter of the system (panel a);
{\it (ii)} $h > h_{c}$, when for some values of $\mu$, the system can host the MBS (panel b -- the MBS supporting regime exist between white arrows).
In the first case  we can find such regions, where ABSs coalesce into ZEBS (red kink in panel a). 
For the latter case, upon varying $\mu$ (or $h$) we can distinguish two regimes: 
supporting (between white arrows) and non-supporting (outside white arrows) emergence of the MBS.
Inside the first region we can see realization of the (asymmetric) resonance of the QD energy levels with the MBS hosted at the ends of the nanowire.
In second region, similar like previously, we can only see a crossing of the ABS in the ZEBS form.
Difference between such resonances has been discussed in previous sections.

\begin{figure}[!t]
\centering
\includegraphics[width=\linewidth]{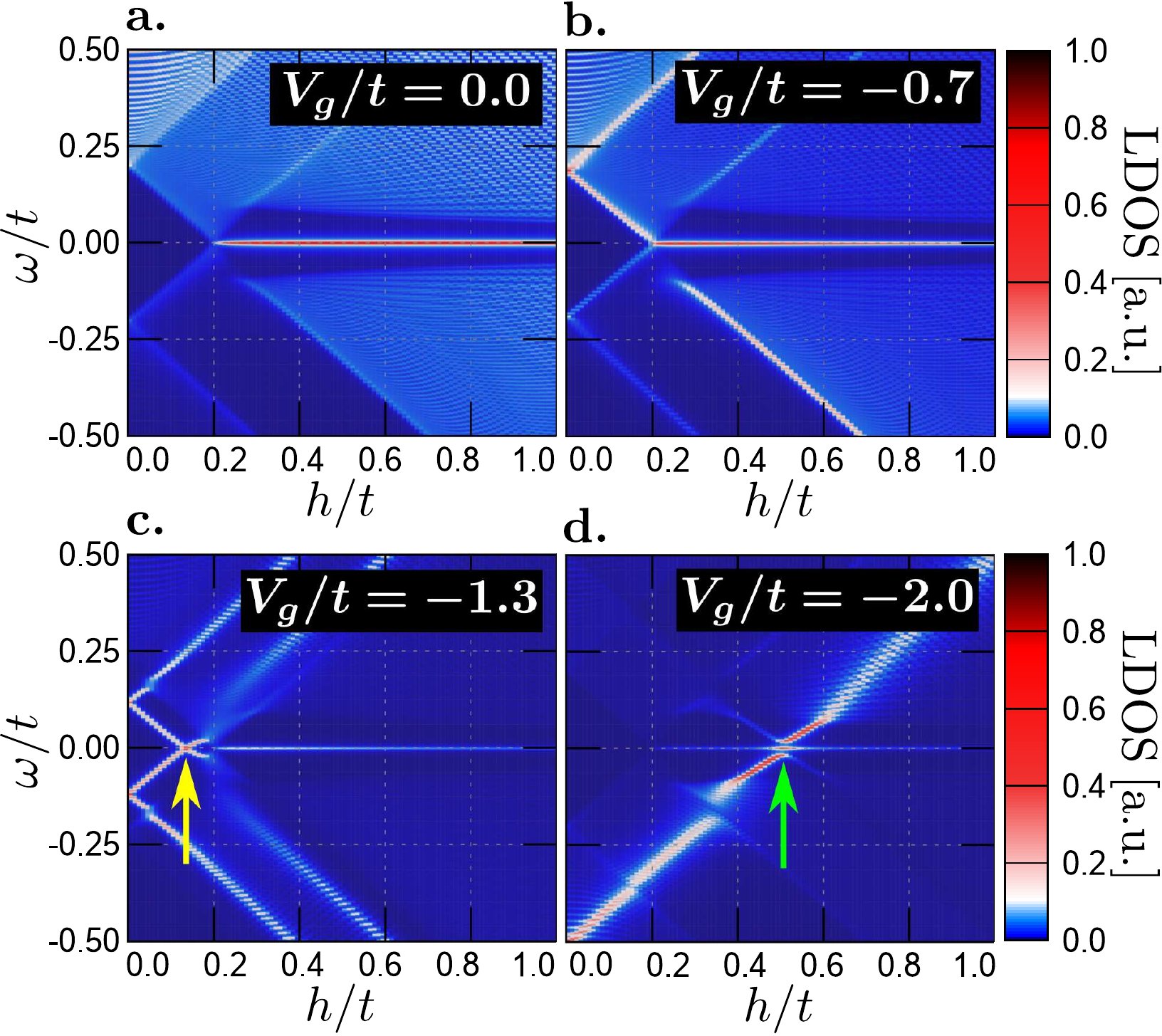}
\caption{
Evolution of the quantum dot spectrum with respect to magnetic field $h$ 
for several gate potentials $V_{g}$, as indicated. Results are obtained for 
$k_{B}T=0$, $\mu = -2t$, $\lambda = 0.15t$ and $\Delta = 0.2t$.
\label{fig.dosdotem}
}
\end{figure}

Following results are discussed for the cross-section of Fig.~\ref{fig.dos1dotmuzero} along 
$V_{g} = -1.125t$ indicated by yellow arrow. 
Fig.~\ref{fig.dos1dotmu} shows the LDOS of the QD as a function of the (global) chemical potential $\mu$.
We can clearly see, that upon varying of $\mu$ the coalescing ABS give rise to ZEBS (panel a -- green arrow).
However, for the non-trivial topological phase (panel b), ZEBS appear only beyond the MBS-supported regime (green arrows outside the region marked by white arrows in panel b).
This results can are complementary to Fig.~\ref{fig.dos1dotmuzero}.b.
Inside this regime there exit the topologically protected Majorana states, while for other parameters the ABSs create new gap around $\omega = 0$.

Further important effects can be seen if we investigate the influence of magnetic field (Fig.~\ref{fig.dosdotem}). 
As we mentioned previously, $h$ detunes the energy levels of the states with opposite spin character.
This is true for whole studied system. 
We remind that in general, for $h < h_{c}$ we can observe the ABS or the ZEBS  coalesced from (i.e. yellow arrow in panel c), what is in agreement with experimental results~\cite{deng.vaitiekenas.16}, whereas the zero-energy MBS can be realized only for $h > h_{c}$.
Similarly to the previous result, increase in $h$ reveals asymmetry in resonance between the QD energy levels with dominant $\sigma$ spin character and MBS, what has been explained in Sec.~\ref{sec.res.qd_mbs}.
In some range of gate potential $V_{g}$ (compare with Fig.~\ref{fig.dosdotvg}), with changed $h$, the dominant $\uparrow$ or $\downarrow$ spin character of the QD energy levels are revealed.
In a case of the energy levels with spin majority character ($\uparrow$), resonance between the QD energy level and MBS is favored by spin-conserving hopping (panels a and b).
For the energy levels with minority character ($\downarrow$), resonance of the QD level and the MBS is more energetically expensive due the fact that the spin-flip hopping $\lambda$ is smaller than spin-conserved hopping $t$.
As a result, we can observe emergence of the ABS in-topological-gap (panel d) and weak resonance with the MBS (green arrow), depending on $\lambda$ (see Fig.~\ref{fig.dotsoc}).
When the QD energy levels penetrate the {\it hard gap} as the ABS (in weak magnetic field, panel c), the ZEBS is formed (yellow arrow).

\section{Double sites quantum dot}
\label{sec.num.2dots}

\begin{figure}[!t]
\centering
\includegraphics[width=\linewidth]{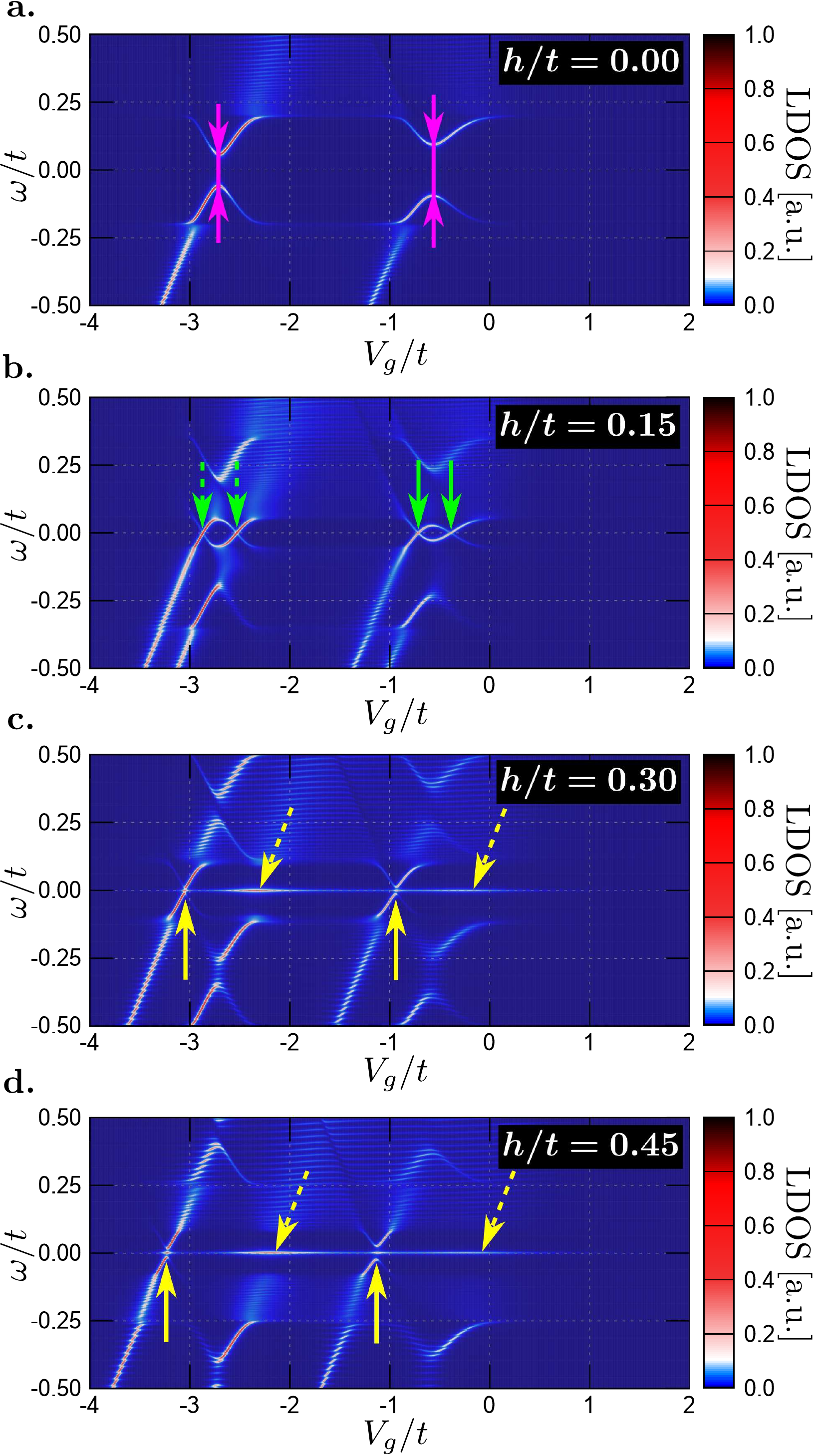}
\caption{
Evolution of the double quantum dot LDOS with respect to the gate voltage $V_{g}$ 
for several magnetic fields, indicated by the white arrows in Fig.~\ref{fig.dosnodot}. 
Results are obtained for $k_{B}T=0t$, $\mu = -2t$, $\lambda = 0.15t$, $\Delta = 0.2t$.
\label{fig.dos2dotvg}
}
\end{figure}

\begin{figure}[!t]
\centering
\includegraphics[width=\linewidth]{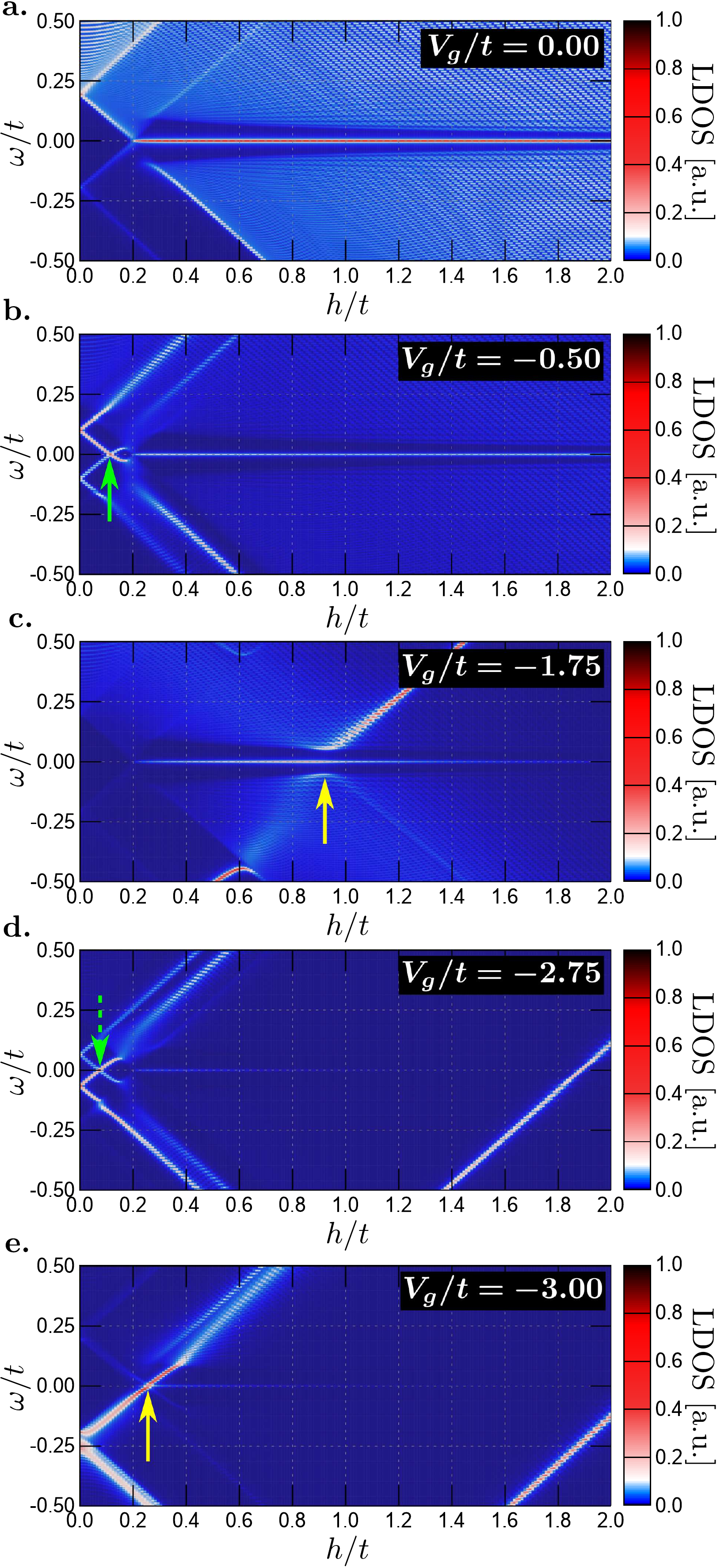}
\caption{
Influnence of the magnetic field $h$ on the LDOS of the double quantum dot for 
different values of the gate voltage $V_{g}$, as indicated. Result are obtained 
for $k_{B}T=0t$, $\mu = -2t$, $\lambda = 0.15t$ and $\Delta = 0.2t$.
\label{fig.dos2dotem}
}
\end{figure}

Similar analysis can be performed for the system comprising two additional sites (double site quantum dot) side-attached to the hybrid-nanowire.
In this case, we observe two pairs of the ABS appearing in the spectrum of such dots (Fig.~\ref{fig.dos2dotvg}). 
We notice, 
that for $h=0$ these pairs of ABSs are split by different energy gaps (indicated by the pink arrows in panel a). 
For this reason, within a range of weak magnetic field $h<h_{c}$ we can observe either one or two pairs of the spin-split ZEBS (marked by the solid and dashed green arrows in panel b).
In the regime of non-trivial topological phase (for $h > h_{c}$) we see emergence of the zero-energy MBS (yellow solid and dashed arrows in panels c and d).
When the MBS (with $\uparrow$ majority spin character of the system) hosted on the wire, coincides with the minority spin character ($\downarrow$) double-site QD energy levels (yellow solid arrows on panels c and d), we can observe its existence in the topological-gap while ABSs do not cross at zero energy level.
In other words, the ABS separate from the zero-energy Majorana mode as a consequence of weak coupling between QD and the strong one  in  the wire due to the spin-flip hopping $\lambda$.
Regarding the case of energy levels with $\uparrow$ spin character (dashed yellow arrows), their bound states do not enter the {\it topological gap} but resonance with the MBS at zero-energy level.
Figure~\ref{fig.dos2dotem} shows the double quantum dot spectrum with as a function of the magnetic field $h$ for several values of the gate voltage $V_{g}$. 
Again, we notice that $V_{g}$ controls the spectral weight of the Majorana mode leaking into the QD region in the non-trivial topological phase (above the critical magnetic field $h > h_{c}$).

\section{Multi-sites quantum dot}
\label{sec.num.multi-dot}

In realistic quantum systems the ABS can sometimes originate from a multitude of the energy levels existing in a subgap regime.
We shall model such situation here, considering a piece of the nanowire (sketched in Fig.~\ref{fig.schem2}) whose energy levels can be identified 
as the finite number of lattice sites in this complex structure.
Such systems can be realized experimentally 
e.g. in the carbon nanotube superconducting device~\cite{pillet.quay.10}. Similar effects can be 
relevant to the experiment reported by M.T. Deng {\it et al.}~\cite{deng.vaitiekenas.16}. Another 
possible realization could refer to the multi-level structure obtained by modern experimental 
technique, designing the quantum dot with atomic precision~\cite{folsch.martinezblanco.14}.
Due to the proximity effect, we can expect appearance of the $N$ pairs
ABS~\cite{pillet.quay.10,chang.albrecht.15,deng.vaitiekenas.16,
albrecht.higginbotham.16}, where N is the number of the sites in the QD region.
In our calculations for multi-level dot we shall focus on $N=10$ sites.

\begin{figure}[!t]
\centering
\includegraphics[width=\linewidth]{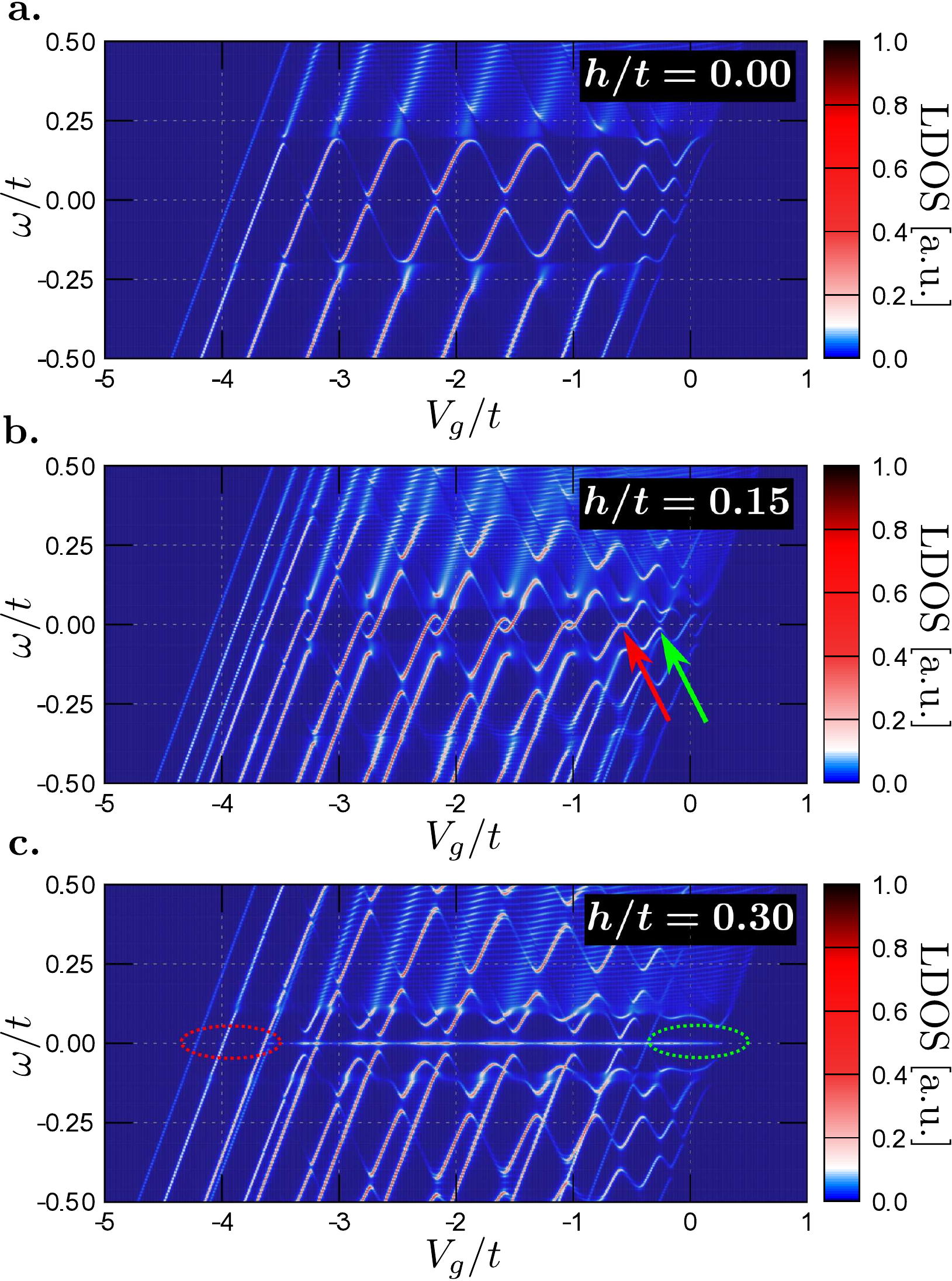}
\caption{
Gate voltage dependence of the LDOS obtained for at zero temperature the quantum dot comprising $10$ sites, using $\mu = -2t$, $\lambda = 0.15t$, $\Delta = 0.2t$.
In panel c, red (green) doted ellipses correspond to region where only the quantum dot energy levels with $\downarrow$ ($\uparrow$) spin character exists.
\label{fig.dos10dotem}
}
\end{figure}

Fig.~\ref{fig.dos10dotem} shows variation of the normal nanowire spectrum with respect to
the gate voltage $V_{g}$ for several magnetic fields $h$, as indicated. For $h=0$ (panel a) 
we observe $N$ {\it quasi}particle branches, which become doubled at low energies (due to particle-hole
mixing). For the weak magnetic field $h<h_{c}$ (panel b) we can observe the Zeeman splitting 
of the initial {\it quasi}particle branches. In a low energy regime these bound states eventually 
reveal either a crossing (red arrow) or avoided crossing (green arrow), depending on the gate
voltage $V_{g}$.
Finally, when hybrid-nanowire transitions to non-trivial topological phase $h > h_{c}$, we can observe resonance of the QD energy levels with MBS hosted in nanowire (panel c), similarly to previous results, but different form for levels with majority or minority spin character.

In consequence of the asymmetric resonances of the QD energy levels with minority and majority spin types, we observe different behavior of this levels in subgap region (panel c).
The QD energy levels with the minority spin ($\downarrow$) character are insensitive to the existence of MBS zero energy level in the wire (red doted ellipse), while in a case of majority spin ($\uparrow$) complete resonance is observed (green doted ellipse).
In a case of the intermediate $V_{g}$ regime (between red and green doted ellipses), the spectral weight of the MBS weakly
oscillates with a varying $V_{g}$ as a consequence of various interplay between the QD energy levels, depending on their dominant spin component.

\begin{figure}[!pt]
\centering
\includegraphics[width=\linewidth]{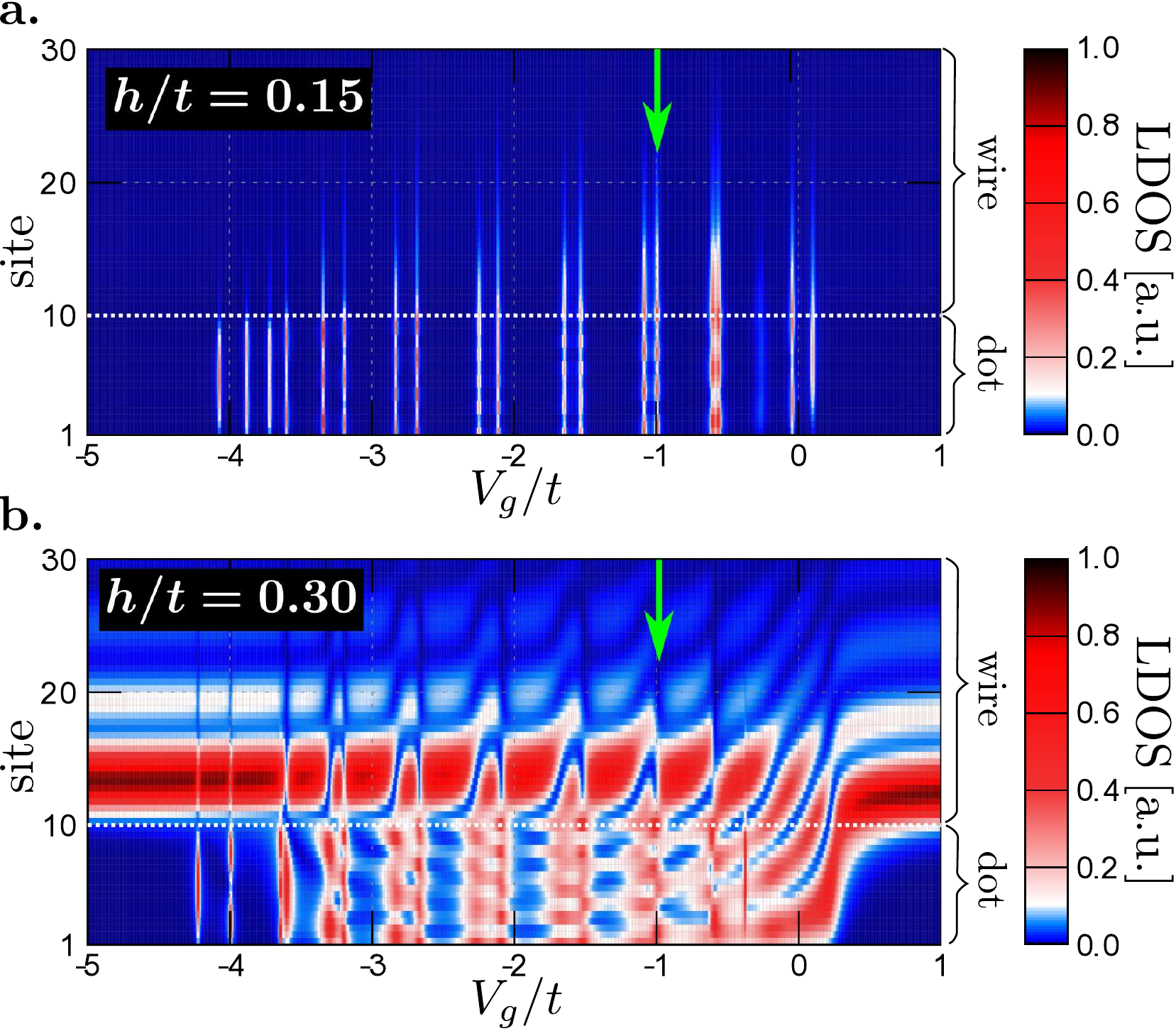}
\caption{Spatial profiles of the ABS and/or MBS obtained at $\omega = 0$ for 
the normal nanowire, comprising 10 sites. Panel a shows the ABS of the trivial
superconducting state ($h < h_{c}$), whereas panel b illustrates spatial profiles 
of the MBS in the topologically nontrivial superconducting state ($h>h_{c}$). 
Results are obtained for $k_{B}T=0t$, $\mu = -2t$, $\lambda = 0.15t$ and $\Delta = 0.2t$.
Doted white line shows the boundary between the quantum dot and wire regions.
\label{fig.10local}
}
\end{figure}

\begin{figure}[!pb]
\centering
\includegraphics[width=\linewidth]{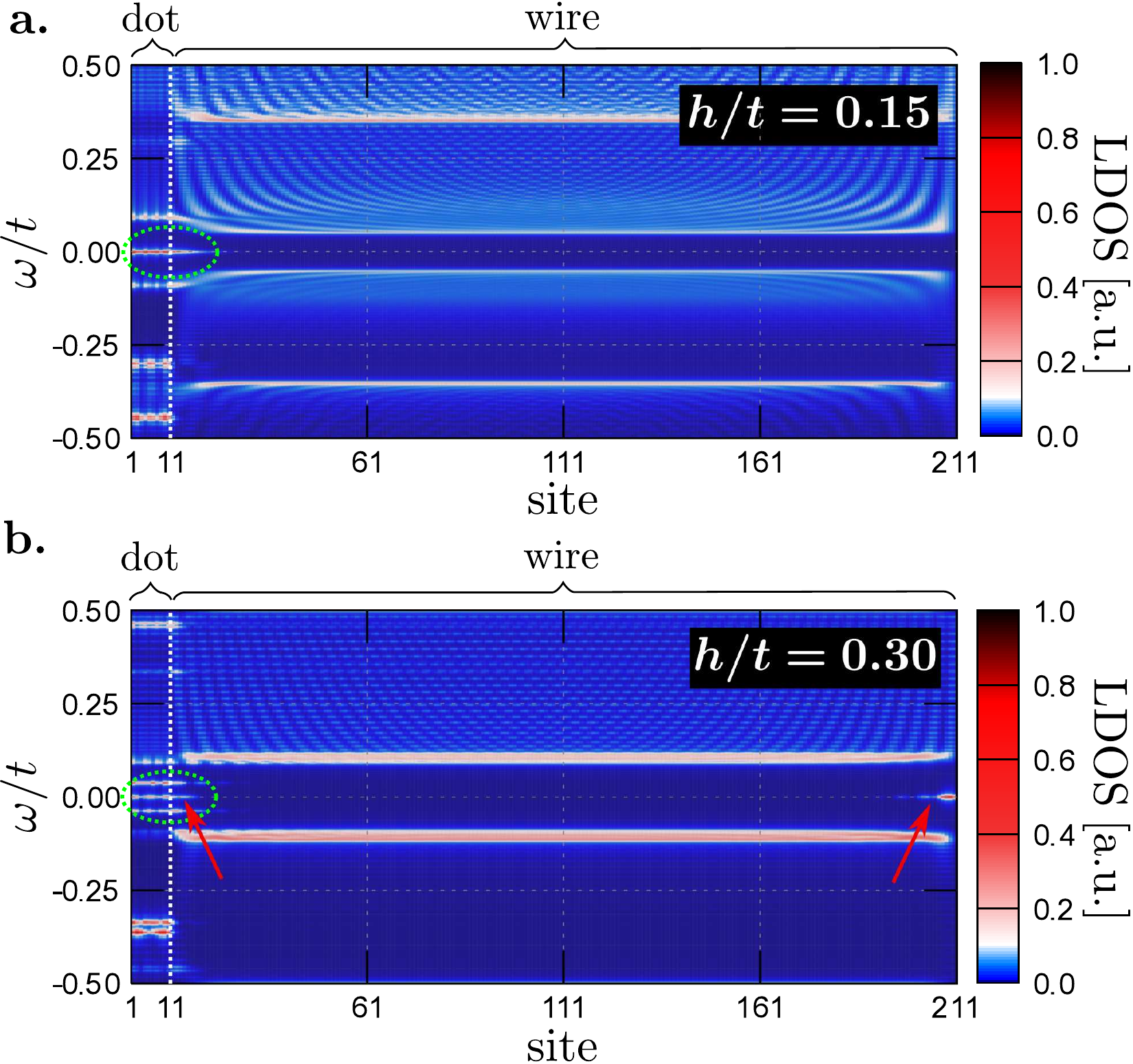}
\caption{
LDOS along the quantum-dot hybrid-nanowire in a cases phase not supported (a) and supported (b) realization of the MBS.
Dot region are localized below 11 site, while superconducting wire above 10 site.
Result for $k_{B}T=0t$, $\mu = -2t$, $\lambda = 0.15t$ and $\Delta = 0.2t$.
Bias voltage are fixed as $V_{g} = -0.99$, what correspond to one of the ABS-resonance level shown in Fig.~\ref{fig.10local} as a green arrows.
Doted white line shows the boundary between the quantum dot and wire regions.
Red arrows shows a pair of the MBS.
\label{fig.10dotdosalong}
}
\end{figure}

From a practical point of view it is important to know what are the spatial profiles of 
the zero-energy bound states of the nanowire, due to their dependence on the magnetic field. 
For $h<h_{c}$ they correspond to crossings of the ZEBS whereas for $h>h_{c}$ they refer to the MBS, respectively.
As mentioned in Sec.~\ref{sec.wires_noQD}, the zero-energy MBS is characterized as the localized, oscillating in space wavefunction formed at the end of wire.
Similarly, the ABS wavefunctions are localized in the QD region of studied system.
In both cases this zero energy bound states can leak from the QD to nanowire region (in a case of the ZEBS) or {\it vice versa} (when MBS is present), via the hybridization between both parts.
Fig.~\ref{fig.10local} presents the spatially dependent spectral weight of the zero energy ($\omega = 0$) quasiparticles.
Let us remark that $i \in \left < 1 , 10 \right>$ in this case correspond to the multi-site QD connected to hybrid-nanowire. 
For some value of the magnetic field smaller than $h_{c}$ (panel a), but bigger than the gap between ABS in the absence of the magnetic field, we can observe several crossings of the ABS (visible as a red lines).
This ZEBS are localized mainly in the QD region and leak into nanowire region.
Situation looks different in a non-trivial topological phase (panel b), where the MBS are present.
In consequence, when QD energy levels change (controlled by gate voltage $V_{g}$), we can observe shift of the MBS initially localized in the end of wire to the dot region.

By inspecting Fig.~\ref{fig.10local} we can also notice spatial oscillations of the zero-energy quasiparticles, 
both in the trivial ($h < h_{c}$) and non-trivial ($h > h_{c}$) topological phases.
This behavior is observable near the edges (spectrum of the entire system is shown in Fig.~\ref{fig.10dotdosalong}). 
In the trivial topological phase (panel a) such oscillations appear mainly in the QD and leak partially to the wire (green doted ellipse).
The situation changes completely for the non-trivial superconducting state (panel b), where the MBS oscillations (red arrows) exist on both sides of the interface and leak to the QD region (green doted ellipse).
In the second case the spatial oscillations are very pronounced, what have been mentioned in Sec.~\ref{sec.wires_noQD}.

\section{Quantum device with tunable Andreev and Majorana bound states}
\label{sec.num.device}

\begin{figure}[!t]
\centering
\includegraphics[width=\linewidth]{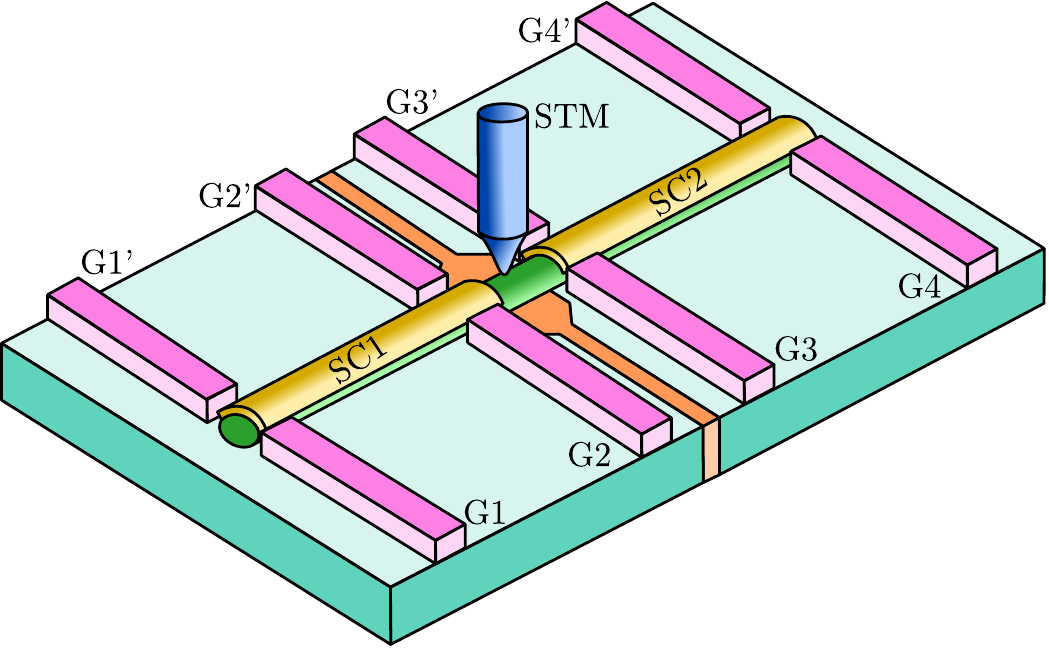}
\caption{
Sketch of the proposed device for a tunable realization of the Andreev/Majorana 
bound states. Semiconducting wire (green) is epitaxially covered by two pieces of 
the superconducting material (SC1 and SC2).
Uncovered part of wire is the multi-site quantum dot, which energy levels are constrained
by the underlying gate (pink).
The side-attached pairs of gates (i.e. G1-G1', G2-G2', etc.) can be used to measure e.g. differential conductance.
Using STM tip (blue) one can detect the bound states present in the quantum dot region.
\label{fig.schemdev}
}
\end{figure}

Finally we propose an experimentally feasible device (sketched in Fig.~\ref{fig.schemdev}) 
for controllable realization of various types of the bound states using electrostatic means.
Motivation to realization of the device in proposed form, is provided by the results from previous sections, which suggest multiple possible outcomes:
{\it (i)} realization and controlling of ZEBS from coalescing ABS (for $h < h_{c}$);
{\it (ii)} ZEBS leakage from the QD to nanowire region (for $h < h_{c}$);
{\it (iii)} MBS leakage from the nanowire to the QD region (for $h > h_{c}$).
In analogy to the setup used 
by M.T. Deng {\it et al}~\cite{deng.vaitiekenas.16} we suggest to use the semiconducting wire 
whose external parts are epitaxially covered by the superconductors (SC1 and SC2).
Such system resembles the typical SNS junction~\cite{cayao.prada.15}, however we omit the phase dependence as superconductors SC1 and SC2 can be taken as made of the same material.
The central piece (which is not covered by superconductors) is treated as the multi-level QD which energy levels can be varied by the gate potential $V_{g}$ (orange region, similarly to Fig.~\ref{fig.schem}). 
Pairs of gates at the ends of the wire (pink), play a crucial role in this setup as they employ the means to measure and verify the existence of zero bias MBS peaks e.g. in a differential conductance discussed in Sec.~\ref{sec.wires_noQD}.
The change of the (global) chemical potential $\mu$, can be realized by changing the voltage at the base (green).
By applying the STM tip to the central QD region, one can probe the different types of the bound states in the differential conductance, for each individual site.
We have in mind that the whole device should be in the external magnetic field, directed along the wire.
Moreover, in generality the SC1 and SC2 may be different materials.
In consequence of this, only in one part of the nanowire part can pass to the non-trivial topological phase, supported realization of the MBS, which should be observe as a zero bias peak in the differential conductance between pairs of the gated, i.e. G1-G1' and G2-G2' (or G3-G3' and G4-G4').
On the other hand, simultaneous measurements carried out by pairs of the gates and the STM can verify the possibility of the bound states leaking from the QD to the nanowire region or {\it vice versa}.

The STM type measurement in the central region of the proposed device, can be also useful to study or check the nature of the realized bound states.
It has been recently emphasized that the Majorana quasiparticles can be distinguished from the usual Andreev states by the spin-polarized spectroscopy called the selective equal spin Andreev reflections 
(SESARs)~\cite{he.ng.14} or spin selective Andreev reflection (SSAR)~\cite{hu.li.16,sun.zhang.16}.
This type of spectroscopy, unambiguously distinguishes between the ''true'' and ''fake'' Majorana quasiparticles~\cite{chirla.moco.16,maska.domanski.17}, which has been used successfully for e.g. the detection of a zero bias peak in Bi$_{2}$Te$_{3}$/NbSe$_{2}$ heterostructure~\cite{xu.wang.15,li.zhou.17} 
or in a case of magnetic atom chain~\cite{jeon.zie.17,feldman.randeria.17,ruby.heinrich.17,li.jeon.17,bjornson.blackschaffer.17}.

\begin{figure}[!t]
\centering
\includegraphics[width=\linewidth]{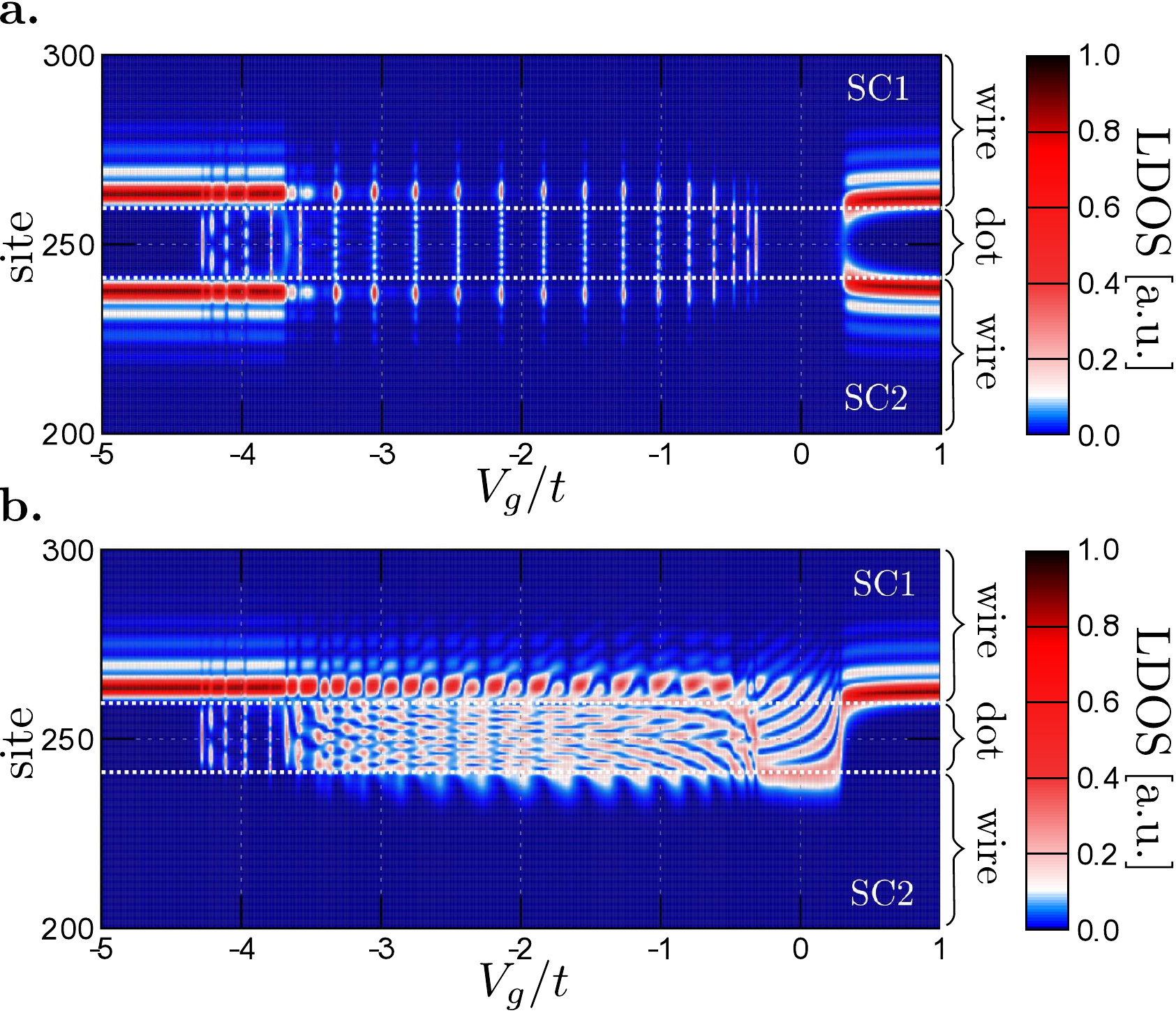}
\caption{
Spectral weight of the zero-energy quasiparticles induced in the multi-level quantum dot 
(sites from 240 to 260) coupled the two nanowires (see Fig.~\ref{fig.schemdev}).
Results are obtained for $k_{B}T=0t$, $\mu = -2t$, $\lambda = 0.15t$, $h=0.3t$ assuming either $\Delta = 0.2t$ for 
both superconducting wires (panel a) or $\Delta = 0.2t$ for the sites $i>260$ and $\Delta = 0.4t$  
for the sites $i<240$, respectively.
Doted white line shows the boundary between the quantum dot and wires regions.
\label{fig.dev}
}
\end{figure}

\begin{figure}[!b]
\centering
\includegraphics[width=\linewidth]{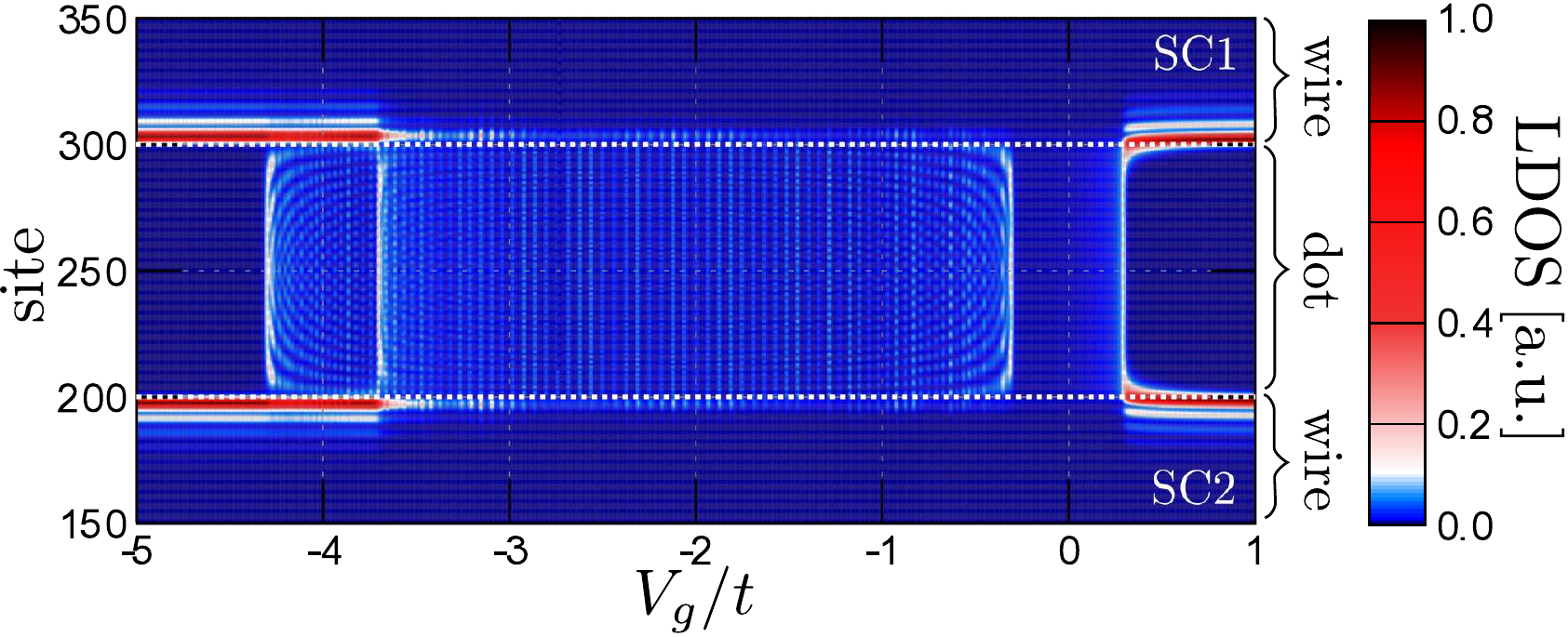}
\caption{
This same as in Fig.~\ref{fig.dev}.a but for the broader central the quantum dot region, comprising 
100 sites ($200\leq i \leq 300$). 
Doted white lines show the boundary between the quantum dot and wires regions.
\label{fig.devbig}
}
\end{figure}

Now, we will show and discuss numerical results, which should be realized in device described above.
We considered the QD comprised of $20$ sites ($240<i<260$).
Fig.~\ref{fig.dev} shows the zero-energy quasiparticles spectrum for two situations: 
{\it (i)} when both nanowires are in the non-trivial topological phase (panel a) 
and 
{\it (ii)} when one part (SC1) is non-trivial topological phase, whereas the other one (SC2) is not (panel b).
In both cases the zero-energy QD levels are available for some discrete values of the gate potential $V_{g}$, approximately in voltage regime of $-4.5 \leq V_{g}/t \leq -0.5$.
What is also important, in both cases, outside this range of $V_{g}$ we can observe a hosting of the MBS in the SC1 region (and in SC2 region in first case).
ZEBS available on the QD and MBS hosted in the wires, can be lead to a resonance between them in a controlled fashion.
In consequence we can check features described in previous sections, a difference in resonance of the MBS with the QD energy levels with majority or minority spin character (asymmetry in panel a around $V_{g}$ equal $-4 t$ and $0 t$).
Other possibility is a experimental study of the leaking the MBS from one part of device to second one via the QD region. 
Moreover, here we have two possibilities:
{\it (i)} when MBS is hosted in both wires  (panel a), what makes realization of {\it interferences} between two different Majorana quasiparticles~\cite{yamakage.sato.14,baranski.kobialka.17} 
and 
{\it (ii)} when only one wire hosts the MBS (panel b), which gives the possibility of studying the  MBS leakage from first to second wire in the ZEBS form (panel b). 
Similar suggestion can be found in Ref.~\cite{setiawan.cole.17}, where authors described expected experimental result of conductance spectroscopy in a nontopological-topological superconductor junctions, which is a building block of our proposed device.
In both cases, measurement of the differential conductance between pairs of gates (i.e. G1-G1', etc.) in described device, can be helpful to verify the realization of the zero-energy type bound states in a form of the zero bias peak or the ABS in a case of non-zero bias.

It should be mentioned that our calculation shows an important role of the finite number of available QD energy levels (compare e.g. Fig.~\ref{fig.dev}.a and Fig.~\ref{fig.devbig}).
Suggested measurement should be more apparent for QD with smaller number of energy levels (which in our case corresponds to number of site in dot region).

\section{Summary}
\label{sec.sum}

Recent experiments suggest possibility of realization of the zero-energy bound state in a hybrid-nanowire structure (Sec.~\ref{sec.intro}), which can be interpreted as a Majorana bound state, with its characteristic features (Sec.~\ref{sec.wires_noQD}).
Motivated by the results obtained by M.T. Deng {\it et al.}~\cite{deng.vaitiekenas.16}, who reported possibility of inducing the bound states in the quantum dot in a controllable way, we described the experimental setup (quantum-dot hybrid-nanowire structure), using the microscopic model (Sec.~\ref{sec.model}) and solving it in real space by the Bogoliubov--de~Gennes technique.

In Sec.~\ref{sec.num.1dot} we studied properties of the system with the one-site quantum dot adjoined to nanowire.
In particular, we analyzed:
{\it (i)} possible influence of gate voltage $V_{g}$ on the bound state realized in the quantum dot
and
{\it (ii)} mutual relation between bound states in the quantum dot and the nanowire region.
We showed that the Andreev bound states, observed for some value of the magnetic field inside hard superconducting gap, can coalesce in controllable way, creating zero energy bound states.
In relation to this, the zero energy Majorana bound states can be realized only when magnetic field is sufficiently large.
Our results are in agreement with those presented in Ref.~\cite{deng.vaitiekenas.16}.

Mutual influence of those two types of bound states is possible as a consequence of shared existence at the zero energy level. 
Therefore, it is possible for bound states to leak from the quantum dot to nanowire region or {\it vice versa}.
Moreover, we showed an asymmetry between resonance of the Majorana bound state and the quantum dot energy levels.
We explained both results as a consequence of:
{\it (i)} change in dominant spin character of quantum dot energy levels by magnetic field (the majority $\uparrow$ and minority $\downarrow$ spin character),
and 
{\it (ii)} different resonance between the Majorana bound states (with $\uparrow$ character) and the quantum dot energy levels (corresponding to $\uparrow$ and $\downarrow$).

This can also be observed as an influence of the spin orbit coupling on the relation between Andreev bound state, which penetrates {\it topological gap}, and zero energy Majorana bound states. 
Increase in the spin orbit coupling leads to an avoided crossing of the 
Andreev bound state with a dominant $\downarrow$ spin character and is accompanied by transfer of a spectral weight to the Majorana bound state. 
This effect is not observed when the quantum dot energy level has $\uparrow$ spin character.

Those results can be also observed in more general structure with multi-site quantum dot (e.g. two site or multi site quantum dot described in Sec.~\ref{sec.num.2dots} and Sec.~\ref{sec.num.multi-dot} respectively).
In this more realistic picture of the quantum dot, we found that the Majorana bound state can resonate with several  quantum dot energy levels with dominant $\uparrow$ spin character, which is visible as a series of the discrete quantum levels in the {\it quasi}particle spectrum.
We showed that the Majorana bound states leak from the wire to quantum dot region and observed the pronounced quantum oscillations in their spatial profiles. 
These effects indicate a tendency towards spatial broadening of the Majorana modes.

In Sec.~\ref{sec.num.device} we proposed a quantum device in a form of a semiconductor nanowire, whose two parts 
are covered by superconductor. The remaining uncovered part can be treated as a quantum dot, with finite number of available energy levels.
This type of device can be used in realization of described properties, i.e. interplay between different types of bound states.
In regime of parameter supporting the realization of the Majorana quasiparticles, presented nanodevice can help to distinguish the differences in resonance between zero energy bound states with different dominant spin character.
We hope that such device would be stimulating for further studies of the Majorana quasiparticles and their {\it interactions} with other kind of the bound states.

Experimental results obtained by M.T. Deng {\it et al.}~\cite{deng.vaitiekenas.16} have been intensively discussed by many groups, studying the tunnelling conductance~\cite{vanheck.lutchyn.16,liu.sau.17,danon.hansen.17,liu.sau.17b}.
However, the zero bias conductance peak does not provide definitive evidence for Majorana zero modes~\cite{liu.sau.17b}.
In relation to this, the zero mode occurring as a consequence of the usual Andreev bound state (in a trivial topological phase) is generally expected to produce zero bias conductance peak of its height varrying between 0 to $4e^{2}/h$. 
We must have in mind, however, influence of additional physical effects (e.g. finite temperature~\cite{liu.sau.17}) leading to reduction of the conductance value, as mentioned in Sec.~\ref{sec.wires_noQD}.
This type of behaviour is important in distinguishing the zero energy feature related to the ''trivial'' Andreev from the ''non-trivial'' Majorana bound states~\cite{liu.sau.17b}.

To this end, let us highlight the main findings of our paper.
Interplay between the quantum dot and nanowire energy levels strongly depend on the {\it topological state} of the system (cf. Fig.~\ref{fig.dos1dotmuzero}).
In a case of the trivial topological phase, the dot energy level creates zero energy bound states via Andreev bound states only for some specific values of the gate potential and magnetic field. 
Contrary to this, in the non-trivial topological phase the Majorana zero energy bound states state can be observed in a wide range of parameters.
We also inspected \textit{leakage} of the Majorana bound states from the nanowire to quantum dot region. 
In relation to the previous work addressing interplay between the quantum dot energy levels with a non-locality of the Majorana zero modes~\cite{prada.aguado.17}, we discussed influence of the coupling in spin-conserved and in spin-flip channels between the quantum dot and nanowire.
We showed that this process strongly depends on the dominant spin component of the quantum dot energy states.
Similar behaviour has been discussed in context of spin dependent coupling between quantum dot and nanowire~\cite{hoffman.cheviallier.17}.
Moreover, we have proposed experimentally feasible device for studying such \textit{leakage} effect and detecting the Majorana {\it quasi}partices. This device can be helpful in experimental verification of the described behavior and in practical realization of the true Majorana qubits ~\cite{guessi.dessotti.17}.
Realizations of here proposed device could be well controlled electrostatically, in which the Majorana bound states could emerge or dissappear in the quantum dot region.
Similar have been previously suggested  for quantum computing based on the Majorana quasiparticles~\cite{hoffman.schrade.2016}.

{\it Additional note:} During the reviewing process of this article, we bacem aware of 
the paper~\cite{chevallier.szumniak.17}, describing detection of the topological phase transition 
in nanowires using the quantum dot analogous to the properties described by us in the Section~\ref{sec.num.1dot}.


\begin{acknowledgments}
We kindly thank Sz. G\l{}odzik, J. Klinovaja, M. P. Nowak, M. M. Ma\'{s}ka, J. Tworzyd\l{}o and D. P. W\'{o}jcik for careful reading of the manuscript, valuable comments and discussions.
We also thank M. T. Deng for consultation and remarks on Section~\ref{sec.num.device}.
This work was supported by the National Science Centre (NCN, Poland) under grants 
UMO-2016/20/S/ST3/00274 (A.P.) and  DEC-2014/13/B/ST3/04451 (A.K. and T.D.).
\end{acknowledgments}


\bibliography{biblio}

\begin{thebibliography}{100}%
\makeatletter
\providecommand \@ifxundefined [1]{%
 \@ifx{#1\undefined}
}%
\providecommand \@ifnum [1]{%
 \ifnum #1\expandafter \@firstoftwo
 \else \expandafter \@secondoftwo
 \fi
}%
\providecommand \@ifx [1]{%
 \ifx #1\expandafter \@firstoftwo
 \else \expandafter \@secondoftwo
 \fi
}%
\providecommand \natexlab [1]{#1}%
\providecommand \enquote  [1]{``#1''}%
\providecommand \bibnamefont  [1]{#1}%
\providecommand \bibfnamefont [1]{#1}%
\providecommand \citenamefont [1]{#1}%
\providecommand \href@noop [0]{\@secondoftwo}%
\providecommand \href [0]{\begingroup \@sanitize@url \@href}%
\providecommand \@href[1]{\@@startlink{#1}\@@href}%
\providecommand \@@href[1]{\endgroup#1\@@endlink}%
\providecommand \@sanitize@url [0]{\catcode `\\12\catcode `\$12\catcode
  `\&12\catcode `\#12\catcode `\^12\catcode `\_12\catcode `\%12\relax}%
\providecommand \@@startlink[1]{}%
\providecommand \@@endlink[0]{}%
\providecommand \url  [0]{\begingroup\@sanitize@url \@url }%
\providecommand \@url [1]{\endgroup\@href {#1}{\urlprefix }}%
\providecommand \urlprefix  [0]{URL }%
\providecommand \Eprint [0]{\href }%
\providecommand \doibase [0]{http://dx.doi.org/}%
\providecommand \selectlanguage [0]{\@gobble}%
\providecommand \bibinfo  [0]{\@secondoftwo}%
\providecommand \bibfield  [0]{\@secondoftwo}%
\providecommand \translation [1]{[#1]}%
\providecommand \BibitemOpen [0]{}%
\providecommand \bibitemStop [0]{}%
\providecommand \bibitemNoStop [0]{.\EOS\space}%
\providecommand \EOS [0]{\spacefactor3000\relax}%
\providecommand \BibitemShut  [1]{\csname bibitem#1\endcsname}%
\let\auto@bib@innerbib\@empty
\bibitem [{\citenamefont {Read}\ and\ \citenamefont
  {Green}(2000)}]{read.green.00}%
  \BibitemOpen
  \bibfield  {author} {\bibinfo {author} {\bibfnamefont {N.}~\bibnamefont
  {Read}}\ and\ \bibinfo {author} {\bibfnamefont {D.}~\bibnamefont {Green}},\
  }\bibfield  {title} {\enquote {\bibinfo {title} {Paired states of fermions in
  two dimensions with breaking of parity and time-reversal symmetries and the
  fractional quantum {Hall} effect},}\ }\href {\doibase
  10.1103/PhysRevB.61.10267} {\bibfield  {journal} {\bibinfo  {journal} {Phys.
  Rev. B}\ }\textbf {\bibinfo {volume} {61}},\ \bibinfo {pages} {10267}
  (\bibinfo {year} {2000})}\BibitemShut {NoStop}%
\bibitem [{\citenamefont {Kitaev}(2001)}]{kitaev.01}%
  \BibitemOpen
  \bibfield  {author} {\bibinfo {author} {\bibfnamefont {A.~Y.}\ \bibnamefont
  {Kitaev}},\ }\bibfield  {title} {\enquote {\bibinfo {title} {Unpaired
  {Majorana} fermions in quantum wires},}\ }\href {\doibase
  10.1070/1063-7869/44/10S/S29} {\bibfield  {journal} {\bibinfo  {journal}
  {Phys.-Usp.}\ }\textbf {\bibinfo {volume} {44}},\ \bibinfo {pages} {131}
  (\bibinfo {year} {2001})}\BibitemShut {NoStop}%
\bibitem [{\citenamefont {Fu}\ and\ \citenamefont {Kane}(2008)}]{fu.kane.08}%
  \BibitemOpen
  \bibfield  {author} {\bibinfo {author} {\bibfnamefont {L.}~\bibnamefont
  {Fu}}\ and\ \bibinfo {author} {\bibfnamefont {C.~L.}\ \bibnamefont {Kane}},\
  }\bibfield  {title} {\enquote {\bibinfo {title} {Superconducting proximity
  effect and {Majorana} fermions at the surface of a topological insulator},}\
  }\href {\doibase 10.1103/PhysRevLett.100.096407} {\bibfield  {journal}
  {\bibinfo  {journal} {Phys. Rev. Lett.}\ }\textbf {\bibinfo {volume} {100}},\
  \bibinfo {pages} {096407} (\bibinfo {year} {2008})}\BibitemShut {NoStop}%
\bibitem [{\citenamefont {Nayak}\ \emph {et~al.}(2008)\citenamefont {Nayak},
  \citenamefont {Simon}, \citenamefont {Stern}, \citenamefont {Freedman},\ and\
  \citenamefont {Das~Sarma}}]{nayak.simon.08}%
  \BibitemOpen
  \bibfield  {author} {\bibinfo {author} {\bibfnamefont {Ch.}\ \bibnamefont
  {Nayak}}, \bibinfo {author} {\bibfnamefont {S.~H.}\ \bibnamefont {Simon}},
  \bibinfo {author} {\bibfnamefont {A.}~\bibnamefont {Stern}}, \bibinfo
  {author} {\bibfnamefont {M.}~\bibnamefont {Freedman}}, \ and\ \bibinfo
  {author} {\bibfnamefont {S.}~\bibnamefont {Das~Sarma}},\ }\bibfield  {title}
  {\enquote {\bibinfo {title} {Non-{Abelian} anyons and topological quantum
  computation},}\ }\href {\doibase 10.1103/RevModPhys.80.1083} {\bibfield
  {journal} {\bibinfo  {journal} {Rev. Mod. Phys.}\ }\textbf {\bibinfo {volume}
  {80}},\ \bibinfo {pages} {1083} (\bibinfo {year} {2008})}\BibitemShut
  {NoStop}%
\bibitem [{\citenamefont {Sau}\ \emph {et~al.}(2010)\citenamefont {Sau},
  \citenamefont {Tewari}, \citenamefont {Lutchyn}, \citenamefont {Stanescu},\
  and\ \citenamefont {Das~Sarma}}]{sau.tewari.10}%
  \BibitemOpen
  \bibfield  {author} {\bibinfo {author} {\bibfnamefont {J.~D.}\ \bibnamefont
  {Sau}}, \bibinfo {author} {\bibfnamefont {S.}~\bibnamefont {Tewari}},
  \bibinfo {author} {\bibfnamefont {R.~M.}\ \bibnamefont {Lutchyn}}, \bibinfo
  {author} {\bibfnamefont {T.~D.}\ \bibnamefont {Stanescu}}, \ and\ \bibinfo
  {author} {\bibfnamefont {S.}~\bibnamefont {Das~Sarma}},\ }\bibfield  {title}
  {\enquote {\bibinfo {title} {Non-{Abelian} quantum order in
  spin-orbit-coupled semiconductors: Search for topological {Majorana}
  particles in solid-state systems},}\ }\href {\doibase
  10.1103/PhysRevB.82.214509} {\bibfield  {journal} {\bibinfo  {journal} {Phys.
  Rev. B}\ }\textbf {\bibinfo {volume} {82}},\ \bibinfo {pages} {214509}
  (\bibinfo {year} {2010})}\BibitemShut {NoStop}%
\bibitem [{\citenamefont {Alicea}\ \emph {et~al.}(2011)\citenamefont {Alicea},
  \citenamefont {Oreg}, \citenamefont {Refael}, \citenamefont {von Oppen},\
  and\ \citenamefont {Fisher}}]{alicea.oreg.11}%
  \BibitemOpen
  \bibfield  {author} {\bibinfo {author} {\bibfnamefont {J.}~\bibnamefont
  {Alicea}}, \bibinfo {author} {\bibfnamefont {Y.}~\bibnamefont {Oreg}},
  \bibinfo {author} {\bibfnamefont {G.}~\bibnamefont {Refael}}, \bibinfo
  {author} {\bibfnamefont {F.}~\bibnamefont {von Oppen}}, \ and\ \bibinfo
  {author} {\bibfnamefont {M.~P.~A.}\ \bibnamefont {Fisher}},\ }\bibfield
  {title} {\enquote {\bibinfo {title} {Non-{Abelian} statistics and topological
  quantum information processing in {1D} wire networks},}\ }\href {\doibase
  10.1038/nphys1915} {\bibfield  {journal} {\bibinfo  {journal} {Nat. Phys.}\
  }\textbf {\bibinfo {volume} {7}},\ \bibinfo {pages} {412} (\bibinfo {year}
  {2011})}\BibitemShut {NoStop}%
\bibitem [{\citenamefont {Rainis}\ and\ \citenamefont
  {Loss}(2012)}]{rainis.loss.12}%
  \BibitemOpen
  \bibfield  {author} {\bibinfo {author} {\bibfnamefont {D.}~\bibnamefont
  {Rainis}}\ and\ \bibinfo {author} {\bibfnamefont {D.}~\bibnamefont {Loss}},\
  }\bibfield  {title} {\enquote {\bibinfo {title} {Majorana qubit decoherence
  by quasiparticle poisoning},}\ }\href {\doibase 10.1103/PhysRevB.85.174533}
  {\bibfield  {journal} {\bibinfo  {journal} {Phys. Rev. B}\ }\textbf {\bibinfo
  {volume} {85}},\ \bibinfo {pages} {174533} (\bibinfo {year}
  {2012})}\BibitemShut {NoStop}%
\bibitem [{\citenamefont {van Heck}\ \emph {et~al.}(2012)\citenamefont {van
  Heck}, \citenamefont {Akhmerov}, \citenamefont {Hassler}, \citenamefont
  {Burrello},\ and\ \citenamefont {Beenakker}}]{vanheck.akhmerov.12}%
  \BibitemOpen
  \bibfield  {author} {\bibinfo {author} {\bibfnamefont {B.}~\bibnamefont {van
  Heck}}, \bibinfo {author} {\bibfnamefont {A.~R.}\ \bibnamefont {Akhmerov}},
  \bibinfo {author} {\bibfnamefont {F.}~\bibnamefont {Hassler}}, \bibinfo
  {author} {\bibfnamefont {M.}~\bibnamefont {Burrello}}, \ and\ \bibinfo
  {author} {\bibfnamefont {C.~W.~J.}\ \bibnamefont {Beenakker}},\ }\bibfield
  {title} {\enquote {\bibinfo {title} {Coulomb-assisted braiding of {Majorana}
  fermions in a {Josephson} junction array},}\ }\href {\doibase
  10.1088/1367-2630/14/3/035019} {\bibfield  {journal} {\bibinfo  {journal}
  {New J. Phys.}\ }\textbf {\bibinfo {volume} {14}},\ \bibinfo {pages} {035019}
  (\bibinfo {year} {2012})}\BibitemShut {NoStop}%
\bibitem [{\citenamefont {Sticlet}\ \emph {et~al.}(2012)\citenamefont
  {Sticlet}, \citenamefont {Bena},\ and\ \citenamefont
  {Simon}}]{sticlet.bena.12}%
  \BibitemOpen
  \bibfield  {author} {\bibinfo {author} {\bibfnamefont {D.}~\bibnamefont
  {Sticlet}}, \bibinfo {author} {\bibfnamefont {C.}~\bibnamefont {Bena}}, \
  and\ \bibinfo {author} {\bibfnamefont {P.}~\bibnamefont {Simon}},\ }\bibfield
   {title} {\enquote {\bibinfo {title} {Spin and {Majorana} polarization in
  topological superconducting wires},}\ }\href {\doibase
  10.1103/PhysRevLett.108.096802} {\bibfield  {journal} {\bibinfo  {journal}
  {Phys. Rev. Lett.}\ }\textbf {\bibinfo {volume} {108}},\ \bibinfo {pages}
  {096802} (\bibinfo {year} {2012})}\BibitemShut {NoStop}%
\bibitem [{\citenamefont {Aasen}\ \emph {et~al.}(2016)\citenamefont {Aasen},
  \citenamefont {Hell}, \citenamefont {Mishmash}, \citenamefont {Higginbotham},
  \citenamefont {Danon}, \citenamefont {Leijnse}, \citenamefont {Jespersen},
  \citenamefont {Folk}, \citenamefont {Marcus}, \citenamefont {Flensberg},\
  and\ \citenamefont {Alicea}}]{aasen.hell.16}%
  \BibitemOpen
  \bibfield  {author} {\bibinfo {author} {\bibfnamefont {D.}~\bibnamefont
  {Aasen}}, \bibinfo {author} {\bibfnamefont {M.}~\bibnamefont {Hell}},
  \bibinfo {author} {\bibfnamefont {R.~V.}\ \bibnamefont {Mishmash}}, \bibinfo
  {author} {\bibfnamefont {A.}~\bibnamefont {Higginbotham}}, \bibinfo {author}
  {\bibfnamefont {J.}~\bibnamefont {Danon}}, \bibinfo {author} {\bibfnamefont
  {M.}~\bibnamefont {Leijnse}}, \bibinfo {author} {\bibfnamefont {T.~S.}\
  \bibnamefont {Jespersen}}, \bibinfo {author} {\bibfnamefont {J.~A.}\
  \bibnamefont {Folk}}, \bibinfo {author} {\bibfnamefont {Ch.~M.}\ \bibnamefont
  {Marcus}}, \bibinfo {author} {\bibfnamefont {K.}~\bibnamefont {Flensberg}}, \
  and\ \bibinfo {author} {\bibfnamefont {J.}~\bibnamefont {Alicea}},\
  }\bibfield  {title} {\enquote {\bibinfo {title} {Milestones toward
  {Majorana}-based quantum computing},}\ }\href {\doibase
  10.1103/PhysRevX.6.031016} {\bibfield  {journal} {\bibinfo  {journal} {Phys.
  Rev. X}\ }\textbf {\bibinfo {volume} {6}},\ \bibinfo {pages} {031016}
  (\bibinfo {year} {2016})}\BibitemShut {NoStop}%
\bibitem [{\citenamefont {Mourik}\ \emph {et~al.}(2012)\citenamefont {Mourik},
  \citenamefont {Zuo}, \citenamefont {Frolov}, \citenamefont {Plissard},
  \citenamefont {Bakkers},\ and\ \citenamefont {Kouwenhoven}}]{mourik.zuo.12}%
  \BibitemOpen
  \bibfield  {author} {\bibinfo {author} {\bibfnamefont {V.}~\bibnamefont
  {Mourik}}, \bibinfo {author} {\bibfnamefont {K.}~\bibnamefont {Zuo}},
  \bibinfo {author} {\bibfnamefont {S.~M.}\ \bibnamefont {Frolov}}, \bibinfo
  {author} {\bibfnamefont {S.~R.}\ \bibnamefont {Plissard}}, \bibinfo {author}
  {\bibfnamefont {E.~P. A.~M.}\ \bibnamefont {Bakkers}}, \ and\ \bibinfo
  {author} {\bibfnamefont {L.~P.}\ \bibnamefont {Kouwenhoven}},\ }\bibfield
  {title} {\enquote {\bibinfo {title} {Signatures of {Majorana} fermions in
  hybrid superconductor-semiconductor nanowire devices},}\ }\href {\doibase
  10.1126/science.1222360} {\bibfield  {journal} {\bibinfo  {journal}
  {Science}\ }\textbf {\bibinfo {volume} {336}},\ \bibinfo {pages} {1003}
  (\bibinfo {year} {2012})}\BibitemShut {NoStop}%
\bibitem [{\citenamefont {Das}\ \emph {et~al.}(2012)\citenamefont {Das},
  \citenamefont {Ronen}, \citenamefont {Most}, \citenamefont {Oreg},
  \citenamefont {Heiblum},\ and\ \citenamefont {Shtrikman}}]{das.ronen.12}%
  \BibitemOpen
  \bibfield  {author} {\bibinfo {author} {\bibfnamefont {A.}~\bibnamefont
  {Das}}, \bibinfo {author} {\bibfnamefont {Y.}~\bibnamefont {Ronen}}, \bibinfo
  {author} {\bibfnamefont {Y.}~\bibnamefont {Most}}, \bibinfo {author}
  {\bibfnamefont {Y.}~\bibnamefont {Oreg}}, \bibinfo {author} {\bibfnamefont
  {M.}~\bibnamefont {Heiblum}}, \ and\ \bibinfo {author} {\bibfnamefont
  {H.}~\bibnamefont {Shtrikman}},\ }\bibfield  {title} {\enquote {\bibinfo
  {title} {Zero-bias peaks and splitting in an {Al-InAs} nanowire topological
  superconductor as a signature of {Majorana} fermions},}\ }\href {\doibase
  10.1038/nphys2479} {\bibfield  {journal} {\bibinfo  {journal} {Nat. Phys.}\
  }\textbf {\bibinfo {volume} {8}},\ \bibinfo {pages} {887} (\bibinfo {year}
  {2012})}\BibitemShut {NoStop}%
\bibitem [{\citenamefont {Deng}\ \emph {et~al.}(2012)\citenamefont {Deng},
  \citenamefont {Yu}, \citenamefont {Huang}, \citenamefont {Larsson},
  \citenamefont {Caroff},\ and\ \citenamefont {Xu}}]{deng.yu.12}%
  \BibitemOpen
  \bibfield  {author} {\bibinfo {author} {\bibfnamefont {M.~T.}\ \bibnamefont
  {Deng}}, \bibinfo {author} {\bibfnamefont {C.~L.}\ \bibnamefont {Yu}},
  \bibinfo {author} {\bibfnamefont {G.~Y.}\ \bibnamefont {Huang}}, \bibinfo
  {author} {\bibfnamefont {M.}~\bibnamefont {Larsson}}, \bibinfo {author}
  {\bibfnamefont {P.}~\bibnamefont {Caroff}}, \ and\ \bibinfo {author}
  {\bibfnamefont {H.~Q.}\ \bibnamefont {Xu}},\ }\bibfield  {title} {\enquote
  {\bibinfo {title} {Anomalous zero-bias conductance peak in a {Nb-InSb}
  nanowire-{Nb} hybrid device},}\ }\href {\doibase 10.1021/nl303758w}
  {\bibfield  {journal} {\bibinfo  {journal} {Nano Lett.}\ }\textbf {\bibinfo
  {volume} {12}},\ \bibinfo {pages} {6414} (\bibinfo {year}
  {2012})}\BibitemShut {NoStop}%
\bibitem [{\citenamefont {Rokhinson}\ \emph {et~al.}(2012)\citenamefont
  {Rokhinson}, \citenamefont {Liu},\ and\ \citenamefont
  {Furdyna}}]{rokhinson.liu.12}%
  \BibitemOpen
  \bibfield  {author} {\bibinfo {author} {\bibfnamefont {L.~P.}\ \bibnamefont
  {Rokhinson}}, \bibinfo {author} {\bibfnamefont {X.}~\bibnamefont {Liu}}, \
  and\ \bibinfo {author} {\bibfnamefont {J.~K.}\ \bibnamefont {Furdyna}},\
  }\bibfield  {title} {\enquote {\bibinfo {title} {The fractional a.c.
  {Josephson} effect in a semiconductor-superconductor nanowire as a signature
  of {Majorana} particles},}\ }\href {\doibase 10.1038/nphys2429} {\bibfield
  {journal} {\bibinfo  {journal} {Nat. Phys.}\ }\textbf {\bibinfo {volume}
  {8}},\ \bibinfo {pages} {795} (\bibinfo {year} {2012})}\BibitemShut {NoStop}%
\bibitem [{\citenamefont {Churchill}\ \emph {et~al.}(2013)\citenamefont
  {Churchill}, \citenamefont {Fatemi}, \citenamefont {Grove-Rasmussen},
  \citenamefont {Deng}, \citenamefont {Caroff}, \citenamefont {Xu},\ and\
  \citenamefont {Marcus}}]{churchill.fatemi.13}%
  \BibitemOpen
  \bibfield  {author} {\bibinfo {author} {\bibfnamefont {H.~O.~H.}\
  \bibnamefont {Churchill}}, \bibinfo {author} {\bibfnamefont {V.}~\bibnamefont
  {Fatemi}}, \bibinfo {author} {\bibfnamefont {K.}~\bibnamefont
  {Grove-Rasmussen}}, \bibinfo {author} {\bibfnamefont {M.~T.}\ \bibnamefont
  {Deng}}, \bibinfo {author} {\bibfnamefont {P.}~\bibnamefont {Caroff}},
  \bibinfo {author} {\bibfnamefont {H.~Q.}\ \bibnamefont {Xu}}, \ and\ \bibinfo
  {author} {\bibfnamefont {C.~M.}\ \bibnamefont {Marcus}},\ }\bibfield  {title}
  {\enquote {\bibinfo {title} {Superconductor-nanowire devices from tunneling
  to the multichannel regime: Zero-bias oscillations and magnetoconductance
  crossover},}\ }\href {\doibase 10.1103/PhysRevB.87.241401} {\bibfield
  {journal} {\bibinfo  {journal} {Phys. Rev. B}\ }\textbf {\bibinfo {volume}
  {87}},\ \bibinfo {pages} {241401} (\bibinfo {year} {2013})}\BibitemShut
  {NoStop}%
\bibitem [{\citenamefont {Finck}\ \emph {et~al.}(2013)\citenamefont {Finck},
  \citenamefont {Van~Harlingen}, \citenamefont {Mohseni}, \citenamefont
  {Jung},\ and\ \citenamefont {Li}}]{finck.vanharlingen.13}%
  \BibitemOpen
  \bibfield  {author} {\bibinfo {author} {\bibfnamefont {A.~D.~K.}\
  \bibnamefont {Finck}}, \bibinfo {author} {\bibfnamefont {D.~J.}\ \bibnamefont
  {Van~Harlingen}}, \bibinfo {author} {\bibfnamefont {P.~K.}\ \bibnamefont
  {Mohseni}}, \bibinfo {author} {\bibfnamefont {K.}~\bibnamefont {Jung}}, \
  and\ \bibinfo {author} {\bibfnamefont {X.}~\bibnamefont {Li}},\ }\bibfield
  {title} {\enquote {\bibinfo {title} {Anomalous modulation of a zero-bias peak
  in a hybrid nanowire-superconductor device},}\ }\href {\doibase
  10.1103/PhysRevLett.110.126406} {\bibfield  {journal} {\bibinfo  {journal}
  {Phys. Rev. Lett.}\ }\textbf {\bibinfo {volume} {110}},\ \bibinfo {pages}
  {126406} (\bibinfo {year} {2013})}\BibitemShut {NoStop}%
\bibitem [{\citenamefont {Feldman}\ \emph {et~al.}(2017)\citenamefont
  {Feldman}, \citenamefont {Randeria}, \citenamefont {Li}, \citenamefont
  {Jeon}, \citenamefont {Xie}, \citenamefont {Wang}, \citenamefont {Drozdov},
  \citenamefont {Andrei~B.},\ and\ \citenamefont
  {Yazdani}}]{feldman.randeria.17}%
  \BibitemOpen
  \bibfield  {author} {\bibinfo {author} {\bibfnamefont {B.~E.}\ \bibnamefont
  {Feldman}}, \bibinfo {author} {\bibfnamefont {M.~T.}\ \bibnamefont
  {Randeria}}, \bibinfo {author} {\bibfnamefont {J.}~\bibnamefont {Li}},
  \bibinfo {author} {\bibfnamefont {S.}~\bibnamefont {Jeon}}, \bibinfo {author}
  {\bibfnamefont {Y.}~\bibnamefont {Xie}}, \bibinfo {author} {\bibfnamefont
  {Z.}~\bibnamefont {Wang}}, \bibinfo {author} {\bibfnamefont {I.~K.}\
  \bibnamefont {Drozdov}}, \bibinfo {author} {\bibfnamefont {B.}~\bibnamefont
  {Andrei~B.}}, \ and\ \bibinfo {author} {\bibfnamefont {A.}~\bibnamefont
  {Yazdani}},\ }\bibfield  {title} {\enquote {\bibinfo {title} {High-resolution
  studies of the {Majorana} atomic chain platform},}\ }\href {\doibase
  10.1038/nphys3947} {\bibfield  {journal} {\bibinfo  {journal} {Nat. Phys.}\
  }\textbf {\bibinfo {volume} {13}},\ \bibinfo {pages} {286} (\bibinfo {year}
  {2017})}\BibitemShut {NoStop}%
\bibitem [{\citenamefont {Nichele}\ \emph {et~al.}(2017)\citenamefont
  {Nichele}, \citenamefont {Drachmann}, \citenamefont {Whiticar}, \citenamefont
  {O'Farrell}, \citenamefont {Suominen}, \citenamefont {Fornieri},
  \citenamefont {Wang}, \citenamefont {Gardner}, \citenamefont {Thomas},
  \citenamefont {Hatke}, \citenamefont {Krogstrup}, \citenamefont {Manfra},
  \citenamefont {Flensberg},\ and\ \citenamefont
  {Marcus}}]{niechele.drachmann.17}%
  \BibitemOpen
  \bibfield  {author} {\bibinfo {author} {\bibfnamefont {F.}~\bibnamefont
  {Nichele}}, \bibinfo {author} {\bibfnamefont {A.~C.~C.}\ \bibnamefont
  {Drachmann}}, \bibinfo {author} {\bibfnamefont {A.~M.}\ \bibnamefont
  {Whiticar}}, \bibinfo {author} {\bibfnamefont {E.~C.~T.}\ \bibnamefont
  {O'Farrell}}, \bibinfo {author} {\bibfnamefont {H.~J.}\ \bibnamefont
  {Suominen}}, \bibinfo {author} {\bibfnamefont {A.}~\bibnamefont {Fornieri}},
  \bibinfo {author} {\bibfnamefont {T.}~\bibnamefont {Wang}}, \bibinfo {author}
  {\bibfnamefont {G.~C.}\ \bibnamefont {Gardner}}, \bibinfo {author}
  {\bibfnamefont {C.}~\bibnamefont {Thomas}}, \bibinfo {author} {\bibfnamefont
  {A.~T.}\ \bibnamefont {Hatke}}, \bibinfo {author} {\bibfnamefont
  {P.}~\bibnamefont {Krogstrup}}, \bibinfo {author} {\bibfnamefont {M.~J.}\
  \bibnamefont {Manfra}}, \bibinfo {author} {\bibfnamefont {K.}~\bibnamefont
  {Flensberg}}, \ and\ \bibinfo {author} {\bibfnamefont {Ch.~M.}\ \bibnamefont
  {Marcus}},\ }\bibfield  {title} {\enquote {\bibinfo {title} {Scaling of
  {Majorana} zero-bias conductance peaks},}\ }\href {\doibase
  10.1103/PhysRevLett.119.136803} {\bibfield  {journal} {\bibinfo  {journal}
  {Phys. Rev. Lett.}\ }\textbf {\bibinfo {volume} {119}},\ \bibinfo {pages}
  {136803} (\bibinfo {year} {2017})}\BibitemShut {NoStop}%
\bibitem [{\citenamefont {Nadj-Perge}\ \emph {et~al.}(2014)\citenamefont
  {Nadj-Perge}, \citenamefont {Drozdov}, \citenamefont {Li}, \citenamefont
  {Chen}, \citenamefont {Jeon}, \citenamefont {Seo}, \citenamefont {MacDonald},
  \citenamefont {Bernevig},\ and\ \citenamefont
  {Yazdani}}]{nadjperge.drozdov.14}%
  \BibitemOpen
  \bibfield  {author} {\bibinfo {author} {\bibfnamefont {S.}~\bibnamefont
  {Nadj-Perge}}, \bibinfo {author} {\bibfnamefont {I.~K.}\ \bibnamefont
  {Drozdov}}, \bibinfo {author} {\bibfnamefont {J.}~\bibnamefont {Li}},
  \bibinfo {author} {\bibfnamefont {H.}~\bibnamefont {Chen}}, \bibinfo {author}
  {\bibfnamefont {S.}~\bibnamefont {Jeon}}, \bibinfo {author} {\bibfnamefont
  {J.}~\bibnamefont {Seo}}, \bibinfo {author} {\bibfnamefont {A.~H.}\
  \bibnamefont {MacDonald}}, \bibinfo {author} {\bibfnamefont {B.~A.}\
  \bibnamefont {Bernevig}}, \ and\ \bibinfo {author} {\bibfnamefont
  {A.}~\bibnamefont {Yazdani}},\ }\bibfield  {title} {\enquote {\bibinfo
  {title} {Observation of {Majorana} fermions in ferromagnetic atomic chains on
  a superconductor},}\ }\href {\doibase 10.1126/science.1259327} {\bibfield
  {journal} {\bibinfo  {journal} {Science}\ }\textbf {\bibinfo {volume}
  {346}},\ \bibinfo {pages} {602} (\bibinfo {year} {2014})}\BibitemShut
  {NoStop}%
\bibitem [{\citenamefont {Krogstrup}\ \emph {et~al.}(2015)\citenamefont
  {Krogstrup}, \citenamefont {Ziino}, \citenamefont {Chang}, \citenamefont
  {Albrecht}, \citenamefont {Madsen}, \citenamefont {Johnson}, \citenamefont
  {Nyg\aa{}rd}, \citenamefont {Marcus},\ and\ \citenamefont
  {Jespersen}}]{krogstrup.ziino.15}%
  \BibitemOpen
  \bibfield  {author} {\bibinfo {author} {\bibfnamefont {P.}~\bibnamefont
  {Krogstrup}}, \bibinfo {author} {\bibfnamefont {N.~L.~B.}\ \bibnamefont
  {Ziino}}, \bibinfo {author} {\bibfnamefont {W.}~\bibnamefont {Chang}},
  \bibinfo {author} {\bibfnamefont {S.~M.}\ \bibnamefont {Albrecht}}, \bibinfo
  {author} {\bibfnamefont {M.~H.}\ \bibnamefont {Madsen}}, \bibinfo {author}
  {\bibfnamefont {E.}~\bibnamefont {Johnson}}, \bibinfo {author} {\bibfnamefont
  {J.}~\bibnamefont {Nyg\aa{}rd}}, \bibinfo {author} {\bibfnamefont {C.~M.}\
  \bibnamefont {Marcus}}, \ and\ \bibinfo {author} {\bibfnamefont {T.~S.}\
  \bibnamefont {Jespersen}},\ }\bibfield  {title} {\enquote {\bibinfo {title}
  {Epitaxy of semiconductor--superconductor nanowires},}\ }\href {\doibase
  10.1038/nmat4176} {\bibfield  {journal} {\bibinfo  {journal} {Nat. Mater.}\
  }\textbf {\bibinfo {volume} {14}},\ \bibinfo {pages} {400} (\bibinfo {year}
  {2015})}\BibitemShut {NoStop}%
\bibitem [{\citenamefont {Chang}\ \emph {et~al.}(2015)\citenamefont {Chang},
  \citenamefont {Albrecht}, \citenamefont {Jespersen}, \citenamefont
  {Kuemmeth}, \citenamefont {Krogstrup}, \citenamefont {Nyg\aa{}rd},\ and\
  \citenamefont {Marcus}}]{chang.albrecht.15}%
  \BibitemOpen
  \bibfield  {author} {\bibinfo {author} {\bibfnamefont {W.}~\bibnamefont
  {Chang}}, \bibinfo {author} {\bibfnamefont {M.~S.}\ \bibnamefont {Albrecht}},
  \bibinfo {author} {\bibfnamefont {S.~T.}\ \bibnamefont {Jespersen}}, \bibinfo
  {author} {\bibfnamefont {F.}~\bibnamefont {Kuemmeth}}, \bibinfo {author}
  {\bibfnamefont {P.}~\bibnamefont {Krogstrup}}, \bibinfo {author}
  {\bibfnamefont {J.}~\bibnamefont {Nyg\aa{}rd}}, \ and\ \bibinfo {author}
  {\bibfnamefont {M.~C.}\ \bibnamefont {Marcus}},\ }\bibfield  {title}
  {\enquote {\bibinfo {title} {Hard gap in epitaxial
  semiconductor--superconductor nanowires},}\ }\href
  {http://doi.org/10.1038/nnano.2014.306} {\bibfield  {journal} {\bibinfo
  {journal} {Nat. Nano.}\ }\textbf {\bibinfo {volume} {10}},\ \bibinfo {pages}
  {232} (\bibinfo {year} {2015})}\BibitemShut {NoStop}%
\bibitem [{\citenamefont {Albrecht}\ \emph {et~al.}(2016)\citenamefont
  {Albrecht}, \citenamefont {Higginbotham}, \citenamefont {Madsen},
  \citenamefont {Kuemmeth}, \citenamefont {Jespersen}, \citenamefont
  {Nyg\aa{}rd}, \citenamefont {Krogstrup},\ and\ \citenamefont
  {Marcus}}]{albrecht.higginbotham.16}%
  \BibitemOpen
  \bibfield  {author} {\bibinfo {author} {\bibfnamefont {S.~M.}\ \bibnamefont
  {Albrecht}}, \bibinfo {author} {\bibfnamefont {A.~P.}\ \bibnamefont
  {Higginbotham}}, \bibinfo {author} {\bibfnamefont {M.}~\bibnamefont
  {Madsen}}, \bibinfo {author} {\bibfnamefont {F.}~\bibnamefont {Kuemmeth}},
  \bibinfo {author} {\bibfnamefont {T.~S.}\ \bibnamefont {Jespersen}}, \bibinfo
  {author} {\bibfnamefont {J.}~\bibnamefont {Nyg\aa{}rd}}, \bibinfo {author}
  {\bibfnamefont {P.}~\bibnamefont {Krogstrup}}, \ and\ \bibinfo {author}
  {\bibfnamefont {C.~M.}\ \bibnamefont {Marcus}},\ }\bibfield  {title}
  {\enquote {\bibinfo {title} {Exponential protection of zero modes in
  {Majorana} islands},}\ }\href {\doibase 10.1038/nature17162} {\bibfield
  {journal} {\bibinfo  {journal} {Nature}\ }\textbf {\bibinfo {volume} {531}},\
  \bibinfo {pages} {206} (\bibinfo {year} {2016})}\BibitemShut {NoStop}%
\bibitem [{\citenamefont {G\"{u}l}\ \emph {et~al.}(2017)\citenamefont
  {G\"{u}l}, \citenamefont {Zhang}, \citenamefont {de~Vries}, \citenamefont
  {van Veen}, \citenamefont {Zuo}, \citenamefont {Mourik}, \citenamefont
  {Conesa-Boj}, \citenamefont {Nowak}, \citenamefont {van Woerkom},
  \citenamefont {Quintero-P\'{e}rez}, \citenamefont {Cassidy}, \citenamefont
  {Geresdi}, \citenamefont {Koelling}, \citenamefont {Car}, \citenamefont
  {Plissard}, \citenamefont {Bakkers},\ and\ \citenamefont
  {Kouwenhoven}}]{gul.zhang.17}%
  \BibitemOpen
  \bibfield  {author} {\bibinfo {author} {\bibfnamefont {\"{O}.}\ \bibnamefont
  {G\"{u}l}}, \bibinfo {author} {\bibfnamefont {H.}~\bibnamefont {Zhang}},
  \bibinfo {author} {\bibfnamefont {F.~K.}\ \bibnamefont {de~Vries}}, \bibinfo
  {author} {\bibfnamefont {J.}~\bibnamefont {van Veen}}, \bibinfo {author}
  {\bibfnamefont {K.}~\bibnamefont {Zuo}}, \bibinfo {author} {\bibfnamefont
  {V.}~\bibnamefont {Mourik}}, \bibinfo {author} {\bibfnamefont
  {S.}~\bibnamefont {Conesa-Boj}}, \bibinfo {author} {\bibfnamefont {M.~P.}\
  \bibnamefont {Nowak}}, \bibinfo {author} {\bibfnamefont {D.~J.}\ \bibnamefont
  {van Woerkom}}, \bibinfo {author} {\bibfnamefont {M.}~\bibnamefont
  {Quintero-P\'{e}rez}}, \bibinfo {author} {\bibfnamefont {M.~C.}\ \bibnamefont
  {Cassidy}}, \bibinfo {author} {\bibfnamefont {A.}~\bibnamefont {Geresdi}},
  \bibinfo {author} {\bibfnamefont {S.}~\bibnamefont {Koelling}}, \bibinfo
  {author} {\bibfnamefont {D.}~\bibnamefont {Car}}, \bibinfo {author}
  {\bibfnamefont {S.~R.}\ \bibnamefont {Plissard}}, \bibinfo {author}
  {\bibfnamefont {E.~P. A.~M.}\ \bibnamefont {Bakkers}}, \ and\ \bibinfo
  {author} {\bibfnamefont {L.~P.}\ \bibnamefont {Kouwenhoven}},\ }\bibfield
  {title} {\enquote {\bibinfo {title} {Hard superconducting gap in {InSb}
  nanowires},}\ }\href {\doibase 10.1021/acs.nanolett.7b00540} {\bibfield
  {journal} {\bibinfo  {journal} {Nano Letters}\ }\textbf {\bibinfo {volume}
  {17}},\ \bibinfo {pages} {2690} (\bibinfo {year} {2017})}\BibitemShut
  {NoStop}%
\bibitem [{\citenamefont {Deng}\ \emph {et~al.}(2016)\citenamefont {Deng},
  \citenamefont {Vaitiekenas}, \citenamefont {Hansen}, \citenamefont {Danon},
  \citenamefont {Leijnse}, \citenamefont {Flensberg}, \citenamefont
  {Nyg\aa{}rd}, \citenamefont {Krogstrup},\ and\ \citenamefont
  {Marcus}}]{deng.vaitiekenas.16}%
  \BibitemOpen
  \bibfield  {author} {\bibinfo {author} {\bibfnamefont {M.~T.}\ \bibnamefont
  {Deng}}, \bibinfo {author} {\bibfnamefont {S.}~\bibnamefont {Vaitiekenas}},
  \bibinfo {author} {\bibfnamefont {E.~B.}\ \bibnamefont {Hansen}}, \bibinfo
  {author} {\bibfnamefont {J.}~\bibnamefont {Danon}}, \bibinfo {author}
  {\bibfnamefont {M.}~\bibnamefont {Leijnse}}, \bibinfo {author} {\bibfnamefont
  {K.}~\bibnamefont {Flensberg}}, \bibinfo {author} {\bibfnamefont
  {J.}~\bibnamefont {Nyg\aa{}rd}}, \bibinfo {author} {\bibfnamefont
  {P.}~\bibnamefont {Krogstrup}}, \ and\ \bibinfo {author} {\bibfnamefont
  {C.~M.}\ \bibnamefont {Marcus}},\ }\bibfield  {title} {\enquote {\bibinfo
  {title} {Majorana bound state in a coupled quantum-dot hybrid-nanowire
  system},}\ }\href {\doibase 10.1126/science.aaf3961} {\bibfield  {journal}
  {\bibinfo  {journal} {Science}\ }\textbf {\bibinfo {volume} {354}},\ \bibinfo
  {pages} {1557} (\bibinfo {year} {2016})}\BibitemShut {NoStop}%
\bibitem [{\citenamefont {Yazdani}\ \emph {et~al.}(1997)\citenamefont
  {Yazdani}, \citenamefont {Jones}, \citenamefont {Lutz}, \citenamefont
  {Crommie},\ and\ \citenamefont {Eigler}}]{yazdani.jones.97}%
  \BibitemOpen
  \bibfield  {author} {\bibinfo {author} {\bibfnamefont {A.}~\bibnamefont
  {Yazdani}}, \bibinfo {author} {\bibfnamefont {B.~A.}\ \bibnamefont {Jones}},
  \bibinfo {author} {\bibfnamefont {C.~P.}\ \bibnamefont {Lutz}}, \bibinfo
  {author} {\bibfnamefont {M.~F.}\ \bibnamefont {Crommie}}, \ and\ \bibinfo
  {author} {\bibfnamefont {D.~M.}\ \bibnamefont {Eigler}},\ }\bibfield  {title}
  {\enquote {\bibinfo {title} {Probing the local effects of magnetic impurities
  on superconductivity},}\ }\href {\doibase 10.1126/science.275.5307.1767}
  {\bibfield  {journal} {\bibinfo  {journal} {Science}\ }\textbf {\bibinfo
  {volume} {275}},\ \bibinfo {pages} {1767} (\bibinfo {year}
  {1997})}\BibitemShut {NoStop}%
\bibitem [{\citenamefont {Pawlak}\ \emph {et~al.}(2016)\citenamefont {Pawlak},
  \citenamefont {Kisiel}, \citenamefont {Klinovaja}, \citenamefont {Meier},
  \citenamefont {Kawai}, \citenamefont {Glatzel}, \citenamefont {Loss},\ and\
  \citenamefont {Meyer}}]{pawlak.kisiel.16}%
  \BibitemOpen
  \bibfield  {author} {\bibinfo {author} {\bibfnamefont {R.}~\bibnamefont
  {Pawlak}}, \bibinfo {author} {\bibfnamefont {M.}~\bibnamefont {Kisiel}},
  \bibinfo {author} {\bibfnamefont {J.}~\bibnamefont {Klinovaja}}, \bibinfo
  {author} {\bibfnamefont {T.}~\bibnamefont {Meier}}, \bibinfo {author}
  {\bibfnamefont {S.}~\bibnamefont {Kawai}}, \bibinfo {author} {\bibfnamefont
  {T.}~\bibnamefont {Glatzel}}, \bibinfo {author} {\bibfnamefont
  {D.}~\bibnamefont {Loss}}, \ and\ \bibinfo {author} {\bibfnamefont
  {E.}~\bibnamefont {Meyer}},\ }\bibfield  {title} {\enquote {\bibinfo {title}
  {Probing atomic structure and {Majorana} wavefunctions in mono-atomic {Fe}
  chains on superconducting {Pb} surface},}\ }\href {\doibase
  10.1038/npjqi.2016.35} {\bibfield  {journal} {\bibinfo  {journal} {npj
  Quantum Information}\ }\textbf {\bibinfo {volume} {2}},\ \bibinfo {pages}
  {16035} (\bibinfo {year} {2016})}\BibitemShut {NoStop}%
\bibitem [{\citenamefont {Ruby}\ \emph {et~al.}(2017)\citenamefont {Ruby},
  \citenamefont {Heinrich}, \citenamefont {Peng}, \citenamefont {von Oppen},\
  and\ \citenamefont {Franke}}]{ruby.heinrich.17}%
  \BibitemOpen
  \bibfield  {author} {\bibinfo {author} {\bibfnamefont {M.}~\bibnamefont
  {Ruby}}, \bibinfo {author} {\bibfnamefont {B.~W.}\ \bibnamefont {Heinrich}},
  \bibinfo {author} {\bibfnamefont {Y.}~\bibnamefont {Peng}}, \bibinfo {author}
  {\bibfnamefont {F.}~\bibnamefont {von Oppen}}, \ and\ \bibinfo {author}
  {\bibfnamefont {K.~J.}\ \bibnamefont {Franke}},\ }\bibfield  {title}
  {\enquote {\bibinfo {title} {Exploring a proximity-coupled {Co} chain on
  {Pb(110)} as a possible {Majorana} platform},}\ }\href {\doibase
  10.1021/acs.nanolett.7b01728} {\bibfield  {journal} {\bibinfo  {journal}
  {Nano Lett.}\ }\textbf {\bibinfo {volume} {17}},\ \bibinfo {pages} {4473}
  (\bibinfo {year} {2017})}\BibitemShut {NoStop}%
\bibitem [{\citenamefont {Chevallier}\ \emph {et~al.}(2012)\citenamefont
  {Chevallier}, \citenamefont {Sticlet}, \citenamefont {Simon},\ and\
  \citenamefont {Bena}}]{chevallier.sticlet.12}%
  \BibitemOpen
  \bibfield  {author} {\bibinfo {author} {\bibfnamefont {D.}~\bibnamefont
  {Chevallier}}, \bibinfo {author} {\bibfnamefont {D.}~\bibnamefont {Sticlet}},
  \bibinfo {author} {\bibfnamefont {P.}~\bibnamefont {Simon}}, \ and\ \bibinfo
  {author} {\bibfnamefont {C.}~\bibnamefont {Bena}},\ }\bibfield  {title}
  {\enquote {\bibinfo {title} {Mutation of {Andreev} into {Majorana} bound
  states in long superconductor-normal and superconductor-normal-superconductor
  junctions},}\ }\href {\doibase 10.1103/PhysRevB.85.235307} {\bibfield
  {journal} {\bibinfo  {journal} {Phys. Rev. B}\ }\textbf {\bibinfo {volume}
  {85}},\ \bibinfo {pages} {235307} (\bibinfo {year} {2012})}\BibitemShut
  {NoStop}%
\bibitem [{\citenamefont {Chevallier}\ \emph {et~al.}(2013)\citenamefont
  {Chevallier}, \citenamefont {Simon},\ and\ \citenamefont
  {Bena}}]{chevallier.simon.13}%
  \BibitemOpen
  \bibfield  {author} {\bibinfo {author} {\bibfnamefont {D.}~\bibnamefont
  {Chevallier}}, \bibinfo {author} {\bibfnamefont {P.}~\bibnamefont {Simon}}, \
  and\ \bibinfo {author} {\bibfnamefont {C.}~\bibnamefont {Bena}},\ }\bibfield
  {title} {\enquote {\bibinfo {title} {From {Andreev} bound states to
  {Majorana} fermions in topological wires on superconducting substrates: A
  story of mutation},}\ }\href {\doibase 10.1103/PhysRevB.88.165401} {\bibfield
   {journal} {\bibinfo  {journal} {Phys. Rev. B}\ }\textbf {\bibinfo {volume}
  {88}},\ \bibinfo {pages} {165401} (\bibinfo {year} {2013})}\BibitemShut
  {NoStop}%
\bibitem [{\citenamefont {Sato}\ and\ \citenamefont
  {Fujimoto}(2009)}]{sato.fujimoto.09}%
  \BibitemOpen
  \bibfield  {author} {\bibinfo {author} {\bibfnamefont {M.}~\bibnamefont
  {Sato}}\ and\ \bibinfo {author} {\bibfnamefont {S.}~\bibnamefont
  {Fujimoto}},\ }\bibfield  {title} {\enquote {\bibinfo {title} {Topological
  phases of noncentrosymmetric superconductors: Edge states, {Majorana}
  fermions, and {non-Abelian} statistics},}\ }\href {\doibase
  10.1103/PhysRevB.79.094504} {\bibfield  {journal} {\bibinfo  {journal} {Phys.
  Rev. B}\ }\textbf {\bibinfo {volume} {79}},\ \bibinfo {pages} {094504}
  (\bibinfo {year} {2009})}\BibitemShut {NoStop}%
\bibitem [{\citenamefont {Sato}\ \emph {et~al.}(2010)\citenamefont {Sato},
  \citenamefont {Takahashi},\ and\ \citenamefont
  {Fujimoto}}]{sato.takahashi.10}%
  \BibitemOpen
  \bibfield  {author} {\bibinfo {author} {\bibfnamefont {M.}~\bibnamefont
  {Sato}}, \bibinfo {author} {\bibfnamefont {Y.}~\bibnamefont {Takahashi}}, \
  and\ \bibinfo {author} {\bibfnamefont {S.}~\bibnamefont {Fujimoto}},\
  }\bibfield  {title} {\enquote {\bibinfo {title} {Non-{Abelian} topological
  orders and {Majorana} fermions in spin-singlet superconductors},}\ }\href
  {\doibase 10.1103/PhysRevB.82.134521} {\bibfield  {journal} {\bibinfo
  {journal} {Phys. Rev. B}\ }\textbf {\bibinfo {volume} {82}},\ \bibinfo
  {pages} {134521} (\bibinfo {year} {2010})}\BibitemShut {NoStop}%
\bibitem [{\citenamefont {Chen}\ \emph {et~al.}(2017)\citenamefont {Chen},
  \citenamefont {Yu}, \citenamefont {Stenger}, \citenamefont {Hocevar},
  \citenamefont {Car}, \citenamefont {Plissard}, \citenamefont {Bakkers},
  \citenamefont {Stanescu},\ and\ \citenamefont {Frolov}}]{chen.yu.16}%
  \BibitemOpen
  \bibfield  {author} {\bibinfo {author} {\bibfnamefont {J.}~\bibnamefont
  {Chen}}, \bibinfo {author} {\bibfnamefont {P.}~\bibnamefont {Yu}}, \bibinfo
  {author} {\bibfnamefont {J.}~\bibnamefont {Stenger}}, \bibinfo {author}
  {\bibfnamefont {M.}~\bibnamefont {Hocevar}}, \bibinfo {author} {\bibfnamefont
  {D.}~\bibnamefont {Car}}, \bibinfo {author} {\bibfnamefont {S.~R.}\
  \bibnamefont {Plissard}}, \bibinfo {author} {\bibfnamefont {E.~P. A.~M.}\
  \bibnamefont {Bakkers}}, \bibinfo {author} {\bibfnamefont {T.~D.}\
  \bibnamefont {Stanescu}}, \ and\ \bibinfo {author} {\bibfnamefont {S.~M.}\
  \bibnamefont {Frolov}},\ }\bibfield  {title} {\enquote {\bibinfo {title}
  {Experimental phase diagram of zero-bias conductance peaks in
  superconductor/semiconductor nanowire devices},}\ }\href {\doibase
  10.1126/sciadv.1701476} {\bibfield  {journal} {\bibinfo  {journal} {Sci.
  Adv.}\ }\textbf {\bibinfo {volume} {3}} (\bibinfo {year} {2017}),\
  10.1126/sciadv.1701476}\BibitemShut {NoStop}%
\bibitem [{\citenamefont {Gazibegovic}\ \emph {et~al.}(2017)\citenamefont
  {Gazibegovic}, \citenamefont {Car}, \citenamefont {Zhang}, \citenamefont
  {Balk}, \citenamefont {Logan}, \citenamefont {de~Moor}, \citenamefont
  {Cassidy}, \citenamefont {Schmits}, \citenamefont {Xu}, \citenamefont {Wang},
  \citenamefont {Krogstrup}, \citenamefont {Op~het Veld}, \citenamefont {Zuo},
  \citenamefont {Vos}, \citenamefont {Shen}, \citenamefont {Bouman},
  \citenamefont {Shojaei}, \citenamefont {Pennachio}, \citenamefont {Lee},
  \citenamefont {van Veldhoven}, \citenamefont {Koelling}, \citenamefont
  {Verheijen}, \citenamefont {Kouwenhoven}, \citenamefont {Palmstr{\o}m},\ and\
  \citenamefont {Bakkers}}]{gazibegovic.car.17}%
  \BibitemOpen
  \bibfield  {author} {\bibinfo {author} {\bibfnamefont {S.}~\bibnamefont
  {Gazibegovic}}, \bibinfo {author} {\bibfnamefont {D.}~\bibnamefont {Car}},
  \bibinfo {author} {\bibfnamefont {H.}~\bibnamefont {Zhang}}, \bibinfo
  {author} {\bibfnamefont {S.~C.}\ \bibnamefont {Balk}}, \bibinfo {author}
  {\bibfnamefont {J.~A.}\ \bibnamefont {Logan}}, \bibinfo {author}
  {\bibfnamefont {M.~W.~A.}\ \bibnamefont {de~Moor}}, \bibinfo {author}
  {\bibfnamefont {M.~C.}\ \bibnamefont {Cassidy}}, \bibinfo {author}
  {\bibfnamefont {R.}~\bibnamefont {Schmits}}, \bibinfo {author} {\bibfnamefont
  {D.}~\bibnamefont {Xu}}, \bibinfo {author} {\bibfnamefont {G.}~\bibnamefont
  {Wang}}, \bibinfo {author} {\bibfnamefont {P.}~\bibnamefont {Krogstrup}},
  \bibinfo {author} {\bibfnamefont {R.~L.~M.}\ \bibnamefont {Op~het Veld}},
  \bibinfo {author} {\bibfnamefont {K.}~\bibnamefont {Zuo}}, \bibinfo {author}
  {\bibfnamefont {Y.}~\bibnamefont {Vos}}, \bibinfo {author} {\bibfnamefont
  {J.}~\bibnamefont {Shen}}, \bibinfo {author} {\bibfnamefont {D.}~\bibnamefont
  {Bouman}}, \bibinfo {author} {\bibfnamefont {B.}~\bibnamefont {Shojaei}},
  \bibinfo {author} {\bibfnamefont {D.}~\bibnamefont {Pennachio}}, \bibinfo
  {author} {\bibfnamefont {J.~S.}\ \bibnamefont {Lee}}, \bibinfo {author}
  {\bibfnamefont {P.~J.}\ \bibnamefont {van Veldhoven}}, \bibinfo {author}
  {\bibfnamefont {S.}~\bibnamefont {Koelling}}, \bibinfo {author}
  {\bibfnamefont {M.~A.}\ \bibnamefont {Verheijen}}, \bibinfo {author}
  {\bibfnamefont {L.~P.}\ \bibnamefont {Kouwenhoven}}, \bibinfo {author}
  {\bibfnamefont {Ch.~J.}\ \bibnamefont {Palmstr{\o}m}}, \ and\ \bibinfo
  {author} {\bibfnamefont {E.~P. A.~M.}\ \bibnamefont {Bakkers}},\ }\bibfield
  {title} {\enquote {\bibinfo {title} {Epitaxy of advanced nanowire quantum
  devices},}\ }\href {http://dx.doi.org/10.1038/nature23468} {\bibfield
  {journal} {\bibinfo  {journal} {Nature}\ }\textbf {\bibinfo {volume} {548}},\
  \bibinfo {pages} {434} (\bibinfo {year} {2017})}\BibitemShut {NoStop}%
\bibitem [{\citenamefont {Reeg}\ \emph {et~al.}(2017)\citenamefont {Reeg},
  \citenamefont {Loss},\ and\ \citenamefont {Klinovaja}}]{reeg.loss.2017}%
  \BibitemOpen
  \bibfield  {author} {\bibinfo {author} {\bibfnamefont {Ch.}\ \bibnamefont
  {Reeg}}, \bibinfo {author} {\bibfnamefont {D.}~\bibnamefont {Loss}}, \ and\
  \bibinfo {author} {\bibfnamefont {J.}~\bibnamefont {Klinovaja}},\ }\bibfield
  {title} {\enquote {\bibinfo {title} {Finite-size effects in a nanowire
  strongly coupled to a thin superconducting shell},}\ }\href {\doibase
  10.1103/PhysRevB.96.125426} {\bibfield  {journal} {\bibinfo  {journal} {Phys.
  Rev. B}\ }\textbf {\bibinfo {volume} {96}},\ \bibinfo {pages} {125426}
  (\bibinfo {year} {2017})}\BibitemShut {NoStop}%
\bibitem [{\citenamefont {Prada}\ \emph {et~al.}(2012)\citenamefont {Prada},
  \citenamefont {San-Jose},\ and\ \citenamefont {Aguado}}]{prada.sanjose.12}%
  \BibitemOpen
  \bibfield  {author} {\bibinfo {author} {\bibfnamefont {E.}~\bibnamefont
  {Prada}}, \bibinfo {author} {\bibfnamefont {P.}~\bibnamefont {San-Jose}}, \
  and\ \bibinfo {author} {\bibfnamefont {R.}~\bibnamefont {Aguado}},\
  }\bibfield  {title} {\enquote {\bibinfo {title} {Transport spectroscopy of
  {NS} nanowire junctions with {Majorana} fermions},}\ }\href {\doibase
  10.1103/PhysRevB.86.180503} {\bibfield  {journal} {\bibinfo  {journal} {Phys.
  Rev. B}\ }\textbf {\bibinfo {volume} {86}},\ \bibinfo {pages} {180503}
  (\bibinfo {year} {2012})}\BibitemShut {NoStop}%
\bibitem [{\citenamefont {Chevallier}\ and\ \citenamefont
  {Klinovaja}(2016)}]{chevallier.klinovaja.16}%
  \BibitemOpen
  \bibfield  {author} {\bibinfo {author} {\bibfnamefont {D.}~\bibnamefont
  {Chevallier}}\ and\ \bibinfo {author} {\bibfnamefont {J.}~\bibnamefont
  {Klinovaja}},\ }\bibfield  {title} {\enquote {\bibinfo {title} {Tomography of
  {Majorana} fermions with {STM} tips},}\ }\href {\doibase
  10.1103/PhysRevB.94.035417} {\bibfield  {journal} {\bibinfo  {journal} {Phys.
  Rev. B}\ }\textbf {\bibinfo {volume} {94}},\ \bibinfo {pages} {035417}
  (\bibinfo {year} {2016})}\BibitemShut {NoStop}%
\bibitem [{\citenamefont {Stenger}\ and\ \citenamefont
  {Stanescu}(2017)}]{stenger.stanescu.17}%
  \BibitemOpen
  \bibfield  {author} {\bibinfo {author} {\bibfnamefont {J.}~\bibnamefont
  {Stenger}}\ and\ \bibinfo {author} {\bibfnamefont {T.~D.}\ \bibnamefont
  {Stanescu}},\ }\href@noop {} {\enquote {\bibinfo {title} {Tunneling
  conductance in semiconductor-superconductor hybrid structures},}\ } (\bibinfo
  {year} {2017}),\ \Eprint {http://arxiv.org/abs/arXiv:1703.02543}
  {arXiv:1703.02543} \BibitemShut {NoStop}%
\bibitem [{\citenamefont {de~Gennes}(1989)}]{degennes.89}%
  \BibitemOpen
  \bibfield  {author} {\bibinfo {author} {\bibfnamefont {P.~G.}\ \bibnamefont
  {de~Gennes}},\ }\href@noop {} {\emph {\bibinfo {title} {Superconductivity of
  metals and alloys}}}\ (\bibinfo  {publisher} {Addison-Wesley},\ \bibinfo
  {year} {1989})\BibitemShut {NoStop}%
\bibitem [{\citenamefont {Matsui}\ \emph {et~al.}(2003)\citenamefont {Matsui},
  \citenamefont {Sato}, \citenamefont {Takahashi}, \citenamefont {Wang},
  \citenamefont {Yang}, \citenamefont {Ding}, \citenamefont {Fujii},
  \citenamefont {Watanabe},\ and\ \citenamefont {Matsuda}}]{matsui.sato.03}%
  \BibitemOpen
  \bibfield  {author} {\bibinfo {author} {\bibfnamefont {H.}~\bibnamefont
  {Matsui}}, \bibinfo {author} {\bibfnamefont {T.}~\bibnamefont {Sato}},
  \bibinfo {author} {\bibfnamefont {T.}~\bibnamefont {Takahashi}}, \bibinfo
  {author} {\bibfnamefont {S.-C.}\ \bibnamefont {Wang}}, \bibinfo {author}
  {\bibfnamefont {H.-B.}\ \bibnamefont {Yang}}, \bibinfo {author}
  {\bibfnamefont {H.}~\bibnamefont {Ding}}, \bibinfo {author} {\bibfnamefont
  {T.}~\bibnamefont {Fujii}}, \bibinfo {author} {\bibfnamefont
  {T.}~\bibnamefont {Watanabe}}, \ and\ \bibinfo {author} {\bibfnamefont
  {A.}~\bibnamefont {Matsuda}},\ }\bibfield  {title} {\enquote {\bibinfo
  {title} {{BCS}-like {Bogoliubov} quasiparticles in high-{T$_{c}$}
  superconductors observed by angle-resolved photoemission spectroscopy},}\
  }\href {\doibase 10.1103/PhysRevLett.90.217002} {\bibfield  {journal}
  {\bibinfo  {journal} {Phys. Rev. Lett.}\ }\textbf {\bibinfo {volume} {90}},\
  \bibinfo {pages} {217002} (\bibinfo {year} {2003})}\BibitemShut {NoStop}%
\bibitem [{\citenamefont {Figgins}\ and\ \citenamefont
  {Morr}(2010)}]{figgins.morr.10}%
  \BibitemOpen
  \bibfield  {author} {\bibinfo {author} {\bibfnamefont {J.}~\bibnamefont
  {Figgins}}\ and\ \bibinfo {author} {\bibfnamefont {D.~K.}\ \bibnamefont
  {Morr}},\ }\bibfield  {title} {\enquote {\bibinfo {title} {Differential
  conductance and quantum interference in {Kondo} systems},}\ }\href {\doibase
  10.1103/PhysRevLett.104.187202} {\bibfield  {journal} {\bibinfo  {journal}
  {Phys. Rev. Lett.}\ }\textbf {\bibinfo {volume} {104}},\ \bibinfo {pages}
  {187202} (\bibinfo {year} {2010})}\BibitemShut {NoStop}%
\bibitem [{\citenamefont {Tersoff}\ and\ \citenamefont
  {Hamann}(1985)}]{tersoff.hamann.85}%
  \BibitemOpen
  \bibfield  {author} {\bibinfo {author} {\bibfnamefont {J.}~\bibnamefont
  {Tersoff}}\ and\ \bibinfo {author} {\bibfnamefont {D.~R.}\ \bibnamefont
  {Hamann}},\ }\bibfield  {title} {\enquote {\bibinfo {title} {Theory of the
  scanning tunneling microscope},}\ }\href {\doibase 10.1103/PhysRevB.31.805}
  {\bibfield  {journal} {\bibinfo  {journal} {Phys. Rev. B}\ }\textbf {\bibinfo
  {volume} {31}},\ \bibinfo {pages} {805} (\bibinfo {year} {1985})}\BibitemShut
  {NoStop}%
\bibitem [{\citenamefont {Klinovaja}\ and\ \citenamefont
  {Loss}(2012)}]{klinovaja.loss.12}%
  \BibitemOpen
  \bibfield  {author} {\bibinfo {author} {\bibfnamefont {J.}~\bibnamefont
  {Klinovaja}}\ and\ \bibinfo {author} {\bibfnamefont {D.}~\bibnamefont
  {Loss}},\ }\bibfield  {title} {\enquote {\bibinfo {title} {Composite
  {Majorana} fermion wave functions in nanowires},}\ }\href {\doibase
  10.1103/PhysRevB.86.085408} {\bibfield  {journal} {\bibinfo  {journal} {Phys.
  Rev. B}\ }\textbf {\bibinfo {volume} {86}},\ \bibinfo {pages} {085408}
  (\bibinfo {year} {2012})}\BibitemShut {NoStop}%
\bibitem [{\citenamefont {Cole}\ \emph {et~al.}(2015)\citenamefont {Cole},
  \citenamefont {Das~Sarma},\ and\ \citenamefont
  {Stanescu}}]{cole.dassarma.15}%
  \BibitemOpen
  \bibfield  {author} {\bibinfo {author} {\bibfnamefont {W.~S.}\ \bibnamefont
  {Cole}}, \bibinfo {author} {\bibfnamefont {S.}~\bibnamefont {Das~Sarma}}, \
  and\ \bibinfo {author} {\bibfnamefont {T.~D.}\ \bibnamefont {Stanescu}},\
  }\bibfield  {title} {\enquote {\bibinfo {title} {Effects of large induced
  superconducting gap on semiconductor {Majorana} nanowires},}\ }\href
  {\doibase 10.1103/PhysRevB.92.174511} {\bibfield  {journal} {\bibinfo
  {journal} {Phys. Rev. B}\ }\textbf {\bibinfo {volume} {92}},\ \bibinfo
  {pages} {174511} (\bibinfo {year} {2015})}\BibitemShut {NoStop}%
\bibitem [{\citenamefont {Liu}\ \emph {et~al.}(2017{\natexlab{a}})\citenamefont
  {Liu}, \citenamefont {Sau}, \citenamefont {Stanescu},\ and\ \citenamefont
  {Das~Sarma}}]{liu.sau.17b}%
  \BibitemOpen
  \bibfield  {author} {\bibinfo {author} {\bibfnamefont {Ch.-X.}\ \bibnamefont
  {Liu}}, \bibinfo {author} {\bibfnamefont {J.~D.}\ \bibnamefont {Sau}},
  \bibinfo {author} {\bibfnamefont {T.~D.}\ \bibnamefont {Stanescu}}, \ and\
  \bibinfo {author} {\bibfnamefont {S.}~\bibnamefont {Das~Sarma}},\ }\bibfield
  {title} {\enquote {\bibinfo {title} {{Andreev} bound states versus {Majorana}
  bound states in quantum dot-nanowire-superconductor hybrid structures:
  Trivial versus topological zero-bias conductance peaks},}\ }\href {\doibase
  10.1103/PhysRevB.96.075161} {\bibfield  {journal} {\bibinfo  {journal} {Phys.
  Rev. B}\ }\textbf {\bibinfo {volume} {96}},\ \bibinfo {pages} {075161}
  (\bibinfo {year} {2017}{\natexlab{a}})}\BibitemShut {NoStop}%
\bibitem [{\citenamefont {Kane}\ and\ \citenamefont
  {Mele}(2005)}]{kane.mele.05}%
  \BibitemOpen
  \bibfield  {author} {\bibinfo {author} {\bibfnamefont {C.~L.}\ \bibnamefont
  {Kane}}\ and\ \bibinfo {author} {\bibfnamefont {E.~J.}\ \bibnamefont
  {Mele}},\ }\bibfield  {title} {\enquote {\bibinfo {title} {${Z}_{2}$
  topological order and the quantum spin {Hall} effect},}\ }\href {\doibase
  10.1103/PhysRevLett.95.146802} {\bibfield  {journal} {\bibinfo  {journal}
  {Phys. Rev. Lett.}\ }\textbf {\bibinfo {volume} {95}},\ \bibinfo {pages}
  {146802} (\bibinfo {year} {2005})}\BibitemShut {NoStop}%
\bibitem [{\citenamefont {Qi}\ and\ \citenamefont {Zhang}(2011)}]{qi.zhang.11}%
  \BibitemOpen
  \bibfield  {author} {\bibinfo {author} {\bibfnamefont {X.-L.}\ \bibnamefont
  {Qi}}\ and\ \bibinfo {author} {\bibfnamefont {S.-Ch.}\ \bibnamefont
  {Zhang}},\ }\bibfield  {title} {\enquote {\bibinfo {title} {Topological
  insulators and superconductors},}\ }\href {\doibase
  10.1103/RevModPhys.83.1057} {\bibfield  {journal} {\bibinfo  {journal} {Rev.
  Mod. Phys.}\ }\textbf {\bibinfo {volume} {83}},\ \bibinfo {pages} {1057}
  (\bibinfo {year} {2011})}\BibitemShut {NoStop}%
\bibitem [{\citenamefont {Zhang}\ and\ \citenamefont
  {Nori}(2016)}]{zhang.nori.16}%
  \BibitemOpen
  \bibfield  {author} {\bibinfo {author} {\bibfnamefont {P.}~\bibnamefont
  {Zhang}}\ and\ \bibinfo {author} {\bibfnamefont {F.}~\bibnamefont {Nori}},\
  }\bibfield  {title} {\enquote {\bibinfo {title} {Majorana bound states in a
  disordered quantum dot chain},}\ }\href {\doibase
  10.1088/1367-2630/18/4/043033} {\bibfield  {journal} {\bibinfo  {journal}
  {New J. Phys.}\ }\textbf {\bibinfo {volume} {18}},\ \bibinfo {pages} {043033}
  (\bibinfo {year} {2016})}\BibitemShut {NoStop}%
\bibitem [{\citenamefont {Oreg}\ \emph {et~al.}(2010)\citenamefont {Oreg},
  \citenamefont {Refael},\ and\ \citenamefont {von Oppen}}]{oreg.refael.10}%
  \BibitemOpen
  \bibfield  {author} {\bibinfo {author} {\bibfnamefont {Y.}~\bibnamefont
  {Oreg}}, \bibinfo {author} {\bibfnamefont {G.}~\bibnamefont {Refael}}, \ and\
  \bibinfo {author} {\bibfnamefont {F.}~\bibnamefont {von Oppen}},\ }\bibfield
  {title} {\enquote {\bibinfo {title} {Helical liquids and {Majorana} bound
  states in quantum wires},}\ }\href {\doibase 10.1103/PhysRevLett.105.177002}
  {\bibfield  {journal} {\bibinfo  {journal} {Phys. Rev. Lett.}\ }\textbf
  {\bibinfo {volume} {105}},\ \bibinfo {pages} {177002} (\bibinfo {year}
  {2010})}\BibitemShut {NoStop}%
\bibitem [{\citenamefont {Reeg}\ and\ \citenamefont
  {Maslov}(2017)}]{reeg.maslov.17}%
  \BibitemOpen
  \bibfield  {author} {\bibinfo {author} {\bibfnamefont {Ch.}\ \bibnamefont
  {Reeg}}\ and\ \bibinfo {author} {\bibfnamefont {D.~L.}\ \bibnamefont
  {Maslov}},\ }\bibfield  {title} {\enquote {\bibinfo {title} {Transport
  signatures of topological superconductivity in a proximity-coupled
  nanowire},}\ }\href {\doibase 10.1103/PhysRevB.95.205439} {\bibfield
  {journal} {\bibinfo  {journal} {Phys. Rev. B}\ }\textbf {\bibinfo {volume}
  {95}},\ \bibinfo {pages} {205439} (\bibinfo {year} {2017})}\BibitemShut
  {NoStop}%
\bibitem [{\citenamefont {Gor'kov}\ and\ \citenamefont
  {Rashba}(2001)}]{gorkov.rashba.01}%
  \BibitemOpen
  \bibfield  {author} {\bibinfo {author} {\bibfnamefont {L.~P.}\ \bibnamefont
  {Gor'kov}}\ and\ \bibinfo {author} {\bibfnamefont {E.~I.}\ \bibnamefont
  {Rashba}},\ }\bibfield  {title} {\enquote {\bibinfo {title} {Superconducting
  {2D} system with lifted spin degeneracy: Mixed singlet-triplet state},}\
  }\href {\doibase 10.1103/PhysRevLett.87.037004} {\bibfield  {journal}
  {\bibinfo  {journal} {Phys. Rev. Lett.}\ }\textbf {\bibinfo {volume} {87}},\
  \bibinfo {pages} {037004} (\bibinfo {year} {2001})}\BibitemShut {NoStop}%
\bibitem [{\citenamefont {Zhang}\ \emph {et~al.}(2008)\citenamefont {Zhang},
  \citenamefont {Tewari}, \citenamefont {Lutchyn},\ and\ \citenamefont
  {Das~Sarma}}]{zhang.tewari.08}%
  \BibitemOpen
  \bibfield  {author} {\bibinfo {author} {\bibfnamefont {Ch.}\ \bibnamefont
  {Zhang}}, \bibinfo {author} {\bibfnamefont {S.}~\bibnamefont {Tewari}},
  \bibinfo {author} {\bibfnamefont {R.~M.}\ \bibnamefont {Lutchyn}}, \ and\
  \bibinfo {author} {\bibfnamefont {S.}~\bibnamefont {Das~Sarma}},\ }\bibfield
  {title} {\enquote {\bibinfo {title} {${p}_{x}+i{p}_{y}$ superfluid from
  $s$-wave interactions of fermionic cold atoms},}\ }\href {\doibase
  10.1103/PhysRevLett.101.160401} {\bibfield  {journal} {\bibinfo  {journal}
  {Phys. Rev. Lett.}\ }\textbf {\bibinfo {volume} {101}},\ \bibinfo {pages}
  {160401} (\bibinfo {year} {2008})}\BibitemShut {NoStop}%
\bibitem [{\citenamefont {Alicea}(2010)}]{alicea.10}%
  \BibitemOpen
  \bibfield  {author} {\bibinfo {author} {\bibfnamefont {J.}~\bibnamefont
  {Alicea}},\ }\bibfield  {title} {\enquote {\bibinfo {title} {Majorana
  fermions in a tunable semiconductor device},}\ }\href {\doibase
  10.1103/PhysRevB.81.125318} {\bibfield  {journal} {\bibinfo  {journal} {Phys.
  Rev. B}\ }\textbf {\bibinfo {volume} {81}},\ \bibinfo {pages} {125318}
  (\bibinfo {year} {2010})}\BibitemShut {NoStop}%
\bibitem [{\citenamefont {Seo}\ \emph {et~al.}(2012)\citenamefont {Seo},
  \citenamefont {Han},\ and\ \citenamefont {S\'a~de Melo}}]{seo.han.12}%
  \BibitemOpen
  \bibfield  {author} {\bibinfo {author} {\bibfnamefont {K.}~\bibnamefont
  {Seo}}, \bibinfo {author} {\bibfnamefont {L.}~\bibnamefont {Han}}, \ and\
  \bibinfo {author} {\bibfnamefont {C.~A.~R.}\ \bibnamefont {S\'a~de Melo}},\
  }\bibfield  {title} {\enquote {\bibinfo {title} {Topological phase
  transitions in ultracold {Fermi} superfluids: The evolution from
  {Bardeen-Cooper-Schrieffer} to {Bose-Einstein}-condensate superfluids under
  artificial spin-orbit fields},}\ }\href {\doibase 10.1103/PhysRevA.85.033601}
  {\bibfield  {journal} {\bibinfo  {journal} {Phys. Rev. A}\ }\textbf {\bibinfo
  {volume} {85}},\ \bibinfo {pages} {033601} (\bibinfo {year}
  {2012})}\BibitemShut {NoStop}%
\bibitem [{\citenamefont {Yu}\ and\ \citenamefont {Wu}(2016)}]{yu.wu.16}%
  \BibitemOpen
  \bibfield  {author} {\bibinfo {author} {\bibfnamefont {T.}~\bibnamefont
  {Yu}}\ and\ \bibinfo {author} {\bibfnamefont {M.~W.}\ \bibnamefont {Wu}},\
  }\bibfield  {title} {\enquote {\bibinfo {title} {Gapped triplet $p$-wave
  superconductivity in strong spin-orbit-coupled semiconductor quantum wells in
  proximity to $s$-wave superconductor},}\ }\href {\doibase
  10.1103/PhysRevB.93.195308} {\bibfield  {journal} {\bibinfo  {journal} {Phys.
  Rev. B}\ }\textbf {\bibinfo {volume} {93}},\ \bibinfo {pages} {195308}
  (\bibinfo {year} {2016})}\BibitemShut {NoStop}%
\bibitem [{\citenamefont {Gibertini}\ \emph {et~al.}(2012)\citenamefont
  {Gibertini}, \citenamefont {Taddei}, \citenamefont {Polini},\ and\
  \citenamefont {Fazio}}]{gibertini.taddei.12}%
  \BibitemOpen
  \bibfield  {author} {\bibinfo {author} {\bibfnamefont {M.}~\bibnamefont
  {Gibertini}}, \bibinfo {author} {\bibfnamefont {F.}~\bibnamefont {Taddei}},
  \bibinfo {author} {\bibfnamefont {M.}~\bibnamefont {Polini}}, \ and\ \bibinfo
  {author} {\bibfnamefont {R.}~\bibnamefont {Fazio}},\ }\bibfield  {title}
  {\enquote {\bibinfo {title} {Local density of states in metal-topological
  superconductor hybrid systems},}\ }\href {\doibase
  10.1103/PhysRevB.85.144525} {\bibfield  {journal} {\bibinfo  {journal} {Phys.
  Rev. B}\ }\textbf {\bibinfo {volume} {85}},\ \bibinfo {pages} {144525}
  (\bibinfo {year} {2012})}\BibitemShut {NoStop}%
\bibitem [{\citenamefont {Liu}\ \emph {et~al.}(2012)\citenamefont {Liu},
  \citenamefont {Potter}, \citenamefont {Law},\ and\ \citenamefont
  {Lee}}]{liu.potter.12}%
  \BibitemOpen
  \bibfield  {author} {\bibinfo {author} {\bibfnamefont {J.}~\bibnamefont
  {Liu}}, \bibinfo {author} {\bibfnamefont {A.~C.}\ \bibnamefont {Potter}},
  \bibinfo {author} {\bibfnamefont {K.~T.}\ \bibnamefont {Law}}, \ and\
  \bibinfo {author} {\bibfnamefont {P.~A.}\ \bibnamefont {Lee}},\ }\bibfield
  {title} {\enquote {\bibinfo {title} {Zero-bias peaks in the tunneling
  conductance of spin-orbit-coupled superconducting wires with and without
  {Majorana} end-states},}\ }\href {\doibase 10.1103/PhysRevLett.109.267002}
  {\bibfield  {journal} {\bibinfo  {journal} {Phys. Rev. Lett.}\ }\textbf
  {\bibinfo {volume} {109}},\ \bibinfo {pages} {267002} (\bibinfo {year}
  {2012})}\BibitemShut {NoStop}%
\bibitem [{\citenamefont {van Heck}\ \emph {et~al.}(2016)\citenamefont {van
  Heck}, \citenamefont {Lutchyn},\ and\ \citenamefont
  {Glazman}}]{vanheck.lutchyn.16}%
  \BibitemOpen
  \bibfield  {author} {\bibinfo {author} {\bibfnamefont {B.}~\bibnamefont {van
  Heck}}, \bibinfo {author} {\bibfnamefont {R.~M.}\ \bibnamefont {Lutchyn}}, \
  and\ \bibinfo {author} {\bibfnamefont {L.~I.}\ \bibnamefont {Glazman}},\
  }\bibfield  {title} {\enquote {\bibinfo {title} {Conductance of a
  proximitized nanowire in the {Coulomb} blockade regime},}\ }\href {\doibase
  10.1103/PhysRevB.93.235431} {\bibfield  {journal} {\bibinfo  {journal} {Phys.
  Rev. B}\ }\textbf {\bibinfo {volume} {93}},\ \bibinfo {pages} {235431}
  (\bibinfo {year} {2016})}\BibitemShut {NoStop}%
\bibitem [{\citenamefont {Liu}\ \emph {et~al.}(2017{\natexlab{b}})\citenamefont
  {Liu}, \citenamefont {Sau},\ and\ \citenamefont {Das~Sarma}}]{liu.sau.17}%
  \BibitemOpen
  \bibfield  {author} {\bibinfo {author} {\bibfnamefont {Ch.-X.}\ \bibnamefont
  {Liu}}, \bibinfo {author} {\bibfnamefont {J.~D.}\ \bibnamefont {Sau}}, \ and\
  \bibinfo {author} {\bibfnamefont {S.}~\bibnamefont {Das~Sarma}},\ }\bibfield
  {title} {\enquote {\bibinfo {title} {Role of dissipation in realistic
  {Majorana} nanowires},}\ }\href {\doibase 10.1103/PhysRevB.95.054502}
  {\bibfield  {journal} {\bibinfo  {journal} {Phys. Rev. B}\ }\textbf {\bibinfo
  {volume} {95}},\ \bibinfo {pages} {054502} (\bibinfo {year}
  {2017}{\natexlab{b}})}\BibitemShut {NoStop}%
\bibitem [{\citenamefont {Danon}\ \emph {et~al.}(2017)\citenamefont {Danon},
  \citenamefont {Hansen},\ and\ \citenamefont {Flensberg}}]{danon.hansen.17}%
  \BibitemOpen
  \bibfield  {author} {\bibinfo {author} {\bibfnamefont {J.}~\bibnamefont
  {Danon}}, \bibinfo {author} {\bibfnamefont {E.~B.}\ \bibnamefont {Hansen}}, \
  and\ \bibinfo {author} {\bibfnamefont {K.}~\bibnamefont {Flensberg}},\
  }\bibfield  {title} {\enquote {\bibinfo {title} {Conductance spectroscopy on
  {Majorana} wires and the inverse proximity effect},}\ }\href {\doibase
  10.1103/PhysRevB.96.125420} {\bibfield  {journal} {\bibinfo  {journal} {Phys.
  Rev. B}\ }\textbf {\bibinfo {volume} {96}},\ \bibinfo {pages} {125420}
  (\bibinfo {year} {2017})}\BibitemShut {NoStop}%
\bibitem [{\citenamefont {Devillard}\ \emph {et~al.}(2017)\citenamefont
  {Devillard}, \citenamefont {Chevallier},\ and\ \citenamefont
  {Albert}}]{devillard.chevallier.17}%
  \BibitemOpen
  \bibfield  {author} {\bibinfo {author} {\bibfnamefont {P.}~\bibnamefont
  {Devillard}}, \bibinfo {author} {\bibfnamefont {D.}~\bibnamefont
  {Chevallier}}, \ and\ \bibinfo {author} {\bibfnamefont {M.}~\bibnamefont
  {Albert}},\ }\bibfield  {title} {\enquote {\bibinfo {title} {Fingerprints of
  majorana fermions in current-correlation measurements from a superconducting
  tunnel microscope},}\ }\href {\doibase 10.1103/PhysRevB.96.115413} {\bibfield
   {journal} {\bibinfo  {journal} {Phys. Rev. B}\ }\textbf {\bibinfo {volume}
  {96}},\ \bibinfo {pages} {115413} (\bibinfo {year} {2017})}\BibitemShut
  {NoStop}%
\bibitem [{\citenamefont {Liu}\ \emph {et~al.}(2017{\natexlab{c}})\citenamefont
  {Liu}, \citenamefont {Setiawan}, \citenamefont {Sau},\ and\ \citenamefont
  {Das~Sarma}}]{liu.setiawan.17}%
  \BibitemOpen
  \bibfield  {author} {\bibinfo {author} {\bibfnamefont {Ch.-X.}\ \bibnamefont
  {Liu}}, \bibinfo {author} {\bibfnamefont {F.}~\bibnamefont {Setiawan}},
  \bibinfo {author} {\bibfnamefont {J.~D.}\ \bibnamefont {Sau}}, \ and\
  \bibinfo {author} {\bibfnamefont {S.}~\bibnamefont {Das~Sarma}},\ }\bibfield
  {title} {\enquote {\bibinfo {title} {Phenomenology of the soft gap, zero-bias
  peak, and zero-mode splitting in ideal {Majorana} nanowires},}\ }\href
  {\doibase 10.1103/PhysRevB.96.054520} {\bibfield  {journal} {\bibinfo
  {journal} {Phys. Rev. B}\ }\textbf {\bibinfo {volume} {96}},\ \bibinfo
  {pages} {054520} (\bibinfo {year} {2017}{\natexlab{c}})}\BibitemShut
  {NoStop}%
\bibitem [{\citenamefont {Setiawan}\ \emph
  {et~al.}(2017{\natexlab{a}})\citenamefont {Setiawan}, \citenamefont {Liu},
  \citenamefont {Sau},\ and\ \citenamefont {Sarma}}]{setiawan.liu.17}%
  \BibitemOpen
  \bibfield  {author} {\bibinfo {author} {\bibfnamefont {F.}~\bibnamefont
  {Setiawan}}, \bibinfo {author} {\bibfnamefont {Ch.-X.}\ \bibnamefont {Liu}},
  \bibinfo {author} {\bibfnamefont {Jay~D.}\ \bibnamefont {Sau}}, \ and\
  \bibinfo {author} {\bibfnamefont {S.~Das}\ \bibnamefont {Sarma}},\
  }\href@noop {} {\enquote {\bibinfo {title} {Electron temperature and tunnel
  coupling dependence of zero-bias and almost-zero-bias conductance peaks in
  {Majorana} nanowires},}\ } (\bibinfo {year} {2017}{\natexlab{a}}),\ \Eprint
  {http://arxiv.org/abs/arXiv:1708.09039} {arXiv:1708.09039} \BibitemShut
  {NoStop}%
\bibitem [{\citenamefont {Prada}\ \emph {et~al.}(2017)\citenamefont {Prada},
  \citenamefont {Aguado},\ and\ \citenamefont {San-Jose}}]{prada.aguado.17}%
  \BibitemOpen
  \bibfield  {author} {\bibinfo {author} {\bibfnamefont {E.}~\bibnamefont
  {Prada}}, \bibinfo {author} {\bibfnamefont {R.}~\bibnamefont {Aguado}}, \
  and\ \bibinfo {author} {\bibfnamefont {P.}~\bibnamefont {San-Jose}},\
  }\bibfield  {title} {\enquote {\bibinfo {title} {Measuring {Majorana}
  nonlocality and spin structure with a quantum dot},}\ }\href {\doibase
  10.1103/PhysRevB.96.085418} {\bibfield  {journal} {\bibinfo  {journal} {Phys.
  Rev. B}\ }\textbf {\bibinfo {volume} {96}},\ \bibinfo {pages} {085418}
  (\bibinfo {year} {2017})}\BibitemShut {NoStop}%
\bibitem [{\citenamefont {Potter}\ and\ \citenamefont
  {Lee}(2010)}]{potter.lee.10}%
  \BibitemOpen
  \bibfield  {author} {\bibinfo {author} {\bibfnamefont {A.~C.}\ \bibnamefont
  {Potter}}\ and\ \bibinfo {author} {\bibfnamefont {P.~A.}\ \bibnamefont
  {Lee}},\ }\bibfield  {title} {\enquote {\bibinfo {title} {Multichannel
  generalization of {Kitaev's} {Majorana} end states and a practical route to
  realize them in thin films},}\ }\href {\doibase
  10.1103/PhysRevLett.105.227003} {\bibfield  {journal} {\bibinfo  {journal}
  {Phys. Rev. Lett.}\ }\textbf {\bibinfo {volume} {105}},\ \bibinfo {pages}
  {227003} (\bibinfo {year} {2010})}\BibitemShut {NoStop}%
\bibitem [{\citenamefont {Das~Sarma}\ \emph {et~al.}(2012)\citenamefont
  {Das~Sarma}, \citenamefont {Sau},\ and\ \citenamefont
  {Stanescu}}]{dassarma.sau.12}%
  \BibitemOpen
  \bibfield  {author} {\bibinfo {author} {\bibfnamefont {S.}~\bibnamefont
  {Das~Sarma}}, \bibinfo {author} {\bibfnamefont {J.~D.}\ \bibnamefont {Sau}},
  \ and\ \bibinfo {author} {\bibfnamefont {T.~D.}\ \bibnamefont {Stanescu}},\
  }\bibfield  {title} {\enquote {\bibinfo {title} {Splitting of the zero-bias
  conductance peak as smoking gun evidence for the existence of the {Majorana}
  mode in a superconductor-semiconductor nanowire},}\ }\href {\doibase
  10.1103/PhysRevB.86.220506} {\bibfield  {journal} {\bibinfo  {journal} {Phys.
  Rev. B}\ }\textbf {\bibinfo {volume} {86}},\ \bibinfo {pages} {220506}
  (\bibinfo {year} {2012})}\BibitemShut {NoStop}%
\bibitem [{\citenamefont {Liu}(2015)}]{liu.15}%
  \BibitemOpen
  \bibfield  {author} {\bibinfo {author} {\bibfnamefont {X.-J.}\ \bibnamefont
  {Liu}},\ }\bibfield  {title} {\enquote {\bibinfo {title} {Soliton-induced
  {Majorana} fermions in a one-dimensional atomic topological superfluid},}\
  }\href {\doibase 10.1103/PhysRevA.91.023610} {\bibfield  {journal} {\bibinfo
  {journal} {Phys. Rev. A}\ }\textbf {\bibinfo {volume} {91}},\ \bibinfo
  {pages} {023610} (\bibinfo {year} {2015})}\BibitemShut {NoStop}%
\bibitem [{\citenamefont {Zhang}\ \emph {et~al.}(2017)\citenamefont {Zhang},
  \citenamefont {G\"{u}l}, \citenamefont {Conesa-Boj}, \citenamefont {Nowak},
  \citenamefont {Wimmer}, \citenamefont {Zuo}, \citenamefont {Mourik},
  \citenamefont {de~Vries}, \citenamefont {van Veen}, \citenamefont {de~Moor},
  \citenamefont {Bommer}, \citenamefont {van Woerkom}, \citenamefont {Car},
  \citenamefont {Plissard}, \citenamefont {Bakkers}, \citenamefont
  {Quintero-P\'{e}rez}, \citenamefont {Cassidy}, \citenamefont {Koelling},
  \citenamefont {Goswami}, \citenamefont {Watanabe}, \citenamefont
  {Taniguchi},\ and\ \citenamefont {Kouwenhoven}}]{zhang.gul.17}%
  \BibitemOpen
  \bibfield  {author} {\bibinfo {author} {\bibfnamefont {H.}~\bibnamefont
  {Zhang}}, \bibinfo {author} {\bibfnamefont {\"{O}.}\ \bibnamefont {G\"{u}l}},
  \bibinfo {author} {\bibfnamefont {S.}~\bibnamefont {Conesa-Boj}}, \bibinfo
  {author} {\bibfnamefont {M.~P.}\ \bibnamefont {Nowak}}, \bibinfo {author}
  {\bibfnamefont {M.}~\bibnamefont {Wimmer}}, \bibinfo {author} {\bibfnamefont
  {K.}~\bibnamefont {Zuo}}, \bibinfo {author} {\bibfnamefont {V.}~\bibnamefont
  {Mourik}}, \bibinfo {author} {\bibfnamefont {F.~K.}\ \bibnamefont
  {de~Vries}}, \bibinfo {author} {\bibfnamefont {J.}~\bibnamefont {van Veen}},
  \bibinfo {author} {\bibfnamefont {M.~W.~A.}\ \bibnamefont {de~Moor}},
  \bibinfo {author} {\bibfnamefont {J.~D.~S.}\ \bibnamefont {Bommer}}, \bibinfo
  {author} {\bibfnamefont {D.~J.}\ \bibnamefont {van Woerkom}}, \bibinfo
  {author} {\bibfnamefont {D.}~\bibnamefont {Car}}, \bibinfo {author}
  {\bibfnamefont {S.~R.}\ \bibnamefont {Plissard}}, \bibinfo {author}
  {\bibfnamefont {E.~P. A.~M.}\ \bibnamefont {Bakkers}}, \bibinfo {author}
  {\bibfnamefont {M.}~\bibnamefont {Quintero-P\'{e}rez}}, \bibinfo {author}
  {\bibfnamefont {M.~C.}\ \bibnamefont {Cassidy}}, \bibinfo {author}
  {\bibfnamefont {S.}~\bibnamefont {Koelling}}, \bibinfo {author}
  {\bibfnamefont {S.}~\bibnamefont {Goswami}}, \bibinfo {author} {\bibfnamefont
  {K.}~\bibnamefont {Watanabe}}, \bibinfo {author} {\bibfnamefont
  {T.}~\bibnamefont {Taniguchi}}, \ and\ \bibinfo {author} {\bibfnamefont
  {L.~P.}\ \bibnamefont {Kouwenhoven}},\ }\bibfield  {title} {\enquote
  {\bibinfo {title} {Ballistic superconductivity in semiconductor nanowires},}\
  }\href {\doibase 10.1038/ncomms16025} {\bibfield  {journal} {\bibinfo
  {journal} {Nat. Commun.}\ }\textbf {\bibinfo {volume} {8}},\ \bibinfo {pages}
  {16025} (\bibinfo {year} {2017})}\BibitemShut {NoStop}%
\bibitem [{\citenamefont {Moore}\ \emph {et~al.}(2016)\citenamefont {Moore},
  \citenamefont {Stanescu},\ and\ \citenamefont {Tewari}}]{more.stanescu.16}%
  \BibitemOpen
  \bibfield  {author} {\bibinfo {author} {\bibfnamefont {Ch.}\ \bibnamefont
  {Moore}}, \bibinfo {author} {\bibfnamefont {T.~D.}\ \bibnamefont {Stanescu}},
  \ and\ \bibinfo {author} {\bibfnamefont {S.}~\bibnamefont {Tewari}},\
  }\href@noop {} {\enquote {\bibinfo {title} {Majorana bound states in
  non-homogeneous semiconductor nanowires},}\ } (\bibinfo {year} {2016}),\
  \Eprint {http://arxiv.org/abs/arXiv:1611.07058} {arXiv:1611.07058}
  \BibitemShut {NoStop}%
\bibitem [{\citenamefont {Cole}\ \emph {et~al.}(2016)\citenamefont {Cole},
  \citenamefont {Sau},\ and\ \citenamefont {Das~Sarma}}]{cole.sau.16}%
  \BibitemOpen
  \bibfield  {author} {\bibinfo {author} {\bibfnamefont {W.~S.}\ \bibnamefont
  {Cole}}, \bibinfo {author} {\bibfnamefont {J.~D.}\ \bibnamefont {Sau}}, \
  and\ \bibinfo {author} {\bibfnamefont {S.}~\bibnamefont {Das~Sarma}},\
  }\bibfield  {title} {\enquote {\bibinfo {title} {Proximity effect and
  {Majorana} bound states in clean semiconductor nanowires coupled to
  disordered superconductors},}\ }\href {\doibase 10.1103/PhysRevB.94.140505}
  {\bibfield  {journal} {\bibinfo  {journal} {Phys. Rev. B}\ }\textbf {\bibinfo
  {volume} {94}},\ \bibinfo {pages} {140505} (\bibinfo {year}
  {2016})}\BibitemShut {NoStop}%
\bibitem [{\citenamefont {Hegde}\ and\ \citenamefont
  {Vishveshwara}(2016)}]{hegde.vishveshwara.16}%
  \BibitemOpen
  \bibfield  {author} {\bibinfo {author} {\bibfnamefont {S.~S.}\ \bibnamefont
  {Hegde}}\ and\ \bibinfo {author} {\bibfnamefont {S.}~\bibnamefont
  {Vishveshwara}},\ }\bibfield  {title} {\enquote {\bibinfo {title} {Majorana
  wave-function oscillations, fermion parity switches, and disorder in {Kitaev}
  chains},}\ }\href {\doibase 10.1103/PhysRevB.94.115166} {\bibfield  {journal}
  {\bibinfo  {journal} {Phys. Rev. B}\ }\textbf {\bibinfo {volume} {94}},\
  \bibinfo {pages} {115166} (\bibinfo {year} {2016})}\BibitemShut {NoStop}%
\bibitem [{\citenamefont {Awoga}\ \emph {et~al.}(2017)\citenamefont {Awoga},
  \citenamefont {Bj\"{o}rnson},\ and\ \citenamefont
  {Black-Schaffer}}]{awoga.bjornson.17}%
  \BibitemOpen
  \bibfield  {author} {\bibinfo {author} {\bibfnamefont {O.~A.}\ \bibnamefont
  {Awoga}}, \bibinfo {author} {\bibfnamefont {K.}~\bibnamefont {Bj\"{o}rnson}},
  \ and\ \bibinfo {author} {\bibfnamefont {A.~M.}\ \bibnamefont
  {Black-Schaffer}},\ }\bibfield  {title} {\enquote {\bibinfo {title} {Disorder
  robustness and protection of {Majorana} bound states in ferromagnetic chains
  on conventional superconductors},}\ }\href {\doibase
  10.1103/PhysRevB.95.184511} {\bibfield  {journal} {\bibinfo  {journal} {Phys.
  Rev. B}\ }\textbf {\bibinfo {volume} {95}},\ \bibinfo {pages} {184511}
  (\bibinfo {year} {2017})}\BibitemShut {NoStop}%
\bibitem [{\citenamefont {Liu}\ and\ \citenamefont
  {Drummond}(2012)}]{liu.drummond.12}%
  \BibitemOpen
  \bibfield  {author} {\bibinfo {author} {\bibfnamefont {X.-J.}\ \bibnamefont
  {Liu}}\ and\ \bibinfo {author} {\bibfnamefont {P.~D.}\ \bibnamefont
  {Drummond}},\ }\bibfield  {title} {\enquote {\bibinfo {title} {Manipulating
  {Majorana} fermions in one-dimensional spin-orbit-coupled atomic {Fermi}
  gases},}\ }\href {\doibase 10.1103/PhysRevA.86.035602} {\bibfield  {journal}
  {\bibinfo  {journal} {Phys. Rev. A}\ }\textbf {\bibinfo {volume} {86}},\
  \bibinfo {pages} {035602} (\bibinfo {year} {2012})}\BibitemShut {NoStop}%
\bibitem [{\citenamefont {Xu}\ \emph {et~al.}(2014)\citenamefont {Xu},
  \citenamefont {Mao}, \citenamefont {Wu},\ and\ \citenamefont
  {Zhang}}]{xu.mao.14}%
  \BibitemOpen
  \bibfield  {author} {\bibinfo {author} {\bibfnamefont {Y.}~\bibnamefont
  {Xu}}, \bibinfo {author} {\bibfnamefont {L.}~\bibnamefont {Mao}}, \bibinfo
  {author} {\bibfnamefont {B.}~\bibnamefont {Wu}}, \ and\ \bibinfo {author}
  {\bibfnamefont {Ch.}\ \bibnamefont {Zhang}},\ }\bibfield  {title} {\enquote
  {\bibinfo {title} {Dark solitons with {Majorana} fermions in
  spin-orbit-coupled {Fermi} gases},}\ }\href {\doibase
  10.1103/PhysRevLett.113.130404} {\bibfield  {journal} {\bibinfo  {journal}
  {Phys. Rev. Lett.}\ }\textbf {\bibinfo {volume} {113}},\ \bibinfo {pages}
  {130404} (\bibinfo {year} {2014})}\BibitemShut {NoStop}%
\bibitem [{\citenamefont {Ma\'{s}ka}\ \emph {et~al.}(2017)\citenamefont
  {Ma\'{s}ka}, \citenamefont {Gorczyca-Goraj}, \citenamefont {Tworzyd\l{}o},\
  and\ \citenamefont {Doma\'{n}ski}}]{maska.gorczycagoraj.17}%
  \BibitemOpen
  \bibfield  {author} {\bibinfo {author} {\bibfnamefont {M.~M.}\ \bibnamefont
  {Ma\'{s}ka}}, \bibinfo {author} {\bibfnamefont {A.}~\bibnamefont
  {Gorczyca-Goraj}}, \bibinfo {author} {\bibfnamefont {J.}~\bibnamefont
  {Tworzyd\l{}o}}, \ and\ \bibinfo {author} {\bibfnamefont {T.}~\bibnamefont
  {Doma\'{n}ski}},\ }\bibfield  {title} {\enquote {\bibinfo {title} {Majorana
  quasiparticles of an inhomogeneous {Rashba} chain},}\ }\href {\doibase
  10.1103/PhysRevB.95.045429} {\bibfield  {journal} {\bibinfo  {journal} {Phys.
  Rev. B}\ }\textbf {\bibinfo {volume} {95}},\ \bibinfo {pages} {045429}
  (\bibinfo {year} {2017})}\BibitemShut {NoStop}%
\bibitem [{\citenamefont {Ptok}\ \emph {et~al.}(2017)\citenamefont {Ptok},
  \citenamefont {Cichy},\ and\ \citenamefont {Doma\'{n}ski}}]{ptok.cichy.17}%
  \BibitemOpen
  \bibfield  {author} {\bibinfo {author} {\bibfnamefont {A.}~\bibnamefont
  {Ptok}}, \bibinfo {author} {\bibfnamefont {A.}~\bibnamefont {Cichy}}, \ and\
  \bibinfo {author} {\bibfnamefont {T.}~\bibnamefont {Doma\'{n}ski}},\
  }\href@noop {} {\enquote {\bibinfo {title} {Quantum engineering of {Majorana}
  quasiparticles in one-dimensional optical lattices},}\ } (\bibinfo {year}
  {2017}),\ \Eprint {http://arxiv.org/abs/arXiv:1706.04155} {arXiv:1706.04155}
  \BibitemShut {NoStop}%
\bibitem [{\citenamefont {Pillet}\ \emph {et~al.}(2010)\citenamefont {Pillet},
  \citenamefont {Quay}, \citenamefont {Morfin}, \citenamefont {Bena},
  \citenamefont {Yeyati},\ and\ \citenamefont {Joyez}}]{pillet.quay.10}%
  \BibitemOpen
  \bibfield  {author} {\bibinfo {author} {\bibfnamefont {J.-D.}\ \bibnamefont
  {Pillet}}, \bibinfo {author} {\bibfnamefont {C.~H.~L.}\ \bibnamefont {Quay}},
  \bibinfo {author} {\bibfnamefont {P.}~\bibnamefont {Morfin}}, \bibinfo
  {author} {\bibfnamefont {C.}~\bibnamefont {Bena}}, \bibinfo {author}
  {\bibfnamefont {A.~L.}\ \bibnamefont {Yeyati}}, \ and\ \bibinfo {author}
  {\bibfnamefont {P.}~\bibnamefont {Joyez}},\ }\bibfield  {title} {\enquote
  {\bibinfo {title} {Andreev bound states in supercurrent-carrying carbon
  nanotubes revealed},}\ }\href {\doibase 10.1038/nphys1811} {\bibfield
  {journal} {\bibinfo  {journal} {Nat. Phys.}\ }\textbf {\bibinfo {volume}
  {6}},\ \bibinfo {pages} {965} (\bibinfo {year} {2010})}\BibitemShut {NoStop}%
\bibitem [{\citenamefont {Dirks}\ \emph {et~al.}(2011)\citenamefont {Dirks},
  \citenamefont {Hughes}, \citenamefont {Lal}, \citenamefont {Uchoa},
  \citenamefont {Chen}, \citenamefont {Chialvo}, \citenamefont {Goldbart},\
  and\ \citenamefont {Mason}}]{dirks.hughes.11}%
  \BibitemOpen
  \bibfield  {author} {\bibinfo {author} {\bibfnamefont {T.}~\bibnamefont
  {Dirks}}, \bibinfo {author} {\bibfnamefont {T.~L.}\ \bibnamefont {Hughes}},
  \bibinfo {author} {\bibfnamefont {S.}~\bibnamefont {Lal}}, \bibinfo {author}
  {\bibfnamefont {B.}~\bibnamefont {Uchoa}}, \bibinfo {author} {\bibfnamefont
  {Y.-F.}\ \bibnamefont {Chen}}, \bibinfo {author} {\bibfnamefont
  {C.}~\bibnamefont {Chialvo}}, \bibinfo {author} {\bibfnamefont {P.~M.}\
  \bibnamefont {Goldbart}}, \ and\ \bibinfo {author} {\bibfnamefont
  {N.}~\bibnamefont {Mason}},\ }\bibfield  {title} {\enquote {\bibinfo {title}
  {Transport through {Andreev} bound states in a graphene quantum dot},}\
  }\href {\doibase 10.1038/nphys1911} {\bibfield  {journal} {\bibinfo
  {journal} {Nat. Phys.}\ }\textbf {\bibinfo {volume} {7}},\ \bibinfo {pages}
  {386} (\bibinfo {year} {2011})}\BibitemShut {NoStop}%
\bibitem [{\citenamefont {Pillet}\ \emph {et~al.}(2013)\citenamefont {Pillet},
  \citenamefont {Joyez}, \citenamefont {\v{Z}itko},\ and\ \citenamefont
  {Goffman}}]{pillet.joyez.13}%
  \BibitemOpen
  \bibfield  {author} {\bibinfo {author} {\bibfnamefont {J.-D.}\ \bibnamefont
  {Pillet}}, \bibinfo {author} {\bibfnamefont {P.}~\bibnamefont {Joyez}},
  \bibinfo {author} {\bibfnamefont {R.}~\bibnamefont {\v{Z}itko}}, \ and\
  \bibinfo {author} {\bibfnamefont {M.~F.}\ \bibnamefont {Goffman}},\
  }\bibfield  {title} {\enquote {\bibinfo {title} {Tunneling spectroscopy of a
  single quantum dot coupled to a superconductor: From {Kondo} ridge to
  {Andreev} bound states},}\ }\href {\doibase 10.1103/PhysRevB.88.045101}
  {\bibfield  {journal} {\bibinfo  {journal} {Phys. Rev. B}\ }\textbf {\bibinfo
  {volume} {88}},\ \bibinfo {pages} {045101} (\bibinfo {year}
  {2013})}\BibitemShut {NoStop}%
\bibitem [{\citenamefont {Lee}\ \emph {et~al.}(2014)\citenamefont {Lee},
  \citenamefont {Jiang}, \citenamefont {Houzet}, \citenamefont {Aguado},
  \citenamefont {Lieber},\ and\ \citenamefont {De~Franceschi}}]{lee.jiang.14}%
  \BibitemOpen
  \bibfield  {author} {\bibinfo {author} {\bibfnamefont {E.~J.~H.}\
  \bibnamefont {Lee}}, \bibinfo {author} {\bibfnamefont {X.}~\bibnamefont
  {Jiang}}, \bibinfo {author} {\bibfnamefont {M.}~\bibnamefont {Houzet}},
  \bibinfo {author} {\bibfnamefont {R.}~\bibnamefont {Aguado}}, \bibinfo
  {author} {\bibfnamefont {Ch.~M.}\ \bibnamefont {Lieber}}, \ and\ \bibinfo
  {author} {\bibfnamefont {S.}~\bibnamefont {De~Franceschi}},\ }\bibfield
  {title} {\enquote {\bibinfo {title} {Spin-resolved {Andreev} levels and
  parity crossings in hybrid superconductor-semiconductor nanostructures},}\
  }\href {\doibase 10.1038/nnano.2013.267} {\bibfield  {journal} {\bibinfo
  {journal} {Nat. Nano}\ }\textbf {\bibinfo {volume} {9}},\ \bibinfo {pages}
  {79} (\bibinfo {year} {2014})}\BibitemShut {NoStop}%
\bibitem [{\citenamefont {Bara\'{n}ski}\ and\ \citenamefont
  {Doma\'{n}ski}(2013)}]{baranski.domanski.13}%
  \BibitemOpen
  \bibfield  {author} {\bibinfo {author} {\bibfnamefont {J.}~\bibnamefont
  {Bara\'{n}ski}}\ and\ \bibinfo {author} {\bibfnamefont {T.}~\bibnamefont
  {Doma\'{n}ski}},\ }\bibfield  {title} {\enquote {\bibinfo {title} {In-gap
  states of a quantum dot coupled between a normal and a superconducting
  lead},}\ }\href {\doibase 10.1088/0953-8984/25/43/435305} {\bibfield
  {journal} {\bibinfo  {journal} {J. Phys.: Condens. Matter}\ }\textbf
  {\bibinfo {volume} {25}},\ \bibinfo {pages} {435305} (\bibinfo {year}
  {2013})}\BibitemShut {NoStop}%
\bibitem [{\citenamefont {Szumniak}\ \emph {et~al.}(2017)\citenamefont
  {Szumniak}, \citenamefont {Chevallier}, \citenamefont {Loss},\ and\
  \citenamefont {Klinovaja}}]{szumniak.chevallier.17}%
  \BibitemOpen
  \bibfield  {author} {\bibinfo {author} {\bibfnamefont {P.}~\bibnamefont
  {Szumniak}}, \bibinfo {author} {\bibfnamefont {D.}~\bibnamefont
  {Chevallier}}, \bibinfo {author} {\bibfnamefont {D.}~\bibnamefont {Loss}}, \
  and\ \bibinfo {author} {\bibfnamefont {J.}~\bibnamefont {Klinovaja}},\
  }\bibfield  {title} {\enquote {\bibinfo {title} {Spin and charge signatures
  of topological superconductivity in {Rashba} nanowires},}\ }\href {\doibase
  10.1103/PhysRevB.96.041401} {\bibfield  {journal} {\bibinfo  {journal} {Phys.
  Rev. B}\ }\textbf {\bibinfo {volume} {96}},\ \bibinfo {pages} {041401}
  (\bibinfo {year} {2017})}\BibitemShut {NoStop}%
\bibitem [{\citenamefont {Folsch}\ \emph {et~al.}(2014)\citenamefont {Folsch},
  \citenamefont {Martinez-Blanco}, \citenamefont {Yang}, \citenamefont
  {Kanisawa},\ and\ \citenamefont {Erwin}}]{folsch.martinezblanco.14}%
  \BibitemOpen
  \bibfield  {author} {\bibinfo {author} {\bibfnamefont {S.}~\bibnamefont
  {Folsch}}, \bibinfo {author} {\bibfnamefont {J.}~\bibnamefont
  {Martinez-Blanco}}, \bibinfo {author} {\bibfnamefont {J.}~\bibnamefont
  {Yang}}, \bibinfo {author} {\bibfnamefont {K.}~\bibnamefont {Kanisawa}}, \
  and\ \bibinfo {author} {\bibfnamefont {S.~C.}\ \bibnamefont {Erwin}},\
  }\bibfield  {title} {\enquote {\bibinfo {title} {Quantum dots with
  single-atom precision},}\ }\href {\doibase 10.1038/nnano.2014.129} {\bibfield
   {journal} {\bibinfo  {journal} {Nat. Nano.}\ }\textbf {\bibinfo {volume}
  {9}},\ \bibinfo {pages} {505} (\bibinfo {year} {2014})}\BibitemShut {NoStop}%
\bibitem [{\citenamefont {Cayao}\ \emph {et~al.}(2015)\citenamefont {Cayao},
  \citenamefont {Prada}, \citenamefont {San-Jose},\ and\ \citenamefont
  {Aguado}}]{cayao.prada.15}%
  \BibitemOpen
  \bibfield  {author} {\bibinfo {author} {\bibfnamefont {J.}~\bibnamefont
  {Cayao}}, \bibinfo {author} {\bibfnamefont {E.}~\bibnamefont {Prada}},
  \bibinfo {author} {\bibfnamefont {P.}~\bibnamefont {San-Jose}}, \ and\
  \bibinfo {author} {\bibfnamefont {R.}~\bibnamefont {Aguado}},\ }\bibfield
  {title} {\enquote {\bibinfo {title} {{SNS} junctions in nanowires with
  spin-orbit coupling: Role of confinement and helicity on the subgap
  spectrum},}\ }\href {\doibase 10.1103/PhysRevB.91.024514} {\bibfield
  {journal} {\bibinfo  {journal} {Phys. Rev. B}\ }\textbf {\bibinfo {volume}
  {91}},\ \bibinfo {pages} {024514} (\bibinfo {year} {2015})}\BibitemShut
  {NoStop}%
\bibitem [{\citenamefont {He}\ \emph {et~al.}(2014)\citenamefont {He},
  \citenamefont {Ng}, \citenamefont {Lee},\ and\ \citenamefont
  {Law}}]{he.ng.14}%
  \BibitemOpen
  \bibfield  {author} {\bibinfo {author} {\bibfnamefont {J.~J.}\ \bibnamefont
  {He}}, \bibinfo {author} {\bibfnamefont {T.~K.}\ \bibnamefont {Ng}}, \bibinfo
  {author} {\bibfnamefont {P.~A.}\ \bibnamefont {Lee}}, \ and\ \bibinfo
  {author} {\bibfnamefont {K.~T.}\ \bibnamefont {Law}},\ }\bibfield  {title}
  {\enquote {\bibinfo {title} {Selective equal-spin {Andreev} reflections
  induced by {Majorana} fermions},}\ }\href {\doibase
  10.1103/PhysRevLett.112.037001} {\bibfield  {journal} {\bibinfo  {journal}
  {Phys. Rev. Lett.}\ }\textbf {\bibinfo {volume} {112}},\ \bibinfo {pages}
  {037001} (\bibinfo {year} {2014})}\BibitemShut {NoStop}%
\bibitem [{\citenamefont {Hu}\ \emph {et~al.}(2016)\citenamefont {Hu},
  \citenamefont {Li}, \citenamefont {Xu}, \citenamefont {Zhou},\ and\
  \citenamefont {Zhang}}]{hu.li.16}%
  \BibitemOpen
  \bibfield  {author} {\bibinfo {author} {\bibfnamefont {L.-H.}\ \bibnamefont
  {Hu}}, \bibinfo {author} {\bibfnamefont {Ch.}\ \bibnamefont {Li}}, \bibinfo
  {author} {\bibfnamefont {D.-H.}\ \bibnamefont {Xu}}, \bibinfo {author}
  {\bibfnamefont {Y.}~\bibnamefont {Zhou}}, \ and\ \bibinfo {author}
  {\bibfnamefont {F.-Ch.}\ \bibnamefont {Zhang}},\ }\bibfield  {title}
  {\enquote {\bibinfo {title} {Theory of spin-selective {Andreev} reflection in
  the vortex core of a topological superconductor},}\ }\href {\doibase
  10.1103/PhysRevB.94.224501} {\bibfield  {journal} {\bibinfo  {journal} {Phys.
  Rev. B}\ }\textbf {\bibinfo {volume} {94}},\ \bibinfo {pages} {224501}
  (\bibinfo {year} {2016})}\BibitemShut {NoStop}%
\bibitem [{\citenamefont {Sun}\ \emph {et~al.}(2016)\citenamefont {Sun},
  \citenamefont {Zhang}, \citenamefont {Hu}, \citenamefont {Li}, \citenamefont
  {Wang}, \citenamefont {Ma}, \citenamefont {Xu}, \citenamefont {Gao},
  \citenamefont {Guan}, \citenamefont {Li}, \citenamefont {Liu}, \citenamefont
  {Qian}, \citenamefont {Zhou}, \citenamefont {Fu}, \citenamefont {Li},
  \citenamefont {Zhang},\ and\ \citenamefont {Jia}}]{sun.zhang.16}%
  \BibitemOpen
  \bibfield  {author} {\bibinfo {author} {\bibfnamefont {H.-H.}\ \bibnamefont
  {Sun}}, \bibinfo {author} {\bibfnamefont {K.-W.}\ \bibnamefont {Zhang}},
  \bibinfo {author} {\bibfnamefont {L.-H.}\ \bibnamefont {Hu}}, \bibinfo
  {author} {\bibfnamefont {Ch.}\ \bibnamefont {Li}}, \bibinfo {author}
  {\bibfnamefont {G.-Y.}\ \bibnamefont {Wang}}, \bibinfo {author}
  {\bibfnamefont {H.-Y.}\ \bibnamefont {Ma}}, \bibinfo {author} {\bibfnamefont
  {Z.-A.}\ \bibnamefont {Xu}}, \bibinfo {author} {\bibfnamefont {Ch.-L.}\
  \bibnamefont {Gao}}, \bibinfo {author} {\bibfnamefont {D.-D.}\ \bibnamefont
  {Guan}}, \bibinfo {author} {\bibfnamefont {Y.-Y.}\ \bibnamefont {Li}},
  \bibinfo {author} {\bibfnamefont {C.}~\bibnamefont {Liu}}, \bibinfo {author}
  {\bibfnamefont {D.}~\bibnamefont {Qian}}, \bibinfo {author} {\bibfnamefont
  {Y.}~\bibnamefont {Zhou}}, \bibinfo {author} {\bibfnamefont {L.}~\bibnamefont
  {Fu}}, \bibinfo {author} {\bibfnamefont {S.-Ch.}\ \bibnamefont {Li}},
  \bibinfo {author} {\bibfnamefont {F.-Ch.}\ \bibnamefont {Zhang}}, \ and\
  \bibinfo {author} {\bibfnamefont {J.-F.}\ \bibnamefont {Jia}},\ }\bibfield
  {title} {\enquote {\bibinfo {title} {Majorana zero mode detected with spin
  selective {Andreev} reflection in the vortex of a topological
  superconductor},}\ }\href {\doibase 10.1103/PhysRevLett.116.257003}
  {\bibfield  {journal} {\bibinfo  {journal} {Phys. Rev. Lett.}\ }\textbf
  {\bibinfo {volume} {116}},\ \bibinfo {pages} {257003} (\bibinfo {year}
  {2016})}\BibitemShut {NoStop}%
\bibitem [{\citenamefont {Chirla}\ and\ \citenamefont
  {Moca}(2016)}]{chirla.moco.16}%
  \BibitemOpen
  \bibfield  {author} {\bibinfo {author} {\bibfnamefont {R.}~\bibnamefont
  {Chirla}}\ and\ \bibinfo {author} {\bibfnamefont {C.~P.}\ \bibnamefont
  {Moca}},\ }\bibfield  {title} {\enquote {\bibinfo {title} {Fingerprints of
  {Majorana} fermions in spin-resolved subgap spectroscopy},}\ }\href {\doibase
  10.1103/PhysRevB.94.045405} {\bibfield  {journal} {\bibinfo  {journal} {Phys.
  Rev. B}\ }\textbf {\bibinfo {volume} {94}},\ \bibinfo {pages} {045405}
  (\bibinfo {year} {2016})}\BibitemShut {NoStop}%
\bibitem [{\citenamefont {Ma\'{s}ka}\ and\ \citenamefont
  {Doma\'{n}ski}(2017)}]{maska.domanski.17}%
  \BibitemOpen
  \bibfield  {author} {\bibinfo {author} {\bibfnamefont {M.~M.}\ \bibnamefont
  {Ma\'{s}ka}}\ and\ \bibinfo {author} {\bibfnamefont {T.}~\bibnamefont
  {Doma\'{n}ski}},\ }\href@noop {} {\enquote {\bibinfo {title} {Spin-polarized
  {Andreev} tunneling through the {Rashba} chain},}\ } (\bibinfo {year}
  {2017}),\ \Eprint {http://arxiv.org/abs/arXiv:1706.01468} {arXiv:1706.01468}
  \BibitemShut {NoStop}%
\bibitem [{\citenamefont {Xu}\ \emph {et~al.}(2015)\citenamefont {Xu},
  \citenamefont {Wang}, \citenamefont {Liu}, \citenamefont {Ge}, \citenamefont
  {Yang}, \citenamefont {Liu}, \citenamefont {Xu}, \citenamefont {Guan},
  \citenamefont {Gao}, \citenamefont {Qian}, \citenamefont {Liu}, \citenamefont
  {Wang}, \citenamefont {Zhang}, \citenamefont {Xue},\ and\ \citenamefont
  {Jia}}]{xu.wang.15}%
  \BibitemOpen
  \bibfield  {author} {\bibinfo {author} {\bibfnamefont {J.-P.}\ \bibnamefont
  {Xu}}, \bibinfo {author} {\bibfnamefont {M.-X.}\ \bibnamefont {Wang}},
  \bibinfo {author} {\bibfnamefont {Z.~L.}\ \bibnamefont {Liu}}, \bibinfo
  {author} {\bibfnamefont {J.-F.}\ \bibnamefont {Ge}}, \bibinfo {author}
  {\bibfnamefont {X.}~\bibnamefont {Yang}}, \bibinfo {author} {\bibfnamefont
  {C.}~\bibnamefont {Liu}}, \bibinfo {author} {\bibfnamefont {Z.~A.}\
  \bibnamefont {Xu}}, \bibinfo {author} {\bibfnamefont {D.}~\bibnamefont
  {Guan}}, \bibinfo {author} {\bibfnamefont {Ch.~L.}\ \bibnamefont {Gao}},
  \bibinfo {author} {\bibfnamefont {D.}~\bibnamefont {Qian}}, \bibinfo {author}
  {\bibfnamefont {Y.}~\bibnamefont {Liu}}, \bibinfo {author} {\bibfnamefont
  {Q.-H.}\ \bibnamefont {Wang}}, \bibinfo {author} {\bibfnamefont {F.-Ch.}\
  \bibnamefont {Zhang}}, \bibinfo {author} {\bibfnamefont {Q.-K.}\ \bibnamefont
  {Xue}}, \ and\ \bibinfo {author} {\bibfnamefont {J.-F.}\ \bibnamefont
  {Jia}},\ }\bibfield  {title} {\enquote {\bibinfo {title} {Experimental
  detection of a {Majorana} mode in the core of a magnetic vortex inside a
  topological insulator-superconductor {Bi$_{2}$Te$_{3}$/NbSe$_{2}$}
  heterostructure},}\ }\href {\doibase 10.1103/PhysRevLett.114.017001}
  {\bibfield  {journal} {\bibinfo  {journal} {Phys. Rev. Lett.}\ }\textbf
  {\bibinfo {volume} {114}},\ \bibinfo {pages} {017001} (\bibinfo {year}
  {2015})}\BibitemShut {NoStop}%
\bibitem [{\citenamefont {Li}\ \emph {et~al.}(2017{\natexlab{a}})\citenamefont
  {Li}, \citenamefont {Zhou}, \citenamefont {He}, \citenamefont {Wang},
  \citenamefont {Zhang}, \citenamefont {Liu}, \citenamefont {Yi}, \citenamefont
  {Wu}, \citenamefont {Law}, \citenamefont {He},\ and\ \citenamefont
  {Wang}}]{li.zhou.17}%
  \BibitemOpen
  \bibfield  {author} {\bibinfo {author} {\bibfnamefont {H.}~\bibnamefont
  {Li}}, \bibinfo {author} {\bibfnamefont {T.}~\bibnamefont {Zhou}}, \bibinfo
  {author} {\bibfnamefont {J.}~\bibnamefont {He}}, \bibinfo {author}
  {\bibfnamefont {H.-W.}\ \bibnamefont {Wang}}, \bibinfo {author}
  {\bibfnamefont {H.}~\bibnamefont {Zhang}}, \bibinfo {author} {\bibfnamefont
  {H.-Ch.}\ \bibnamefont {Liu}}, \bibinfo {author} {\bibfnamefont
  {Y.}~\bibnamefont {Yi}}, \bibinfo {author} {\bibfnamefont {Ch.}\ \bibnamefont
  {Wu}}, \bibinfo {author} {\bibfnamefont {K.~T.}\ \bibnamefont {Law}},
  \bibinfo {author} {\bibfnamefont {H.}~\bibnamefont {He}}, \ and\ \bibinfo
  {author} {\bibfnamefont {J.}~\bibnamefont {Wang}},\ }\bibfield  {title}
  {\enquote {\bibinfo {title} {Origin of bias-independent conductance plateaus
  and zero-bias conductance peaks in {Bi$_{2}$Se$_{3}$/NbSe$_{2}$} hybrid
  structures},}\ }\href {\doibase 10.1103/PhysRevB.96.075107} {\bibfield
  {journal} {\bibinfo  {journal} {Phys. Rev. B}\ }\textbf {\bibinfo {volume}
  {96}},\ \bibinfo {pages} {075107} (\bibinfo {year}
  {2017}{\natexlab{a}})}\BibitemShut {NoStop}%
\bibitem [{\citenamefont {Jeon}\ \emph {et~al.}(2017)\citenamefont {Jeon},
  \citenamefont {Xie}, \citenamefont {Li}, \citenamefont {Wang}, \citenamefont
  {Bernevig},\ and\ \citenamefont {Yazdani}}]{jeon.zie.17}%
  \BibitemOpen
  \bibfield  {author} {\bibinfo {author} {\bibfnamefont {S.}~\bibnamefont
  {Jeon}}, \bibinfo {author} {\bibfnamefont {Y.}~\bibnamefont {Xie}}, \bibinfo
  {author} {\bibfnamefont {J.}~\bibnamefont {Li}}, \bibinfo {author}
  {\bibfnamefont {Z.}~\bibnamefont {Wang}}, \bibinfo {author} {\bibfnamefont
  {B.~A.}\ \bibnamefont {Bernevig}}, \ and\ \bibinfo {author} {\bibfnamefont
  {A.}~\bibnamefont {Yazdani}},\ }\bibfield  {title} {\enquote {\bibinfo
  {title} {Distinguishing a {Majorana} zero mode using spin-resolved
  measurements},}\ }\href {\doibase 10.1126/science.aan3670} {\bibfield
  {journal} {\bibinfo  {journal} {Science}\ } (\bibinfo {year} {2017}),\
  10.1126/science.aan3670}\BibitemShut {NoStop}%
\bibitem [{\citenamefont {Li}\ \emph {et~al.}(2017{\natexlab{b}})\citenamefont
  {Li}, \citenamefont {Jeon}, \citenamefont {Xie}, \citenamefont {Yazdani},\
  and\ \citenamefont {Bernevig}}]{li.jeon.17}%
  \BibitemOpen
  \bibfield  {author} {\bibinfo {author} {\bibfnamefont {J.}~\bibnamefont
  {Li}}, \bibinfo {author} {\bibfnamefont {S.}~\bibnamefont {Jeon}}, \bibinfo
  {author} {\bibfnamefont {Y.}~\bibnamefont {Xie}}, \bibinfo {author}
  {\bibfnamefont {A.}~\bibnamefont {Yazdani}}, \ and\ \bibinfo {author}
  {\bibfnamefont {B.~A.}\ \bibnamefont {Bernevig}},\ }\href@noop {} {\enquote
  {\bibinfo {title} {The {Majorana} spin in magnetic atomic chain systems},}\ }
  (\bibinfo {year} {2017}{\natexlab{b}}),\ \Eprint
  {http://arxiv.org/abs/arXiv:1709.05967} {arXiv:1709.05967} \BibitemShut
  {NoStop}%
\bibitem [{\citenamefont {Bj\"{o}rnson}\ and\ \citenamefont
  {Black-Schaffer}(2017)}]{bjornson.blackschaffer.17}%
  \BibitemOpen
  \bibfield  {author} {\bibinfo {author} {\bibfnamefont {K.}~\bibnamefont
  {Bj\"{o}rnson}}\ and\ \bibinfo {author} {\bibfnamefont {A.~M.}\ \bibnamefont
  {Black-Schaffer}},\ }\href@noop {} {\enquote {\bibinfo {title} {Probing
  chiral edge states in topological superconductors through spin-polarized
  local density of state measurements},}\ } (\bibinfo {year} {2017}),\ \Eprint
  {http://arxiv.org/abs/arXiv:1709.09061} {arXiv:1709.09061} \BibitemShut
  {NoStop}%
\bibitem [{\citenamefont {Yamakage}\ and\ \citenamefont
  {Sato}(2014)}]{yamakage.sato.14}%
  \BibitemOpen
  \bibfield  {author} {\bibinfo {author} {\bibfnamefont {A.}~\bibnamefont
  {Yamakage}}\ and\ \bibinfo {author} {\bibfnamefont {M.}~\bibnamefont
  {Sato}},\ }\bibfield  {title} {\enquote {\bibinfo {title} {Interference of
  {Majorana} fermions in {NS} junctions},}\ }\href {\doibase
  10.1016/j.physe.2013.08.030} {\bibfield  {journal} {\bibinfo  {journal}
  {Phys. E}\ }\textbf {\bibinfo {volume} {55}},\ \bibinfo {pages} {13}
  (\bibinfo {year} {2014})}\BibitemShut {NoStop}%
\bibitem [{\citenamefont {Bara\'{n}ski}\ \emph {et~al.}(2017)\citenamefont
  {Bara\'{n}ski}, \citenamefont {Kobia\l{}ka},\ and\ \citenamefont
  {Doma\'{n}ski}}]{baranski.kobialka.17}%
  \BibitemOpen
  \bibfield  {author} {\bibinfo {author} {\bibfnamefont {J.}~\bibnamefont
  {Bara\'{n}ski}}, \bibinfo {author} {\bibfnamefont {A.}~\bibnamefont
  {Kobia\l{}ka}}, \ and\ \bibinfo {author} {\bibfnamefont {T.}~\bibnamefont
  {Doma\'{n}ski}},\ }\bibfield  {title} {\enquote {\bibinfo {title}
  {Spin-sensitive interference due to {Majorana} state on the interface between
  normal and superconducting leads},}\ }\href {\doibase
  10.1088/1361-648X/aa5214} {\bibfield  {journal} {\bibinfo  {journal} {J.
  Phys.: Condens. Matter}\ }\textbf {\bibinfo {volume} {29}},\ \bibinfo {pages}
  {075603} (\bibinfo {year} {2017})}\BibitemShut {NoStop}%
\bibitem [{\citenamefont {Setiawan}\ \emph
  {et~al.}(2017{\natexlab{b}})\citenamefont {Setiawan}, \citenamefont {Cole},
  \citenamefont {Sau},\ and\ \citenamefont {Das~Sarma}}]{setiawan.cole.17}%
  \BibitemOpen
  \bibfield  {author} {\bibinfo {author} {\bibfnamefont {F.}~\bibnamefont
  {Setiawan}}, \bibinfo {author} {\bibfnamefont {W.~S.}\ \bibnamefont {Cole}},
  \bibinfo {author} {\bibfnamefont {J.~D.}\ \bibnamefont {Sau}}, \ and\
  \bibinfo {author} {\bibfnamefont {S.}~\bibnamefont {Das~Sarma}},\ }\bibfield
  {title} {\enquote {\bibinfo {title} {Conductance spectroscopy of
  nontopological-topological superconductor junctions},}\ }\href {\doibase
  10.1103/PhysRevB.95.020501} {\bibfield  {journal} {\bibinfo  {journal} {Phys.
  Rev. B}\ }\textbf {\bibinfo {volume} {95}},\ \bibinfo {pages} {020501}
  (\bibinfo {year} {2017}{\natexlab{b}})}\BibitemShut {NoStop}%
\bibitem [{\citenamefont {Hoffman}\ \emph {et~al.}(2017)\citenamefont
  {Hoffman}, \citenamefont {Chevallier}, \citenamefont {Loss},\ and\
  \citenamefont {Klinovaja}}]{hoffman.cheviallier.17}%
  \BibitemOpen
  \bibfield  {author} {\bibinfo {author} {\bibfnamefont {S.}~\bibnamefont
  {Hoffman}}, \bibinfo {author} {\bibfnamefont {D.}~\bibnamefont {Chevallier}},
  \bibinfo {author} {\bibfnamefont {D.}~\bibnamefont {Loss}}, \ and\ \bibinfo
  {author} {\bibfnamefont {J.}~\bibnamefont {Klinovaja}},\ }\bibfield  {title}
  {\enquote {\bibinfo {title} {Spin-dependent coupling between quantum dots and
  topological quantum wires},}\ }\href {\doibase 10.1103/PhysRevB.96.045440}
  {\bibfield  {journal} {\bibinfo  {journal} {Phys. Rev. B}\ }\textbf {\bibinfo
  {volume} {96}},\ \bibinfo {pages} {045440} (\bibinfo {year}
  {2017})}\BibitemShut {NoStop}%
\bibitem [{\citenamefont {Guessi}\ \emph {et~al.}(2017)\citenamefont {Guessi},
  \citenamefont {Dessotti}, \citenamefont {Marques}, \citenamefont {Ricco},
  \citenamefont {Pereira}, \citenamefont {Menegasso}, \citenamefont
  {de~Souza},\ and\ \citenamefont {Seridonio}}]{guessi.dessotti.17}%
  \BibitemOpen
  \bibfield  {author} {\bibinfo {author} {\bibfnamefont {L.~H.}\ \bibnamefont
  {Guessi}}, \bibinfo {author} {\bibfnamefont {F.~A.}\ \bibnamefont
  {Dessotti}}, \bibinfo {author} {\bibfnamefont {Y.}~\bibnamefont {Marques}},
  \bibinfo {author} {\bibfnamefont {L.~S.}\ \bibnamefont {Ricco}}, \bibinfo
  {author} {\bibfnamefont {G.~M.}\ \bibnamefont {Pereira}}, \bibinfo {author}
  {\bibfnamefont {P.}~\bibnamefont {Menegasso}}, \bibinfo {author}
  {\bibfnamefont {M.}~\bibnamefont {de~Souza}}, \ and\ \bibinfo {author}
  {\bibfnamefont {A.~C.}\ \bibnamefont {Seridonio}},\ }\bibfield  {title}
  {\enquote {\bibinfo {title} {Encrypting {Majorana} fermion qubits as bound
  states in the continuum},}\ }\href {\doibase 10.1103/PhysRevB.96.041114}
  {\bibfield  {journal} {\bibinfo  {journal} {Phys. Rev. B}\ }\textbf {\bibinfo
  {volume} {96}},\ \bibinfo {pages} {041114} (\bibinfo {year}
  {2017})}\BibitemShut {NoStop}%
\bibitem [{\citenamefont {Hoffman}\ \emph {et~al.}(2016)\citenamefont
  {Hoffman}, \citenamefont {Schrade}, \citenamefont {Klinovaja},\ and\
  \citenamefont {Loss}}]{hoffman.schrade.2016}%
  \BibitemOpen
  \bibfield  {author} {\bibinfo {author} {\bibfnamefont {S.}~\bibnamefont
  {Hoffman}}, \bibinfo {author} {\bibfnamefont {C.}~\bibnamefont {Schrade}},
  \bibinfo {author} {\bibfnamefont {J.}~\bibnamefont {Klinovaja}}, \ and\
  \bibinfo {author} {\bibfnamefont {D.}~\bibnamefont {Loss}},\ }\bibfield
  {title} {\enquote {\bibinfo {title} {Universal quantum computation with
  hybrid spin-majorana qubits},}\ }\href {\doibase 10.1103/PhysRevB.94.045316}
  {\bibfield  {journal} {\bibinfo  {journal} {Phys. Rev. B}\ }\textbf {\bibinfo
  {volume} {94}},\ \bibinfo {pages} {045316} (\bibinfo {year}
  {2016})}\BibitemShut {NoStop}%
\bibitem [{\citenamefont {Chevallier}\ \emph {et~al.}(2017)\citenamefont
  {Chevallier}, \citenamefont {Szumniak}, \citenamefont {Hoffman},
  \citenamefont {Loss},\ and\ \citenamefont
  {Klinovaja}}]{chevallier.szumniak.17}%
  \BibitemOpen
  \bibfield  {author} {\bibinfo {author} {\bibfnamefont {D.}~\bibnamefont
  {Chevallier}}, \bibinfo {author} {\bibfnamefont {P.}~\bibnamefont
  {Szumniak}}, \bibinfo {author} {\bibfnamefont {S.}~\bibnamefont {Hoffman}},
  \bibinfo {author} {\bibfnamefont {D.}~\bibnamefont {Loss}}, \ and\ \bibinfo
  {author} {\bibfnamefont {J.}~\bibnamefont {Klinovaja}},\ }\href@noop {}
  {\enquote {\bibinfo {title} {Topological phase detection in {Rashba}
  nanowires with a quantum dot},}\ } (\bibinfo {year} {2017}),\ \Eprint
  {http://arxiv.org/abs/arXiv:1710.05576} {arXiv:1710.05576} \BibitemShut
  {NoStop}%
\end{thebibliography}%

\end{document}